\documentclass[12pt, preprint]{aastex}

\usepackage{natbib}
\usepackage{apjfonts}

\slugcomment{VERSION: 07 March 2002; Accepted to ApJ}

\shortauthors{Muench et al.}
\shorttitle{On the Trapezium Cluster IMF}

\newcommand{\solarmass}{\ensuremath{ \mbox{M}_{\sun} }}
\newcommand{\jupmass}{\ensuremath{ \mbox{M}_{Jup} }}
\newcommand{\av}{\ensuremath{ \mbox{A}_{V} }}
\newcommand{\Ks}{\ensuremath{ \mbox{K}_{\mbox{s}} }}
\newcommand{\mavlim}{ \ensuremath{ \mbox{M}-\av }}
\newcommand{\cotracer}{\ensuremath{ \mbox{C}^{18}\mbox{O} }}

\begin{document}

\title{The Luminosity and Mass Function of the Trapezium Cluster:\\
From B Stars to the Deuterium Burning Limit.}

\author{
August A. Muench\altaffilmark{} and 
Elizabeth A. Lada\altaffilmark{} 
}
\affil{Department of Astronomy, University of Florida,
  Gainesville, FL 32611}
\email{muench@astro.ufl.edu, lada@astro.ufl.edu}
 
\author{Charles J. Lada\altaffilmark{1} }
\affil{Harvard-Smithsonian Center for Astrophysics,
  Cambridge, MA 02138}
\email{clada@cfa.harvard.edu}
 
\and
 
\author{Jo\~{a}o Alves\altaffilmark{1} }
\affil{European Southern Observatory, Karl-Schwartzschild-Strasse 2, 85748
Garching Germany}
\email{jalves@eso.org}

\altaffiltext{1}{Visiting Astronomer, European Southern Observatory}

\begin{abstract}

We use the results of a new, multi-epoch, multi-wavelength, near-infrared census
of the Trapezium Cluster in Orion to construct and to analyze the structure of
its infrared (K band) luminosity function. Specifically, we employ an improved 
set of model luminosity functions to derive this cluster's underlying
Initial Mass Function (IMF) across the entire range of mass from OB stars
to sub-stellar objects down to near the deuterium burning limit.  We derive an IMF
for the Trapezium Cluster that rises with decreasing mass, having a Salpeter-like IMF
slope until near $\sim\,0.6\,\solarmass$ where the IMF flattens and forms a broad peak 
extending to the hydrogen burning limit, below which the IMF declines into the sub-stellar
regime.  Independent of the details, we find that sub-stellar objects account for no more
than $\sim\,22\%$ of the total number of likely cluster members.  Further, the sub-stellar 
Trapezium IMF breaks from a steady power-law decline and forms a significant
secondary peak at the lowest masses (10-20 times the mass of Jupiter).
This secondary peak may contain as many as $\sim\,30\%$ of the sub-stellar objects
in the cluster.  Below this sub-stellar IMF peak, our KLF modeling requires a subsequent
sharp decline toward the planetary mass regime. 
Lastly, we investigate the robustness of pre-main sequence luminosity evolution
as predicted by current evolutionary models, and we discuss possible origins for
the IMF of brown dwarfs.
\end{abstract}


\section{Introduction}
\label{sec:intro}

Little is known about the similarities or differences between the star formation process that 
created the first generation of stars in the universe and the process that is forming stars in nearby 
stellar nurseries today.  One important diagnostic for studying any evolution of the star formation 
process is the statistical distribution of stellar masses at birth, or the stellar initial mass 
function (IMF). Comparisons of the mass functions of stars in globular clusters,  in the field, 
in  intermediate age open clusters such as the Pleiades and in extremely young clusters embedded 
in  nearby molecular clouds might reveal similarities or differences that would test the 
notion of a universal mass function (or a dominant star formation process) or that could 
bring about a better understanding of its stochastic nature \citep[see for example, ][ for recent 
discussions]{pk01, gil01}.   Because  the initial mass function is an intrinsically
statistical quantity, such comparisons require numerous measurements of the IMF in a variety 
of environments, in turn, requiring tools that can probe the IMF over a large volume of space 
and time.

Very young embedded clusters are particularly valuable for IMF studies. A simple photometric 
census of the members of a young embedded cluster yields a statistically significant population 
of stars sharing a common heritage (e.g., age and metallicity). 
Further such a census is relatively complete because young
clusters have not lost significant numbers of members to either dynamical or stellar evolution; 
hence, the observed mass function is the cluster's {\em initial} mass function.  
Because very young clusters are still embedded within their parental molecular cloud, an infrared
photometric census is often necessary to identify a complete cluster population.
One direct product of such an IR census is the young cluster's stellar infrared luminosity function,
which can be used as a tool for studying a cluster's IMF.
This may be a particularly effective tool for studying the low mass end of a cluster's 
IMF because infrared luminosities are relatively easy to derive for brown dwarf cluster 
members  since such  intrinsically red sub-stellar sources are at brighter luminosities
than at any subsequent point in their evolution. Further, modern infrared cameras on modest sized
telescopes can efficiently survey numerous young clusters, deriving infrared luminosities
for complete populations, and, thus, potentially sampling the IMF of the current epoch over
a relatively large volume of the local galaxy.

In our initial work, \citet[hereafter MLL2000]{mll00}, we  performed a series  of numerical  
experiments aimed at testing the usefulness of infrared luminosity functions for deriving
cluster IMFs. By varying the different fundamental physical properties of a very young cluster, 
e.g., the cluster's  underlying initial mass function (IMF),  the cluster's star forming history 
(SFH) and the appropriate (fixed metallicity) mass-luminosity relation (M-L relation) for cluster
members, we showed that a  cluster's luminosity function is considerably more sensitive to the
form of the underlying  initial mass function than to any other parameter. In particular, we found 
that differences between various theoretical pre-main sequence evolutionary calculations
produced only negligible differences in the resulting model luminosity functions and
that these differences would not likely be observable.  As we show here, this is the result of the 
fact that canonical PMS luminosity evolution is a relatively robust prediction of current PMS models.

In MLL2000, we applied these results to the infrared (K band) luminosity function (KLF)
of the famous Trapezium Cluster in Orion. It is the richest young cluster within the
local kiloparsec \citep{lah97}; therefore, it provides a nearby statistically significant sample
of sources for which infrared luminosities can be derived, including even the faintest planetary mass 
members. This is an ideal laboratory for applying and for testing our model luminosity function
method.  Using the cluster's mean age and age spread taken from spectroscopic observations of 
optically visible cluster members, we used our method to derive the Trapezium Cluster IMF from its
KLF.  In MLL2000, we found the cluster's IMF rose steeply in  number with decreasing mass until just 
less  than one solar mass  where the IMF then flattened and peaked near the hydrogen 
burning limit before falling  throughout the sub-stellar regime with a brown dwarf IMF
slope\footnote{In all cases we will refer to the mass function not the mass spectrum
     of stars. The mass function is the number of stars per unit volume per unit {\it log mass}
     The \cite{sal55} mass function would be a power-law 
     $\xi(\:\log(m_{\star})\:)\:=\:m^{\Gamma}$ with an index $\Gamma\:=\: -1.35$ 
     in this definition.} 
of $\:\sim\:+1$.
We also found good agreement between the form of the Trapezium IMF derived from its KLF 
and that IMF derived by \citet{lah97} for this cluster using a spectroscopic analysis of the 
optically visible members. Further, luminosity function modeling enabled the derivation of the 
sub-stellar IMF, which was not possible from the optical/spectroscopic analysis.

In addition to the three fundamental quantities (age, IMF, M-L relation) that govern the structure 
of a young cluster's  luminosity function, there are a number of observational characteristics 
that may affect the conversion of a cluster's infrared luminosity function into the cluster's IMF.
In particular, very young stars are reddened due to extinction from the parental  molecular 
cloud and from excess infrared emission arising from a circumstellar disk.  Cluster membership 
is often poorly  known, requiring a statistical estimate of the interloping field star population.
We did not include these observational effects into our MLL2000 derivation of the Trapezium 
IMF  because we had access to only a monochromatic KLF taken from the literature with no color or
completeness information.  In this current paper, we account for these observational effects by
employing a new multi-wavelength near-infrared census of the Trapezium Cluster 
that we have constructed over a three year period of  observations. 
Using these deeper, more complete observations, we study the contribution of background field
stars to the Trapezium Cluster's near-infrared luminosity function, expand our model luminosity
function efforts to account for source reddening, and explore the Trapezium KLF at fainter
magnitudes than was possible in MLL2000.  We re-derive the Trapezium Cluster IMF from its luminosity 
function, providing a detailed range of possible IMF parameters.
We discuss the relative degeneracy of the theoretical mass-luminosity relations derived
from different sets of evolutionary models, and we compare the results of several published
derivations of the Trapezium IMF that rely upon infrared photometry.
Lastly, we discuss possible origins for a derived feature in the
Trapezium sub-stellar IMF near the planetary mass regime.

\section{The Trapezium Cluster K band Luminosity Function}
\label{sec:lf}

\subsection{The Near-Infrared Census}
\label{sec:lf:census}

To derive a complete multi-wavelength census of the sources in the Trapezium Cluster,
we performed infrared observations during 1997 December,  1998 November and 2000 March 
using two telescopes:  the 1.2m telescope at the Fred  Lawrence Whipple Observatory (FLWO) 
at Mt. Hopkins, Arizona (USA) and the ESO 3.5m New Technology Telescope (NTT) 
in La Silla, Chile.  These observations yielded the multi-epoch, multi-wavelength FLWO-NTT 
infrared catalog that contains $\sim\;1000$ sources (see  Appendix \ref{app:catalog} for 
catalog  details). Subsets of this  catalog have been previously used in the \citet{lada00} and 
\citet{aam01} studies of the frequency of  circumstellar disks around stars and brown dwarfs in 
the Trapezium Cluster.

\placefigure{fig:trap_color}

We restrict our subsequent analysis of the cluster's luminosity and mass function to the 
area surveyed by our deeper NTT observations. We display a JH\Ks~color composite image
of the $\:5\arcmin\:\times\:5\arcmin\:$ NTT region in figure \ref{fig:trap_color}.
Our observations detected  749 sources within this region. The completeness 
of this  sample at the faintest magnitudes is difficult to quantify because of the spatially variable 
nebular background.   The formal $10\,\sigma$ detection limits of our NTT images are  19.75  at 
J, 18.75 at H   and 18.10 at \Ks~based  upon the pixel to pixel noise in non-nebulous off-cluster 
observations  that were taken adjacent in  time to the on-cluster  images.  
To estimate our actual completeness limits, we  performed artificial star 
experiments by constructing a stellar PSF for each of  our images and using  the IRAF 
ADDSTAR  routine to place synthetic stars in both the off-cluster and the nebulous on-cluster 
images. A small number of synthetic stars (30-70) with a range of input 
magnitudes were randomly added  across each image and were then recovered using the  
DAOFIND routine.  This  was repeated a large number of times (40-200) to achieve sufficient 
signal to noise for these  tests.  In off-cluster images, the derived 90\% completeness limits
agreed well with the estimated $10\,\sigma$ detection limits.
In the on-cluster images, the completeness limits were reduced to 90\% completeness limits of 
$\mbox{J}~ \sim\,18.15$, $\mbox{H}~ \sim\,17.8$, and $\Ks~ \sim\,17.5$ with slightly brighter 
limits in the dense central core (0.5\arcmin~radius from $\theta^{1}\mbox{C}$ Orionis). 
We also carefully compared our source list to those published by other recent surveys for the
NTT region. To our resolution limit, we detected all the sources found by the \citet[][ hereafter, 
HC2000]{hc2000} Keck survey except for one, all but two sources from the \citet{luh2000} NICMOS 
survey, but  we could not identify nine sources listed in \citet[][ hereafter, LR2000]{lr2000} UKIRT 
survey.  Further, it is our finding of  58  new sources within our NTT region and unreported  by 
prior catalogs that adds support to the deep and very sensitive nature of our census.

\subsection{Constructing Infrared Luminosity Function(s)}
\label{sec:lf:construct}

The FLWO and NTT observations overlap considerably in dynamic range with 504 stars having 
multi-epoch photometry.  For our analysis, we preferentially adopt infrared luminosities 
from the NTT photometry because it has higher angular resolution and it is an essentially 
simultaneous  set of  near-infrared  data. For 123 stars that are saturated in one or more bands 
on  the  NTT images, the  FLWO  photometry was  used.  This transition from NTT to FLWO 
photometry  is at  approximately J = 11.5, H = 11.0, and K = 11.0.  For the brightest 5 OB stars, 
saturated on all  our  images,  we used  JHK photometry from  the \citet{lah98} catalog.
Photometric differences between the FLWO and NTT datasets are small (see Appendix 
\ref{app:catalog})  and will not affect our construction of the Trapezium Cluster infrared
 luminosity function(s).

In Figure \ref{fig:jhklf}, we present the raw infrared Trapezium Luminosity Functions (LFs).
We use relatively wide bins (0.5 magnitudes) that are much larger than the photometric errors
(see Appendix \ref{app:catalog:phot_comp}).   In Figure \ref{fig:jhklf}(a), we compare 
the J and H band LFs for stars in  this region.  In the Figure \ref{fig:jhklf}(b), we compare
the K band LF of the NTT region to that K band LF  constructed in MLL2000.  As was observed in
previous studies of the Trapezium Cluster, the cluster's infrared luminosity function  (J, H, or K)
rises steeply toward  fainter magnitudes,  before flattening and forming a broad peak. 
The LF steadily declines in number below this peak but then rapidly tails off a full 
magnitude above our completeness limits.

\placefigure{fig:jhklf}

For our current derivation of the Trapezium IMF, we use the Trapezium K band Luminosity Function,
rather than the J or H LFs. We do so in order to minimize the effects of extinction
on the luminosities of cluster members (see section \ref{sec:imf:recipe_av}), to maximize our 
sensitivity to intrinsically red, low luminosity brown  dwarf members of  this cluster, and to make 
detailed comparisons to our work in MLL2000.  For example, the new FLWO-NTT Trapezium 
KLF contains significantly more stars at faint  ($\mbox{K}\,>\,14$) magnitudes than the KLF 
used in MLL2000.  Interestingly, a secondary peak near K = 15 seen in the MLL2000 KLF 
\citep[originally ][]{mcc95} is much more significant and peaks at K=15.5 in the new 
FLWO-NTT KLF. Similar peaks are not apparent in the J or H band LFs constructed here, 
though \citet{lr2000} report a strong secondary peak in their Trapezium JLF. Such secondary 
peaks in young cluster luminosity  functions have often been evidence of a  background field star 
population contributing to the source counts  \citep[e.g., ][]{luh98, luh99a}.  

\placefigure{fig:off_klf}

To account for the possible field star contamination, we systematically obtained images of
control fields away from the cluster and off of the Orion Molecular Cloud. 
The FLWO off-cluster field(s) were centered at approximately  R.A. = $05^{h}26^{m}$;
DEC. = $-06\degr00\arcmin$ (J2000) and 
were roughly twice the area of the NTT off-fields.  The NTT off-cluster region was 
centered at  R.A. = $05^{h}37^{m}43\fs7$; DEC.= $-01\degr55\arcmin42\farcs7$ (J2000).  
Figure \ref{fig:off_klf}(a) displays the two off-field KLFs  (scaled to the same area) from 
these observations and in the inset,  their (H - K) distributions. The relatively narrow (H - K) 
distributions indicate that the two off-fields sample similar populations and that they are 
un-reddened.  We constructed an observed field star KLF by averaging these luminosity 
functions,  weighting (by area) toward the FLWO off-fields for K brighter than 16th magnitude, the
completeness limit of that dataset, and toward the more  sensitive NTT off-fields for fainter
than K = 16.  In Figure \ref{fig:off_klf}(b), we compare the  resulting  field star KLF to the
Trapezium KLF of the  NTT region. It is plainly apparent from the raw control field observations
that while field stars may contribute to the Trapezium Cluster IR luminosity function over a range 
of magnitudes, their numbers peak at magnitudes fainter than the secondary peak of the on-cluster
KLF and do not appear sufficient in number to explain it.

\subsection{Defining a Complete Cluster KLF}
\label{sec:lf:mavlim}

We determine the completeness of our FLWO-NTT Trapezium Cluster KLF by constructing and by
analyzing the cluster's infrared  (H - K) versus K  color-magnitude diagram.
For the  purposes of our analysis, we adopt the same cluster parameters as we used in MLL2000,  
e.g., a  cluster mean age of 0.8 Myrs \citep{lah97} and a cluster distance of $400\,\mbox{pc}$.
As seen in figure \ref{fig:mavlim}(a), the luminosities of the Trapezium sources form a continuously
populated  sequence from the bright OB members ($\mbox{K}\,\sim\,5$) through sources detected below
our completeness limits.

\placefigure{fig:mavlim}

To interpret this diagram,  we compare the locations of the FLWO-NTT sources
in color-magnitude space to the cluster's mean age isochrone as derived from
theoretical pre-main sequence (PMS) calculations.  Because the \citet[][ hereafter, DM97]{dm97}
models include masses and ages representative of the Trapezium Cluster
and inorder to provide continuity with our work in MLL2000, we will use
these tracks to define a complete cluster sample from figure \ref{fig:mavlim}(a).
Differences among pre-main sequence tracks should not have significant effect upon
our analysis of the color-magnitude diagram (see section \ref{sec:discuss:comp_trap}). 
It is clear from this diagram that the cluster sources are reddened 
away from the theoretical 0.8 Myr isochrone, which forms a satisfactory left hand boundary 
to the sources in this color-magnitude space. This isochrone, however, does not span the 
full luminosity range of the observations and a number ($\sim~40$) sources lie below
the faint end of the DM97 isochrone.  As a result, our subsequent analysis that makes use
of the DM97 models will be restricted to considering only those sources whose luminosities,
after correction for extinction, would correspond to masses greater than the mass limit of the
DM97 tracks, i.e., 0.017~\solarmass~or roughly 17 times the mass of Jupiter ($\jupmass$).
Despite the lower mass limit imposed by these PMS tracks, our infrared census spans nearly three 
orders of  magnitude in mass, illustrating the  utility of studying the mass function of such
rich young  clusters.

Extinction acts to redden and to dim sources of a given mass to a brightness below our detection
limits.  To determine our ability to detect extincted stars as a function of mass, we draw a
reddening vector from the luminosity (and color) of a particular mass star on the mean age isochrone
until it intersects the $10\sigma$ sensitivity limit of our census.  We can detect the 1 Myr old
Sun seen through  $\av\;\sim\;60$ magnitudes of extinction or a PMS star
at the hydrogen burning limit seen  through $\sim\;35$ magnitudes. For very young brown dwarfs at
our lower mass limit ($17\;\jupmass$), we probe the cloud to $\av\,=\,17$ magnitudes. 
We use this latter reddening vector as a  boundary to which we are  complete in mass,
and we draw a mass and extinction ($\mavlim$) limited subset  of sources bounded by the mean age
isochrone and the $\av\,=\,17$ 
reddening  vector and mark  these as filled circles in  figure  \ref{fig:mavlim}(a).  Our 
$\mavlim$ limits probe the vast majority of the cluster population, including 
81\% of the sources the color-magnitude diagram.

In figure \ref{fig:mavlim}(b) we present the  $\mavlim$ limited KLF, containing 583 sources. 
Thirty-two sources, detected only at K band (representing only 4\% of our catalog), were also
excluded from our further analysis.  The median K magnitude of these sources is K = 15, and 
we expect that these are likely heavily reddened objects.  We compare the $\mavlim$~limited KLF
to the unfiltered Trapezium KLF.  Clearly, heavily reddened sources contributed to  the cluster 
KLF at all magnitudes and their removal results in a narrower cluster KLF. 
However, the structure  (e.g., peak, slope, inflections,
etc.)  of the KLF remains largely unchanged.  The secondary peak of the cluster  KLF between 
$\mbox{K}\,\sim\,14\,-\,17$ seems to be real since it is present in both  the raw and the 
$\mavlim$ limited KLFs, though we have not yet corrected for background field stars.

There are at least three possible sources of incompleteness in our mass/extinction limited sample.
The first arises because sources that are formally within our mass and extinction limits may 
be  additionally reddened by  infrared excess from circumstellar disks and, hence, be left out 
of our  analysis.  However, this bias will affect sources of all masses equally because infrared 
excess appears to be a property of  the young Trapezium sources over the entire luminosity range 
\citep{aam01}.  Second, the Trapezium Cluster is not fully coeval and our use of  the cluster's 
mean age to draw the $\mavlim$ sample means that cluster members at our lower mass limit 
($17\,\jupmass$) but older than the cluster's mean age ($\tau\,>\,0.8\,\mbox{Myrs}$) will be 
fainter than the lower boundary and left out of our sample. Further, sources younger than
the mean age but below $17\,\jupmass$~will be included into the sample. This ``age bias'' will
affect the lowest mass sources, i.e., $<\;20\,\jupmass$.  Third, because of the strong nebular
background, our true completeness limit (see section \ref{sec:lf:census}) is brighter than our
formal $10\sigma$ sensitivity for approximately 60\% of the area surveyed.
The resulting sample incompleteness only affects our sensitivity to sources
less than $30\,\jupmass$ and with $\av\,>\,10$.  We do not correct the Trapezium KLF to
account for these effects or biases.

\subsection{Field Star Contamination to the KLF}
\label{sec:lf:fieldstar}

The lack of specific membership criteria for the embedded sources in the Trapezium Cluster 
requires an estimate of the number of interloping non-cluster field stars in our sample.
Some published studies, for example LR2000 and \citet{luh2000}, assume that the 
parental molecular  cloud acts as a shield to background field stars; whereas 
HC2000 suggests that the background contribution is non-negligible. 
HC2000 estimates the field star contribution using an empirical model of the infrared field
star population and convolving this model with a local extinction map derived from a molecular
line map of the region. 
This approach may suffer from its dependence upon a field star model that is not
calibrated to these faint magnitudes and that does not include very low mass field stars.
As we show, there are also considerable uncertainties in the conversion 
of a molecular line map to an extinction map.  For our current study, we use our observed K band 
field star luminosity function (see figure  \ref{fig:off_klf}) to test these prior methodologies
and to correct for the field star contamination.  We point out that no such estimate can account
for contamination due to young, low mass members of the foreground Orion OB1 association.

\placefigure{fig:off-fraction}

We compare in figure \ref{fig:off-fraction} the effects of six different extinction models upon our 
observed field star KLF.  In panels A and B, we tested simple Gaussian distributions of extinction
centered respectively at $\av\,=\,10\;\mbox{and}\;25$ magnitudes with $\sigma\:=\:5\:$
magnitudes. In both cases, the reddened field star KLF contains significant counts above our 
completeness limit  and ``background extinction shields'' such as these do not prevent the 
infiltration of field stars  into our counts. In the second pair of reddened off-fields (panels C 
and D), we followed the HC2000 prescription for estimating background field stars by convolving
our observed field star KLF with the $\cotracer$ map from \citet{gold97} converted from column
density to dust extinction. We note that there is substantial uncertainty in the conversion from
$\cotracer$ column density to dust extinction.  There is at least a factor of 2 variation in this
conversion value in  the literature, where \citet{frer82} derived a range from  
$0.7\:-\:2.4\:(\mbox{in units of}\;10^{14}\;\mbox{cm}^{2}\,\mbox{mag}^{-1})$
and \citet{gold97} estimated a range of values from $1.7\:-\:3$. Either the result of measurement
uncertainty or the product of different environmental conditions,  this variation produces
a factor of 2  uncertainty in the extinction estimates from the $\cotracer$ map.
In short, we find that a $\cotracer\:\rightarrow\:\av$ ratio of 3.0 (panel C) results in 
twice as many interloping background field stars as  would a  value of 1.7 (panel D; equivalent 
to that used by HC2000). 

In panels E and F of figure \ref{fig:off-fraction} we derive the same reddened off-field KLFs
as in the prior pair, but they have been filtered to estimate the actual contribution of 
field-stars to our $\mavlim$ limited sample.
These filters, which were based upon on the K brightness of the lower mass limit
of our PMS models and on the derived extinction limit of the $\mavlim$~sample,
were applied during the convolution of the field star KLF with the cloud extinction
model such that only reddened field stars that would have $\av\:<20$ and unreddened K
magnitudes $\:<\:16$ would be counted into filtered reddened off-field KLF.  
The extinction limit was expanded from 17 to 20 magnitudes to account for the
dispersion of the H-K distribution of un-reddened field-stars ($\sim0.2$).
A factor of 2 uncertainty remains.
\citet{alv99} derive a more consistent estimate of the  $\cotracer\:\rightarrow\:\av$ ratio 
from near-infrared extinction mapping of dark clouds,  suggesting a median ratio 
of $\,2.1$.  Adopting $\cotracer\:\rightarrow\:\av\,=\,2.1$,  we estimate there are 
$\sim\;20\;\pm\,10$ field stars in our $\mavlim$ limited KLF. From these experiments,
we find, however, that both the raw and reddened off-field KLFs always peak at 
{\em fainter magnitudes} than the secondary peak of the Trapezium KLF, and that the subtraction
of these field-star corrections from the Trapezium KLF do not remove this secondary peak. 
These findings suggest that the secondary KLF peak is a real feature in the Trapezium Cluster's
infrared luminosity function.

\section{The Trapezium Cluster Initial Mass Function}
\label{sec:imf}

We analyze the Trapezium Cluster's K band luminosity function constructed in section 
\ref{sec:lf} using our model luminosity function algorithm first described in MLL2000.   
Following our work in MLL2000, our goal is to derive the underlying mass function or set
of mass functions  whose  model luminosity functions best fit the Trapezium Cluster KLF.  
We have improved our modeling algorithm by including statistical distributions
of the reddening properties of the cluster. We have also improved our analysis by
applying the background field star correction from section \ref{sec:lf:fieldstar} and 
by employing improved fitting  techniques for deriving IMF parameters and confidence intervals. 
Before deriving the cluster IMF, we use the extensive color information available from the
FLWO-NTT catalog to explore the reddening (extinction and infrared excess) properties of the 
Trapezium sources.  In section \ref{sec:imf:reddening}, we use this  information to create recipes 
for deriving the probability distributions functions of extinction and excess which can be folded 
back into our modeling algorithm during our derivation of the Trapezium IMF.  We present the 
new model luminosity  functions and fitting techniques in section \ref{sec:imf:model} and 
summarize the derived IMF in section \ref{sec:imf:summary}.

\placefigure{fig:hk_jhk}

\subsection{Deriving Distributions of Reddening}
\label{sec:imf:reddening}

\subsubsection{Extinction Probability Distribution Function}
\label{sec:imf:recipe_av}

We use the extensive color information provided by our FLWO-NTT catalog  to construct a 
probability distribution function of the intra-cluster extinctions (hereafter referred to
as the Extinction Probability Distribution Function or EPDF) based upon the color excesses of 
individual Trapezium sources.  Because the stellar  photospheric (H - K) color has a  very narrow 
distribution of intrinsic photospheric values it should be the  ideal color from which to derive
line of sight extinction estimates, as shown, for example, in the \citet{alv98} study of the
structure of molecular clouds.  In figure \ref{fig:hk_jhk}(a) we show the histogram of observed
(H - K) color for all our Trapezium Cluster  sources. This histogram peaks at (H - K) = 0.5 and is 
quite broad especially when compared to the narrow unreddened photospheric (field-star) (H - K) 
distributions seen in figure \ref{fig:off_klf}(a). This broad distribution may be in part the 
result of extinction; however, as recently shown in  \citet{lada00} and \citet{aam01}, 
approximately 50\% of the these Trapezium Cluster sources, independent of luminosity, display 
infrared excess indicative of emission from circumstellar disks. This is illustrated in figure
\ref{fig:hk_jhk}(b) where it  is clear that there are both heavily reddened sources 
($\av\;\sim\;35$) and sources with large infrared excesses (falling to the right of the 
reddening band for main sequence objects).   If  the (H - K) color excess were assumed  to be 
produced by extinction alone without accounting for disk emission, the resulting extinction 
estimates would be too large.

\citet{mch97} showed that the intrinsic infrared colors of star with disks are confined to a
locus (the classical T-Tauri star locus or CTTS locus) in the (H - K)/(J - H) color-color diagram.  
We derive individual $\av$ estimates for sources in the (H - K)/(J - H) color-color diagram by
dereddening these stars back to this CTTS locus along a reddening vector defined by the
\citet{coh81} reddening law.  Sources without J  comprise $\sim20\%$ of the catalog and as shown
in figure \ref{fig:hk_jhk}, their (H - K) colors appear to sample a more heavily embedded population, 
implying  extinctions as high as  $\av\:\sim\:60$.  $\av$ estimates are derived for these sources
by  assigning a typical star-disk (H - K) color = 0.5 magnitudes, and de-reddening that 
source. Sources near to but below the CTTS locus are assigned an $\av\:=\:0$. The individual 
extinctions are binned into an extinction probability distribution  function (EPDF) as shown in 
figure \ref{fig:av_ixex}(a).  Also shown are the effects of changing the typical star-disk (H - K) 
color assumed for those stars without J band. Little change is seen. Compared to the cloud extinction
distribution  function, which was integrated over area from the $\cotracer$~map, the cluster 
EPDF is very non-gaussian and peaks at relatively low extinctions, $\av\:=\:2.5$, having a 
median $\av\:=\:4.75$ and a mean $\av\:=\:9.2$.  Further, the cluster EPDF is not well separated 
from  the reddening distribution provided by the molecular cloud.  Rather the cluster population 
significantly extends to extinctions as high as $\av\:=\:10\:-\:25$,  near and beyond the peak of 
the cloud extinction function.  Ancillary evidence of this significant  population of heavily 
reddened stars is seen in the  color-color diagram (figure  \ref{fig:hk_jhk}b) which clearly 
illustrates the extension  of the cluster to regions of  the molecular cloud with $\av\:>\:10$. 
Lastly, it is clear that the deep nature of our survey has allowed us to sample both the majority 
of the embedded cluster, and the cloud over the full range of density.

\placefigure{fig:av_ixex}

In our revised model luminosity function algorithm, we randomly sample the cluster's EPDF to 
assign an $\av$ to each artificial star in the model LF.  
The effect of the EPDF on the model luminosity function is wavelength and reddening 
law  \citep[ in this case]{coh81} dependent. In figure \ref{fig:lf_av} we construct model 
I, J, H, and K luminosity functions, reddening each by the Trapezium Cluster EPDF.
The effect of the EPDF on the intrinsic I and J band LFs is profound, rendering the 
reddened I band LF almost unrecognizable. Yet at longer wavelengths, specifically at K band,  
the effects of extinction are minimized. We note that the overall form of the reddened
model K band luminosity function has not been changed by the Trapezium EPDF in a significant way,
e.g., the peak of the model KLF is not significantly blurred and the faint slope of the KLF has not
been changed  from falling  to flat.  This suggests that our  modeling of the Trapezium KLF 
in MLL2000, which did not account for reddening due to extinction, is generally correct.
However, we likely derived too low of a turnover mass for the Trapezium IMF 
because reddening shifted the intrinsic LF to fainter magnitudes. 

\placefigure{fig:lf_av}

\subsubsection{Infrared Excess Probability Distribution Function}
\label{sec:imf:recipe_irex}

Because we wish to use the Trapezium K band LF to  minimize the effects of extinction,
we must also account for the effects of circumstellar disk emission at K band. 
The frequency distribution of the resulting excess infrared flux is not a well known quantity,
and when previously derived, it has depended significantly upon additional information derived
from the spectral classification of cluster members \citep{hc2000,lah98}.
One of the goals of this present work is to construct a recipe for deriving the
K band excess distribution directly from the infrared colors of the cluster members.

To derive a first-order infrared excess probability distribution function (IXPDF) for the  
Trapezium Cluster sources, we simply assume that any excess (H - K) color (above the photosphere, 
after removing the effects of extinction) reflects {\em an excess at K band alone}, realizing this
may underestimate the infrared excess of  individual sources.  We only use the sources having 
JHK measurements and lying above the CTTS locus in the color-color diagram. 
We remove the effects of extinction from each source's observed (H - K) color using the same 
method described above, i.e., dereddening back to the CTTS locus. However, the 
photospheric (H - K) color for each star cannot be discreetly removed from this data alone. 
The photospheric infrared colors of pre-main sequence stars appear to be mostly dwarf-like 
\citep{luh99b}, and therefore, we used the observed field star (H - K) distribution shown in figure  
\ref{fig:off_klf}(a) as a probability distribution of photospheric values. We derive the  IXPDF by 
binning the  de-reddened (H - K) colors into a probability function and then subtracting the 
distribution of  photospheric colors using a Monte Carlo integration.

The Trapezium Cluster IXPDF is shown in figure \ref{fig:av_ixex}(b). The IXPDF peaks near 
0.2 magnitudes with a mean = 0.37, a median = 0.31, and a maximum excess of $\sim\,2.0$ 
magnitudes. Probabilities of negative excesses were ignored. The IXPDF is similar to the (H - K) 
excess distribution shown in HC2000 and derived in \citet{lah98} yet extends to somewhat 
larger excess values. Each artificial star in our models is randomly assigned a K band 
excess (in magnitudes) drawn from the IXPDF.

\subsection{Modeling the Trapezium Cluster KLF}
\label{sec:imf:model}

To model the Trapezium Cluster KLF, we apply the appropriate field star correction derived in section 
\ref{sec:lf:fieldstar} to the $\mavlim$~limited KLF constructed in section \ref{sec:lf:mavlim}.
We  fix the Trapezium Cluster's star forming history and distance to be identical to that used
in MLL2000.  Specifically, these are a distance of  
400pc (m-M=8.0) and a star forming history characterized by constant star formation
from 1.4 to 0.2 Myrs ago, yielding a cluster mean age of 0.8 Myrs \citep{lah97} and an age 
spread of 1.2 Myrs.  We adopt our standard set of merged theoretical pre-main sequence tracks
from MLL2000\footnote{Our standard set of  theoretical tracks are described in Appendix B of MLL2000:
  they are a merger of evolutionary calculations including a theoretical Zero Age Main 
  Sequence (ZAMS) from \citet{sch92}, a set of intermediate mass (1-5 \solarmass)
  ``accretion-scenario'' PMS tracks from \citet{ber96}, and the low mass standard deuterium
  abundance PMS models from \citet{dm97} for masses from 1 to 0.017 \solarmass}.
Our merged standard set of tracks span a mass range from 60 to 0.017 \solarmass, allowing 
us to construct a continuous IMF within this range.
We incorporated the cluster's reddening distributions derived in section \ref{sec:imf:reddening} into 
our modeling algorithm and chose a functional form of the cluster IMF similar to that used in 
MLL2000; specifically, an IMF constructed of power-law segments, $\Gamma_{i}$ connected 
at break masses, $m_{j}$.  As in MLL2000, we found that an underlying 3 power-law IMF produced
model KLFs that fit the observations over most of the luminosity range, corresponding
to masses from 5 to 0.03~\solarmass. 
In section \ref{sec:imf:model:best}, we utilize our $\chi^{2}$ minimization
routine to identify those 3 power-law IMFs that best fit the observed KLF within this mass range, and
we estimate confidence intervals for these IMF parameters in section \ref{sec:imf:model:range}.
We find that the faint Trapezium brown dwarf KLF, corresponding to masses less than 0.03~\solarmass,
contains structure and a secondary peak that are not well fit by the 3 power-law IMF models.
In section \ref{sec:imf:model:bdpeak} we model this secondary KLF peak using a corresponding
break and secondary feature in the cluster brown dwarf IMF between 0.03 and 0.01~\solarmass.

\placefigure{fig:fit-best}

\subsubsection{Results of $\chi^{2}$ Fitting: Best Fit Three Power-Law IMFs}
\label{sec:imf:model:best}

Our $\chi^{2}$ minimization  procedure  calculates the reduced $\chi^{2}$ statistic and 
probability  for a particular model KLF fit to the Trapezium KLF over a range of magnitude bins.
Parameters for the underlying three power-law IMF are taken from the best fit model KLFs, and 
we fit both reddened and unreddened model KLFs. The 3 power-law IMF derived from these fits
is summarized in table \ref{tab:imf_pars}. We found that the results of our model fits were dependent
upon the dynamic range of K magnitude bins over which the models were minimized.
Specifically, we find that our results are very sensitive to the formation of a secondary peak in
the Trapezium KLF at K = 15.5, which remains despite the subtraction of the field star KLF.

We derive good model KLF fits $(\chi^{2}\:\mbox{prob}\:\sim\:1)$ when fitting between the
$\mbox{K}\,=\,7.5$ bin and the $\mbox{K}\,=\,14.5$ bin (see figure \ref{fig:fit-best}a), the 
same luminosity range we modeled in MLL2000.  Within the this fit range we find an optimal 
Trapezium IMF nearly identical to that found in  MLL2000 even after accounting for reddening.
The derived IMF rises  steeply from the most  massive stars with $\Gamma_{1}\:=\:-1.3$ before 
breaking to a shallower  IMF slope  of  $\Gamma_{2}\:=\:-0.2$  at  $0.6\:\solarmass$ 
($\log\,m_{1}\:\sim\:-0.2$).   The derived IMF peaks  near the hydrogen burning  limit  
($0.10 - 0.08\:\solarmass$  or  $\log\,m_{2}\:=\:-1.0 - 1.1$) and  then breaks and falls  steeply 
throughout the brown dwarf   regime with  $\Gamma_{3}\:\sim\:+1.0$.  We also derive good fits 
to K=15 (just before the  secondary peak in the cluster KLF), with the resulting  IMF peaking at 
slightly higher masses  ($0.13\;-\;0.10\,\solarmass$) and falling with a slightly shallower slope, 
$\Gamma_{3}\:\sim\:+0.7\;\mbox{to}\;0.8$.   The unreddened luminosities of this fit range 
correspond to a mass range from 5.0 to 0.03~\solarmass.

However, we cannot produce models KLFs based upon a three power-law IMF that adequately fit
the secondary peak in the Trapezium KLF.  For example, our best fit to the secondary peak in
figure \ref{fig:fit-best}(b) is inconsistent with the overall form of the faint KLF, being unable to
replicate both the falling KLF at K = 14.5 nor the secondary peak at K = 15.5.
Such structure in the faint Trapezium KLF implies similar non-power law structure in the underlying
IMF, while our current models based upon a three power-law IMF essentially assign a single power-law IMF
slope for the entire brown dwarf regime. We will explore this structure in the faint brown dwarf
KLF and IMF in section \ref{sec:imf:model:bdpeak}, but first we examine the confidence intervals
for the derived 3 power-law IMFs.

\placefigure{fig:fit-range}

\subsubsection{Results of $\chi^{2}$ Fitting: Range of Permitted Three Power-Law IMFs}
\label{sec:imf:model:range}

Our $\chi^{2}$ fitting routine also allows us to investigate the range  of permitted cluster IMFs 
from modeling the cluster KLF. We illustrate the range of IMFs and the effects of source 
reddening on our fits in figure  \ref{fig:fit-range} and summarize the corresponding constraints
on the IMF parameters in table \ref{tab:imf_pars}.  In each panel, we plot the contours of 
$\chi^{2}$ probability for two of the 5 dependent IMF parameters while restricting the other 
three parameters to a best fit model. In each panel we also display contours for fits with (solid)
and without (dashed) source reddening, and we examine the dependence of these parameters for 
models fit to the K=14.5 and K=15.0 bins.

In all our fitting experiments (here and MLL2000), the high mass slope of the cluster IMF, 
$\Gamma_{1}$, was well constrained with  slopes measured between -1.0 and -1.3. Based on 
this result, we fix $\Gamma_{1}$ to equal -1.3. Panels (a) - (c) in figure \ref{fig:fit-range} 
display the  ranges of the other 4 IMF parameters when fitting to a $\mbox{K limit}\,=\,14.5$.
Panel (a) plots the dependence of the two break masses, $m_{1}\,\mbox{and}\,m_{2}$.
The fits for these parameters are well behaved with 90\% contours have a typical width of 0.1-0.2 
dex  in units of log mass.  Source reddening has two clear effects upon our fit results. When 
source  reddening is included, the high mass break, $m_{1}$, decreases and the low mass break, 
$m_{2}$, increases. The second effect is that the size of the 90\% confidence contour increases
when source reddening is included into the model fits. Panel (b) displays the dependence of
the low mass break, $m_{2}$, on the middle  power-law slope, $\Gamma_{2}$.
$\Gamma_{2}$ is fairly well constrained to be slightly rising to lower masses, and the 
permitted range  of  $m_{2}$ is again roughly 0.1 - 0.2 dex, centered near 
0.1 \solarmass~($\log{m}\,\sim\,-1$).  Accounting for source reddening again shifts the
low-mass break to slightly higher masses, increases the size of the 90\% 
contour, and in this case, flattens the central power-law. 

Panel (c) displays the dependence of $\Gamma_{3}$ upon the second break mass, $m_{2}$.
Though $m_{2}$ is fairly well constrained to have values between $0.13\;\mbox{and}\:0.08\:\solarmass$,
the low mass power-law slope, $\Gamma_{3}$, has a large range of possible slopes from 0.50 to
1.50 within the 90\% $\chi^{2}$ contour for models with source reddening.
Panel (d) plots the same parameters as panel (c) but for fits to the 
$\mbox{K limit}\,=\,15$.  These fits give somewhat flatter $\Gamma_{3}$ slopes
and somewhat higher mass $m_{2}$ breaks, but are actually slightly better constrained.
As discussed in the previous section, our model KLFs employing a 3 power-law IMF do not
provide good fits to the secondary peak in the KLF. As the fit range shifts to fainter
magnitudes, $\Gamma_{3}$ flattens, but the total $\chi^{2}$ confidence depreciates
due to the secondary peak. We explore the IMF parameters necessary to fit this secondary
peak in the next section.

\subsubsection{Fitting the Secondary Peak in the Trapezium Cluster KLF}
\label{sec:imf:model:bdpeak}

In contrast to our expectations in MLL2000, the departure from a power-law decline and the
formation of a secondary peak at the faint end of the Trapezium KLF remains after correcting for
reddened background field stars.  When we attempt to fit the faint KLF using an underlying three
power-law IMF, we find that our model KLFs, while producing excellent fits over the majority of the
Trapezium KLF, could not simultaneously reproduce the formation of the secondary peak. 
Since there is no known corresponding feature in the mass-luminosity relation
(see section \ref{sec:discuss:ml_comp}), we hypothesize that the KLF's break from a single
continuous declining slope at  $\mbox{K}\,>\,14.5$ ($\mbox{M}\;<\;30\,\jupmass$) and the formation of
a secondary KLF peak directly imply a similar break and feature in the cluster IMF.
Further, the rapid tailing off of the cluster KLF below this secondary peak also directly
implies a similar rapid decline or truncation in the underlying IMF, as was also
discussed in LR2000.

We modeled the secondary KLF peak by adding a fourth, truncated, power-law
segment, $\Gamma_{4}$, to the three power-law IMFs derived in section
\ref{sec:imf:model:best}. The truncation of the fourth power-law segment enabled
us to model the rapid tailing off of the cluster KLF below the secondary peak,
but was also dictated by the artificial low mass cut off present in the adopted merged
PMS tracks, which for the sub-stellar regime come from DM97.
As such, the truncation mass of the model IMF was arbitrarily set to 
0.017 \solarmass.  We found that this 4 power-law truncated IMF produced
good $\chi^{2}$ model KLF fits to the secondary KLF peak. The best fit model KLF
shown in figure \ref{fig:imf_derived}(a) has an underlying brown dwarf IMF that breaks
from a steady decline at $m_{3}\,=\,0.025\,\solarmass$ and then rises steeply with 
$\Gamma_{4}\,\sim\,-5$ before truncating at the lower mass limit.
Examination of the confidence intervals for the $m_{3}$ and $\Gamma_{4}$ parameters showed
that higher mass breaks ($m_{3}\,\sim\,0.035$) required flatter $\Gamma_{4}$ slopes, but they
had worse $\chi^{2}$ and peaked before the observed peak at K=15.5. 
This suggests that were the slope of the mass-luminosity relation continuous
(and constant) toward lower masses, the exact location of the secondary IMF peak would
shift to somewhat lower masses than we can derive using the truncated PMS tracks.

\placefigure{fig:imf_derived}

\subsection{Derived Trapezium Cluster IMF}
\label{sec:imf:summary}

\subsubsection{The Overall Cluster IMF}
\label{sec:imf:summary:all}

Figure \ref{fig:imf_derived}(b) shows our overall best fit Trapezium IMF and graphically displays the
range of cluster IMFs permitted by our modeling of the Trapezium KLF using our standard set of merged
PMS tracks. We adopt the following four power-law function with a truncation at the lowest masses
for the underlying IMF of the Trapezium:

\begin{equation} 
\frac{d\,N}{d\,\log{m}}\,=\,
\left\{ \begin{array}{l@{\quad;\quad}l}
M^\Gamma_\star & \Gamma\:=\:\left\{ \begin{array} {l@{\quad:\quad}l}
                                     -1.21 &\:M_\star\:>\:0.600\:\solarmass \\
                                     -0.15 &\:0.600\:\solarmass\:>\:M_\star\:>\:0.120\:\solarmass \\
                                     +0.73 &\:0.120\:\solarmass\:>\:M_\star\:>\:0.025\:\solarmass \\
                                     -5.00 &\:0.025\:\solarmass\:>\:M_\star\:>\:0.017\:\solarmass \\
                                    \end{array} \right. \\[1cm]
   0           & M_\star\:=\:0.017\:\solarmass
\end{array} \right. 
\label{eq:best_imf_dm97}
\end{equation}

We find that despite the use of deeper, more complete observations, the application of detailed mass
and extinction limits to the cluster sample, the inclusion of source reddening into the model
luminosity function algorithm and the correction of our infrared census for reddened field
stars, our derived Trapezium IMF is not a significant revision over that found in MLL2000. 
We find that  the inclusion of source reddening into our modeling algorithm, while
providing a more accurate  representation of the  cluster properties, results in cluster IMFs 
that have very similar power-law slopes and break masses as fits without source reddening,
especially when fit to the same luminosity range.
Source reddening does indeed blur the precision of the IMFs  we can derive. For example, our 
MLL2000 IMF derived without accounting for source reddening is somewhat broader and peaks 
to slightly  lower mass than the IMF derived here with source reddening. 
Though the overall derived IMF has not significantly changed from our findings in MLL2000,
our more complete infrared census and improved understanding of the field-star
population do allow us to explore the Trapezium IMF at lower masses than  MLL2000.
We find that the secondary peak of the observed Trapezium KLF is not the result of background
field stars, and we derive a corresponding secondary peak in the low mass brown dwarf IMF between
10 and 30~\jupmass. However, because of the restriction imposed by the low mass limit of the
PMS tracks, both the precise location and amplitude of the secondary peak and the precise form of the
IMF below $17\,\jupmass$~are somewhat uncertain. 

\subsubsection{The Sub-Stellar IMF, A Closer Look}
\label{sec:imf:summary:bur97}

To better define the secondary peak in the Trapezium IMF, we consider 
only the sub-stellar regime of the Trapezium KLF ($\mbox{K}\,\ge\,13$), 
where we can employ a different set of PMS tracks that cover the corresponding
brown dwarf regime but also extend to masses less than the limit of our standard
merged PMS tracks.  The \citet[][ hereafter, B97]{bur97} PMS models are available 
from 0.12 to 0.001 \solarmass~(1 \jupmass), and for the relevant age range 
of the Trapezium. While the mass to K luminosity relation is relatively 
robust between different sets of PMS tracks, the B97 PMS models do display a 
somewhat flatter mass-luminosity relation for sub-stellar objects than do DM97.  
We rederived the cluster's $\mavlim$~limited, background corrected Trapezium KLF
following our prescription in section \ref{sec:lf:mavlim} but using the B97 isochrone. 
Since the nebular background decreases our survey's completeness to heavily
reddened ($\av\,>\,10$), very low mass ($\mbox{M}\,<\,0.03\,\solarmass$) brown dwarfs,
we draw this $\mavlim$~sample to our completeness limit rather than our $10\sigma$
sensitivity limit to ensure the precision of the sub-stellar IMF.
The resulting $\mavlim$~limited KLF from the extended B97 isochrone 
samples the cluster population to a predicted mass limit of $0.01\,\solarmass$~and
an $\av\,=\,9$.

\placefigure{fig:bur97_fit}
\placetable{tab:ssimf_pars}

New model KLF fits (see figure \ref{fig:bur97_fit}a) that employ the B97 tracks
and use a three power-law underlying sub-stellar IMF yield a power-law brown dwarf
IMF falling with a similar but somewhat steeper slope than our standard tracks
(see summarized IMF parameters in table \ref{tab:ssimf_pars}) \footnote{
     Fiting the sub-stellar sample drawn to an $\av\,<\,17$ with model LFs using B97 tracks
     yielded a power-law decline closer to that derived from our standard merged models}.
Similar to DM97, the B97 tracks require the presence of a significant secondary peak
that departs from the power-law function at $m_{3}\,\sim\,0.02\,\solarmass$ and rises 
very steeply with $\Gamma_{4}\,=\,-5$ as shown in figure \ref{fig:bur97_fit}b.  
Further, the extended mass range of the B97 tracks allows us to resolve 
the location of the secondary peak: our fits require a peak near the 
deuterium burning limit, i.e., $13-14\,\jupmass$, followed by a rapidly 
declining IMF with a slope of $\Gamma_{5}\,=\,+5$ down to $10\,\jupmass$.  
The sharp decline in the sub-stellar IMF below this peak is {\em not} the
result of the application of the off-field correction. Fits to
Trapezium sub-stellar KLF without correcting for background field stars
yeild nearly identical cluster IMFs (see figure \ref{fig:bur97_fit}a and
table \ref{tab:ssimf_pars}).  Further, the sharp decline in our derived sub-stellar
IMF below the deuterium burning limit and independent of the background correction confirms
a similar, straightforward interpretation of the observed rapid turn down in
the cluster KLF and dearth of sources in the (H - K)/K color magnitude diagram
nearly a full magnitude above our completeness limits (see figure \ref{fig:mavlim}).  

\placefigure{fig:ssimf}

Lastly, this feature in the IMF appears to be a statistically significant
departure from the power-law decline of the brown dwarf IMF as was implied
by our $\chi^2$ fitting in section \ref{sec:imf:model:best}. We ran a Monte
Carlo simulation of the derived Trapezium sub-stellar IMF for a 
population of 150 brown dwarfs. In figure \ref{fig:ssimf} we show the
resulting histogram form of the average simulated cluster IMF. Using
equally sized bins in log mass units, we calculated the statistical variation
in an IMF bin as a function of 50 drawn samples. From these plotted $1\sigma$
error bars, it is clear that the derived secondary peak is a significant
statistical result. In addition, these results imply that a statistically
significant identification of such a feature at the tail of the IMF requires
the examination of a rich sub-stellar population such as that provided by
the Trapezium Cluster.

\section{Discussion} \label{sec:discuss}

\subsection{The Structure of the Trapezium KLF and IMF}
\label{sec:discuss:imf_klf}

\subsubsection{The Stellar Regime}
\label{sec:discuss:imf_klf:stellar}

From our careful construction and improved modeling of the Trapezium Cluster KLF,
we are able to derive the underlying Trapezium Cluster IMF, spanning the entire mass
range from OB stars to sub-stellar objects near the deuterium burning limit.
We find that the stellar Trapezium IMF first rises steeply with a  Salpeter-like 
power-law slope from high mass stars to near 0.6~\solarmass~where the IMF flattens 
and forms a broad peak extending to the hydrogen burning limit. There the IMF turns 
over and declines into the brown dwarf regime.  From our modeling experiments in 
MLL2000, we knew that where an underlying IMF has a power-law form, the young cluster's 
model KLF also has a power-law  form.  Further we found that peaks in the model KLFs 
can arise both due to peaks in the underlying IMF and from features in the M-L relation. 
From our current modeling of the Trapezium KLF, we  find that these  conclusions 
about the  relationship between the structure of the KLF and IMF are unchanged by 
the presence of  source reddening.  The power-law slope of the bright end
($\mbox{K}\,<\,11.5$) of the cluster KLF reflects the power-law slope of the derived IMF. 
The formation of the primary KLF peak is also similar to the  structure of the underlying
IMF we derive.  The broad main peak  of  the Trapezium KLF is formed by a combination 
of a peak in the underlying stellar IMF and a feature in the theoretical mass-luminosity
relation due to deuterium burning.
Moreover, we find that our KLF modeling  has allowed us to disentangle these two effects.
The main KLF peak at  $\mbox{K}\,=\,11\,-\,11.5$ corresponds to PMS stars between 0.4 and 
0.2~\solarmass, which according to the DM97 PMS models are undergoing deuterium burning at
the mean age of this cluster, while the derived IMF has a broad peak at somewhat lower  masses 
(0.2-0.08 \solarmass) than the KLF peak would to first order imply.  
Lastly, our detailed KLF modeling has determined that the turn-over and decline in the
cluster KLF does reflect a similar turn-over and decline of the Trapezium IMF across
the hydrogen burning limit and is not solely a product of the deuterium burning
spike \citep[e.g.,][]{zin93}.

\subsubsection{The Sub-Stellar Regime}
\label{sec:discuss:imf_klf:substar}

As in our work in MLL2000, our KLF modeling technique has permitted us to derive the Trapezium
sub-stellar IMF, while the improved depth of our IR census has allowed us to extend this derivation
from $0.03\,\solarmass$ down to near the deuterium burning limit. Our KLF modeling that now includes
source reddening confirms that the steady decline of the cluster KLF between K = 12 and K = 15
reflects a steady power-law decline in the sub-stellar IMF.  Independent of our modeling results,
however, no more than $22_{\,-2}^{\,+4}\,\%$ of the sources are sub-stellar
objects\footnote{Error based upon the uncertainty in the hydrogen burning
                 limit due to distance and to cluster mean age.}.

The secondary peak in the brown dwarf regime of the cluster KLF at K=15.5 and the subsequent rapid
decline of the cluster KLF, however, do not correspond to any known features in the theoretical
mass-luminosity relations we have examined (see section \ref{sec:discuss:ml_comp} and
figure \ref{fig:ml_comp}).  Moreover, detailed KLF modeling using two different sets of PMS tracks
require both the presence of a break from a single power-law decline of the Trapezium brown dwarf 
IMF around $0.02-0.03\,\solarmass$ and the formation of a significant, secondary IMF peak. 
Using the \citet{bur97} tracks, this IMF peak is located near the deuterium burning limit, 
13-14 \jupmass, and is followed by a rapid decline to lower masses.  Although both sets of 
PMS tracks suggest the presence of a secondary peak, the precise details (e.g., location and 
amplitude) may be track dependent. For example, in the Trapezium sub-stellar IMF found using 
the B97 tracks, 36\% of the brown dwarfs in the cluster are found in the secondary IMF peak 
while 64\% have their mass distribution governed by the power-law regime. For the IMF found 
using our standard merged tracks, these number are 15 and 85\%, respectively, however the 
truncation of the tracks at the lowest masses will slightly skew these latter percentages.

One proviso to the derivation of a significant IMF peak at the deuterium burning limit is
the contamination of our IR census by non-cluster members. 
Though, we have accounted for the background field stars contribution to the  cluster KLF, 
we have also shown that there is reasonable uncertainty in the cloud extinction properties.
Additionally, the large beamsize of the \cotracer~map may mask low extinction holes in the
molecular cloud.  Since there are $\sim\,75$ sources in the secondary peak of the~\mavlim~limited
KLF before background subtraction, our current background field star estimate would have
to be off more than a factor of two to remove any feature from the IMF at these low masses; it
would have to be off by a factor of 4, however, to account for all of the brown dwarf members.
Alternately, our IR census may be contaminated by the presence of low mass members 
from the intervening but only slightly older Orion OB1c association.   Though spectroscopic follow 
up of a few of these faint sources would separate out background stars (and provide a good test 
of the $\cotracer\,\rightarrow\,\av$~ conversion), members of the foreground OB1 association 
would be difficult  to spectroscopically separate from actual Trapezium cluster  members.  
However, as we have shown, the derived turn down in the ``sub-brown dwarf'' IMF below
the deuterium burning limit appears independent of background correction.

We conclude, therefore, that if the mass-luminosity relation for low mass brown dwarfs is reasonably
robust and does not contain a previously unidentified feature, and our estimate of the contamination
of our infrared census by non-cluster members is accurate, then the existing structure of the faint 
cluster KLF can only be created by a break from a single declining power-law brown dwarf IMF,
the formation of a corresponding peak in the underlying Trapezium IMF near the deuterium burning
limit, and a rapid decline of the IMF into the planetary mass regime.

\subsection{Sensitivity of Results to Theoretical PMS Models}
\label{sec:discuss:ml_comp}

The accuracy of an IMF derived for a young stellar cluster is intrinsically dependent upon
the robustness of the conversion from observables to a mass function (or individual masses)
provided by the theoretical evolutionary models.  In MLL2000, we came
to the somewhat surprising conclusion that model KLFs were fairly insensitive
to differences in the evolutionary PMS models from which the mass-luminosity
relations were drawn.  This was despite that fact that the detailed physics
(e.g., opacities, model atmospheres, internal convection theory, and initial conditions)
involved with calculating the theoretical  PMS  evolutionary models are poorly 
constrained and that changes in the assumed physics of these models have been shown to
produce significant differences in the  locations of evolutionary  tracks and isochrones
on the theoretical HR diagram \citep{dm94, dm97, bcah98, dant98, sdf00, bcah02}. Our findings
in MLL2000 would also appear to disagree with recent summaries of the IMF in young clusters
which concluded, based upon the track variations in the HR diagram, that the accuracy
of current PMS models are the primary uncertainty to the form of the derived
IMF \citep{mmpp4}.  Therefore, we explore in more detail the dependence of the 
theoretical mass-luminosity relation relevant for luminosity function
modeling upon the different PMS tracks.

\placefigure{fig:ml_comp}
\placetable{tab:ml_comp}

In figure \ref{fig:ml_comp} we compare the theoretical mass-infrared luminosity (K magnitude) 
relations converted from six sets of theoretical PMS models for a progressive series of young 
cluster mean ages.  In table \ref{tab:ml_comp} we summarize the different input physics and 
parameters used by various PMS models.  The sets of theoretical PMS models were taken from 
literature sources and converted to observables using a single set of bolometric corrections
(see MLL2000).  Remarkably, the theoretical mass-K magnitude relations are fairly degenerate 
between the different PMS models, and those differences that do exist are the largest at very 
young ages ($\tau\,<\,1$ Myrs), agreeing with the recent analysis of \citep{bcah02}.  
Consequently, this implies that for the Trapezium Cluster there will be some uncertainty in 
our derived mass function due to the PMS tracks. On the other hand, while the models of
\citet{bur97} and \citet{sdf00} display the most significant variations in their predicted M-L 
relations, we have shown in section \ref{sec:imf:summary:bur97} that the sub-stellar Trapezium 
IMF derived from KLF modeling using the B97 tracks is not significantly different than that 
found using the DM97 tracks.  As we concluded in MLL2000, most differences in the mass-luminosity 
relations due to differences in input physics are much smaller than we could ever observe and 
will not impact our modeling results. This result may be understood by considering the fact 
that the luminosity of a PMS star is determined by very basic physics, simply the conversion 
of gravitational potential energy to radiant luminosity during the Kevin-Helmholtz contraction. 
And this primarily depends on the general physical conditions in the stellar interior (e.g., 
whether the interior is radiative or partially to fully convective).  The luminosity evolution 
at the youngest ages ($<\,1$~Myrs) will depend, however, on the initial conditions of the 
contracting PMS star as it leaves its proto-stellar stage, though these differences are
quickly erased \citep{bcah02}.

This is significantly different than the situation for mass functions derived by placing the stars
on the theoretical HR diagram using spectroscopic and photometric observations.  
Because most young stars have late type K-M spectral types, they are on nearly vertical Hayashi
contraction tracks in the HR diagram. As a result  a star's mass derived from the HR diagram
is primarily a function of its assigned effective temperature, i.e., its observed spectral type.
We illustrate this dependence in figure \ref{fig:teff_comp}(a) where we plot the predicted effective
temperatures as a function of mass for stars in a 1 Myr old cluster. In contrast to the quite similar
mass-luminosity relations, a star's mass derived based upon its spectral type is uncertain due
to differences in the PMS models by factors of 3 (or more).
The conversion from spectral type to mass is made worse by the uncertain conversion of spectral type
to effective temperatures for late type sources, resulting from their sub-giant gravities
\citep{luh99b}.
Such uncertainties will undoubtedly result in spectroscopically derived IMFs that vary substantially
as a function of PMS tracks used \citep[compare, for example, the ][ derivation of the Trapezium IMF
from DM97 and BCAH98 tracks]{luh2000}.
Further, these track differences, while decreasing with time, are not resolved by 5 Myrs as shown in
figure \ref{fig:teff_comp}(b).  In summary, the uncertainties in the PMS models primarily manifest
themselves in variations in the predicted effective temperatures of the young stars rather than the
predicted luminosities.

We do not surmise that luminosity function modeling, which employs mass-luminosity relations,
is free from systematic dependencies.
As concluded in MLL2000, a cluster's mean age must be known in order to derive a cluster's 
initial mass function  from its luminosity function; this can only be derived from placing
the stars on the theoretical HR diagram. In general, however, age is a function of luminosity
for low mass stars on the HR diagram and will be more or less similar when derived from these
PMS models.  The exception again occurs at the youngest ages, where the definition of a star's
age may differ if the models include the proto-stellar lifetimes.  Even in the case of
the \citet[][hereafter, PS99]{ps99} models, which begin as protostars accreting along
an initial mass-radius relationship or birthline in the HR diagram \citep{stah85}, the
mass-luminosity relations are not substantially divergent from canonical theoretical models
except at the very youngest ages.

\subsection{Comparison of Trapezium IMFs Based on Infrared Observations}
\label{sec:discuss:comp_trap}

In addition to our initial modeling in MLL2000 and our present study, a number of other
authors have recently derived Trapezium IMFs based upon deep infrared observations. 
While all of these derivations make use of the same set of theoretical pre-main sequence models
for converting observations to mass (functions), they use somewhat different cluster parameters
and employ a variety of different methodologies.  Systematic uncertainties might arise due to
varying of cluster parameters such as distance, due to different assumptions about the cluster
population such as field star contamination or from simple  observational effects such as survey
area or the wavelength regime analyzed. Further, it is not understood how closely different
methods can arrive at the same IMF.

\placefigure{fig:imf_comp}
\placetable{tab:imf_comp}

In figure  \ref{fig:imf_comp} we compare the IMFs derived by MLL2000,  LR2000, HC2000  
and  \citet{luh2000}. Globally, these IMFs are remarkably similar.  They all have Salpeter-like 
high mass slopes, all reach a broad peak at sub-solar masses and all decline in frequency with 
decreasing mass below the hydrogen burning limit with brown dwarf IMF slopes between +1 and +0.5.
After inspecting the different methods and cluster parameters used by these authors,
which we summarize in table \ref{tab:imf_comp}, this result should be in part expected. 
When different methods use the same PMS tracks (in this case DM97; $\mbox{M}\,<\,1\,\solarmass$)
and essentially the same star forming histories, the resulting IMFs should basically agree.

The cluster parameters used by these workers are not exactly homogeneous.  Further, there
are slight variations between these IMFs that might be due in part to observational  effects.
For example, the truncation or turn down in the high mass end of the LR2000 and HC2000 IMFs is
due to bright source saturation in these surveys, not to a real IMF feature. At the low mass end,
the IMF derivations  appear to diverge below 30 \jupmass~(-1.5 in log solar mass units) with
a ``spike'' in the LR2000 IMF but no feature in the Luhman et al study.
Because LR2000 surveys the largest area while the \citet{luh2000} study surveys the smallest area,
one might suspect that this difference is due to an increase in field star contamination or,
perhaps, counting statistics for the smaller study.  The latter is the most likely explanation
since both our study and that of HC2000 survey similar large areas and apply field star
corrections while finding sub-stellar IMFs that contain either a secondary peak or a
plateau at the lowest masses.

Lastly, the methodologies employed range from a purely statistical approach (e.g., the LF 
modeling of MLL2000 and this paper) to the derivation of individual masses of the stars 
via a hybrid combination of spectroscopy and infrared colors (e.g., Luhman et al.). 
It is unclear how to make detailed comparisons of these methods, however, in general, 
the LR2000, HC2000 and MLL2000 primarily depend upon the theoretical mass-luminosity 
relation extracted from the PMS tracks. The stellar portion of the Luhman et al IMF 
depends upon the theoretical HR diagram, while the sub-stellar depends upon the predicted 
infrared colors and magnitudes.  One apparent difference between the resulting IMFs that 
might be related to the different methods is the exact location of  the  IMF's ``peak,'' 
or what is sometimes termed the  ``characteristic'' mass.  This ``peak'' mass varies 
between IMF derivations by  0.7 dex in log solar mass units.  It is not immediately 
apparent that internal uncertainties in the IMF derivations could cause this scatter. 
For example, the ``peak'' of the MLL2000 IMF is revised only 0.1 dex by the inclusions 
of source reddening.  For methods that depend upon mass-luminosity relations, the resulting 
IMF will be dependent upon the assumed cluster distance and age; modest changes in these 
parameters should result in slightly different M-L relations and slightly different IMFs.  
However, there is no strict correlation between ``peak'' mass and the cluster age or 
distance used.  Hence, we conclude that specific IMF details such as the exact location 
of ``peak'' mass cannot be securely identified by these methods; although, we can conclude 
that the Trapezium IMF peaks at sub-solar masses somewhere between 0.3~\solarmass~and the 
hydrogen burning limit.

\subsection{The Trapezium and the Global IMF}
\label{sec:discuss:imf_origin}

From our current work and by comparison of our work to that of other authors,
the general form of the Trapezium IMF is readily apparent: a continuous IMF that
rises with a relatively steep slope toward sub-solar masses, flattens and forms a
broad peak between 0.3 \solarmass~and the hydrogen burning limit before turning over and
declining into the brown dwarf regime. This IMF structure is roughly half-gaussian, though,
not exactly log-normal (see figure \ref{fig:imf_derived}b ), and it is quite consistent with
current derivations of other star cluster IMFs. Open cluster IMFs such as
the Pleiades \citep{bouv98} and M35 \citep{barr01} rise with similar power-law slopes and
form broad peaks at sub-solar masses before apparently rolling over and declining into
the brown dwarf regime. The color-magnitude
diagrams of the very luminous clusters NGC 2362 \citep{moit01},
NGC 3603 \citep{bran99}, and NGC 6231 \citep{baume99} all display evidence
of IMFs~that peak at sub-solar masses, though more complete discussion awaits detailed derivations
of their sub-solar and sub-stellar IMFs.  Further, this apparent general form of the IMF
has also been found for the IMFs of globular clusters \citep{par99} and for the field star
IMF \citep{ms79, ktg93, scalo98, pk01}, though again, the sub-stellar IMFs are not yet
robustly known. The continuity of the IMF across so many environments suggests
that a single star formation process may be responsible for producing the majority of the
mass spectrum.  Indeed, evidence that stars {\em and} brown dwarfs
form with similar initial frequencies of circumstellar disks \citep{aam01} and
similar disk properties \citep{natta01} suggests that even sub-stellar objects form
via the same mechanism as stars.
While various physical processes might influence the fine details (the IMF's
``peak'' mass, for example) of the IMF's final form \citep[ e.g., ][]{adams96},
the original fragmentation distribution function of a turbulent molecular cloud probably
dominates the final form of the stellar IMF \citep{klessen01}.

It is not clear, however, the extent to which this documented continuity
between the Trapezium, open cluster and field star IMFs extends across the
entire brown dwarf regime. Very young stellar clusters appear to have
consistent low mass IMFs, i.e., IMFs that decline across the hydrogen burning
limit \citep{ntc00, luh2000}.  As we have already discussed, LR2000 and our
work here find that there is reasonable evidence in the Trapezium luminosity
function(s) that the brown dwarf Trapezium IMF departs at the lowest masses
(e.g., $<\,30\,\jupmass$; see section  \ref{sec:discuss:comp_trap}) from a
single power-law decline for the brown dwarf regime and forms a strong
secondary peak in the IMF near the deuterium burning limit. While a similar
IMF feature has yet to be identified in the sub-stellar regime of the well
studied IC 348 young cluster \citep[see][]{ntc00}, the recent work of
\citet{bejar01} suggests a mass function for the young $\sigma$~Orionis
cluster \citep{wal97} that is slowly falling throughout the brown dwarf regime
but may rise toward the deuterium burning limit.

Were a secondary peak in the IMF of the lowest mass brown 
dwarfs confirmed, then it may provide evidence for a secondary, competing
formation mechanism for these low mass objects.  Indeed, the transition in the sub-stellar IMF
at $30\,\jupmass$ from a steady power-law decline to the secondary peak at the deuterium
burning limit may represent the transition from the formation 
of brown dwarfs as individual fragments of the molecular cloud to their formation, for example, 
as truncated stellar embryos that were dynamically ejected from hierarchical proto-stellar 
systems before they had a chance to accrete into higher mass objects \citep{rc01}. Yet the 
subsequent rapid decline in the IMF below the deuterium burning limit constrains the mass 
range over which this secondary mechanism is operating. The frequency and characteristics 
of circumstellar disks around these very low mass brown dwarfs and planetary mass objects
may provide an essential test of their formation from individual pre-{\em sub}-stellar
cores or via some an entirely different mechanism.

\section{Conclusions}
\label{sec:conclusions}

Using a new and very complete near-infrared census of the Trapezium Cluster, we have
performed a detailed analysis of the Trapezium Cluster's K band luminosity function
and its underlying mass function. Following our earlier work in \citet{mll00}, we expanded
our luminosity function modeling to include the effects of source reddening, and we studied
in detail the field star contribution to the cluster KLF.  We applied our new models to
the Trapezium KLF to explore its structure and to derive the cluster's initial mass function. 
From this analysis we draw the following conclusion(s):

\noindent
1.\  The Trapezium Cluster IMF rises in number with decreasing mass and forms a broad peak at 
sub-solar masses between 0.3 \solarmass~ and the hydrogen burning limit before declining
into the brown dwarf regime.  Independent of modeling details, no more than $\sim\,22\%$ 
of the young sources fall below the hydrogen burning limit, placing a strict limit on the
brown dwarf population in this cluster.

\noindent
2.\  The Trapezium Cluster sub-stellar IMF breaks from a single declining power-law slope
between 0.02 - 0.03 \solarmass~and forms a significant secondary peak
near the deuterium burning limit ($\sim\,13\,\jupmass$).  We derive these results
through detailed analysis of the likely field star contamination and from our modeling of the
cluster's faint KLF using two different sets of theoretical mass-luminosity relations, although
the precise details of this peak do depend upon the PMS models.
Regardless, this peak may contain between 15 and 36\% of all the sub-stellar objects in this cluster.
Below this peak the sub-stellar IMF declines rapidly toward lower masses suggesting that the
yield of freely floating, planetary mass objects during the formation of the Trapezium Cluster
was extremely low.

\noindent
3.\  We find that source reddening (due to infrared excess and extinction) has only modest 
effects upon our modeling of the Trapezium cluster's luminosity function.  Source reddening
tends to broaden the IMFs derived and blur the precision with which we can derive IMF parameters.
However, the Trapezium IMF we derive here after accounting for source reddening and field stars
is not a substantial revision over that Trapezium IMF we derived in MLL2000.

\noindent 
4.\  Pre-main sequence luminosity evolution and the resulting age dependent mass-luminosity
relations are relatively robust results of most modern PMS evolutionary 
models, except at the very youngest ages where the models are affected by initial conditions.  
Conversely, the predicted effective temperatures, hence predicted spectral types, are 
considerably less robust.  This suggests that modeling a cluster's K band luminosity function
is likely to produce a faithful representation of the true IMF of the cluster. 
Further, we find that the different published methodologies used for deriving the 
Trapezium IMF from near-infrared photometry produce nearly identical results, although the 
precise location of a ``peak'' or characteristic mass for the Trapezium cannot be securely 
identified.

\noindent
5.\  The globally consistent form of the stellar IMF suggests a single physical mechanism may 
dominate the star formation process in galactic clusters. If the secondary peak in the Trapezium 
Cluster KLF and the corresponding strong secondary peak in the sub-stellar IMF near the 
deuterium burning limit were proven to be real cluster characteristics, then this implies 
that a secondary physical mechanism may be responsible for the formation of very low mass 
brown dwarfs.  Whether this secondary mechanism could be the ejection of very low mass
(10-20~\jupmass) brown dwarfs from hierarchical proto-stellar systems, (e.g., Reipurth \& Clarke), 
or some other process remains unclear.

\acknowledgments

We would particularly like to thank Alyssa Goodman, Kevin Luhman and John Stauffer for helpful 
discussions, and Pavel Kroupa for especially productive suggestions regarding the modeling 
algorithms.  We acknowledge Ted Bergin for providing the \cotracer~and dust continuim maps 
of the Trapezium Region and thank the anonymous referee for making a number of suggestions 
that improved the results of the paper.  AAM was supported by a Smithsonian Predoctoral  
Fellowship and by the NASA Graduate Student Research Program (grant NTG5-50233).  
EAL and AAM acknowledge support from a Research Corporation Innovation Award and Presidential 
Early Career Award for Scientists and Engineers (NSF AST 9733367) to the University of Florida.  
This publication makes use of data products from the Two Micron All Sky Survey, which is a 
joint project of the University of Massachusetts and the Infrared Processing and Analysis 
Center/California Institute of Technology, funded by the National Aeronautics and Space 
Administration and the National Science Foundation.

\appendix

\section{FLWO-NTT Near-Infrared Catalog of the Trapezium Cluster}
\label{app:catalog}

We summarize in table \ref{tab:obs} the characteristics of the three observing runs used to
obtain the infrared photometry that comprise the FLWO-NTT Near-Infrared Catalog of the
Trapezium Cluster.  We compare the  area(s) covered by the FLWO-NTT catalog to those of other
recent IR surveys in figure \ref{fig:trap_area}.  The reduction and analysis of the 1997-1998 FLWO
observations are fully summarized in  \citet{lada00}, and we do not discuss these further here.
Preliminary analysis of the NTT  observations was  presented   in \citet{aam01}, and we detail these
observations and their reduction below.

\placetable{tab:obs}

\subsection{ NTT 3.5m JH\Ks~observations}
\label{app:catalog:ntt}

Our NTT images of the Trapezium Cluster were obtained under conditions of superb seeing 
($\sim\:0.5\arcsec$  FWHM) on 14 March 2000  using the SOFI infrared spectrograph and 
imaging camera. The NTT telescope uses an active optics platform to achieve ambient  seeing 
and high image quality, and the SOFI camera employs a large  format $1024\:\times\:1024$ pixel 
Hawaii HgCdTe  array.  To obtain a single wide field image of the Trapezium Cluster, we 
configured SOFI to have  a $\:4\farcm95\:\times\:4\farcm95\:$ field of view with a plate scale of 
0\farcs29 /pixel.  Each exposure consisted of  9 separate dithers each randomly  falling within 
$20\arcsec$ of the  observation center.  Each individual dither was the co-average of eight 1.2 
second  exposures, yielding an total effective integration time of  86.4 seconds for each combined 
image.

We observed the Trapezium Cluster with identical sequential pairs of on and off-cluster dithered 
images. During one hour on 14 March 2000, we obtained four image pairs of the Trapezium 
Cluster and off-cluster positions. These were, in temporal order, at 
$\Ks\;(2.162\;\mu\mbox{m})$, $\mbox{H}\;(1.65\;\mu\mbox{m})$, 
$\mbox{J}\;(1.25\;\mu\mbox{m})$  and again at \Ks,~and the on-cluster images
 had FWHM estimates of  0.53\arcsec, 0.55\arcsec,  0.61\arcsec and 0.78\arcsec.
Seeing estimates of stars in the paired non-nebulous off-cluster image(s)  yielded similar 
if not marginally higher  resolution point spread functions.  Observations were taken near
transit with a very small range  of  airmass  ( $1.138\:<\:\sec(z)\:<\:1.185$).  

Data reduction of the NTT images was performed using routines in the Image Reduction and 
Analysis Facility (IRAF)  and Interactive Data Language (IDL).   Our standard data reduction 
algorithm  was described in \citet{lada00} for the FLWO images, and it was subsequently used 
for the  NTT images. Simply, individual dithered frames were reduced using sky and flat field 
images derived from the non-nebulous off-cluster dithered images which were interspersed with 
the on-cluster images.  Each set of reduced dithered frames  were then combined using a standard 
``shift-and- add'' technique.   While all the FLWO data was linearized after dark-subtraction
using a system supplied linearity correction, linearization coefficients were not obtained for the 
NTT data.  ``Sky'' flat-fields constructed from the NTT images were compared to system 
flat-fields which are regularly taken and monitored by the NTT staff.  While the NTT system 
flat-fields were  found to vary by only 2-3\% over long periods of  time, when we compared our 
sky flat-fields  to the system flat-fields, significant small scale variations (5-10\%) were revealed 
across  the array. We concluded this was due to our relatively short NTT integration times which 
results in poor sampling of the intrinsically non-flat SOFI  array.  Therefore, we substituted the 
system supplied flat-fields into our  reduction procedure. The high resolution of our NTT images 
results in moderate under-sampling of the point spread  functions;  we tested to see if sub-pixel 
linear reconstruction (drizzling) of our images would  improve our data quality.  
Since our images have only a few dithers (9), the drizzle algorithm did not improve our result 
over standard integer ``shift-and-add.''

The 2000 NTT images had FWHM estimates ranging between 1.8 and 2.1 pixels, and these 
images are therefore marginally under-sampled and  not easily suitable for PSF photometry. 
Further, the SOFI field of view suffers from coma-like geometric  distortions on the northern 10-
15\% of the array. For these two reasons, we decided to perform only aperture photometry on
the NTT images.   Multi-aperture photometry was performed on sources detected in the 
NTT image using annuli with radii from 2 to 10 pixels. The sky was measured from the mode of 
the distribution of  pixel values in an annuli from 10 to 20 pixels. From inspection of the curves
of growth of both isolated and nebulous  sources, we chose a 3 pixel radius ($1.8\arcsec$ beam) for
most of our NTT sources. Additionally, the choice of small apertures allowed us to minimize
the effects of nebular contamination and crowding on the stellar PSF. For faint sources in 
very confused or highly nebulous regions,  we repeated the photometry with a 2 pixel aperture 
and a sky annulus  from 7 to 12 pixels.  The change in sky annulus does not significantly
affect our photometry because the fraction of the stellar PSF beyond 7 pixels contains less than 
5\% of the flux, and the errors resulting from including this flux in the sky estimate are smaller 
than the errors introduced from using too distant a sky annulus on the nebulous background.

Aperture corrections were derived for our data by choosing $\sim\,15$ relatively bright stars as free
of nebular contamination as possible. We performed multi-aperture photometry on them and using
the IRAF MKAPFILE routine to visually inspect the stellar curves of growth and calculate corrections.
Since small apertures were used to minimize the effects of the bright nebular background,
the resulting corrections which constituted a somewhat substantial fraction of the stellar flux.
Aperture corrections were carefully checked by comparing the corrections derived for on (nebulous)
and off-cluster positions, which are interspersed in time with the on-cluster frames, and found
to agree or to correlate with changes in seeing.  The typical 3 pixel aperture correction was
-0.14 magnitudes and for those stars photometered using a 2 pixel aperture, a correction of
-0.34 was used.

\subsection{Photometric Comparisons of Datasets}
\label{app:catalog:phot_comp}

We report in the electronically published catalog all the photometry from the FLWO and NTT
observations.
We explored any photometric differences between the FLWO and NTT observations.
These  differences include the filter systems, the methods and effective  beamsizes of the 
photometry and the epochs of the  observations.  We tested if any color terms were present due to 
differing photometric (filter) systems, we compared the magnitudes and colors of 504 sources 
common to both the NTT and FLWO  photometry.We compared the (J - H) and (H - K) colors of 
the NTT photometry to the FLWO photometry and fit these  comparisons with linear relations. 
The (J-H) colors were well fit by a linear relation  (slope $\sim$ 1); however, we found an offset, 
$\Delta(J-H)\:\simeq\:0.10$ magnitudes between the two systems. A similar comparison to the 
photometry of sources in the 2MASS catalog  
     \footnote{A current  un-restricted search of the 2MASS  First  and Second  Incremental  Point 
     Source and Extended Source Catalogs currently returns only 171 sources. }
indicated this offset was at J band and was restricted  to the FLWO sources. Comparison of  
2MASS photometry  to the NTT photometry revealed no systematic offsets.  A comparison of the 
FLWO and NTT  (H - K) colors was also well fit by a linear relation  (slope $\sim$ 0.97) though 
this slope suggests that  for the reddest sources,  the  NTT (H - \Ks)  color is bluer than the FLWO 
(H - K) color. 

Further, it was evident from these comparisons that while the global filter systems are  quite 
similar,  the difference in  the NTT and FLWO photometry of individual sources was larger than 
expected  from {\em formal} photometric errors 
\footnote{the quadratic sum of uncertainties from aperture corrections, zeropoint and airmass
    corrections, flat fielding error and sky noise}.  
From our fake star experiments  and from the photometry of sources in overlap regions on 
mosaicked frames, we determined our {\em measured} photometric error is 5\% for the majority 
of our sources increasing up to 15\% for the sources at our completeness limit. However, when 
comparing sources common to both the FLWO and NTT data (well above our completeness 
limit), we  derived $1\sigma$  standard  deviations of  $\sim\:0.22$ for magnitudes and  
$\sim\:0.18$ for  colors. Very similar dispersions were derived when  comparing our  FLWO 
photometry to the \citet{lah98} or 2MASS catalogs or when comparing our NTT data to the 
\citet{hc2000} H and  K band dataset.   We attribute a portion of this additional photometric
noise between the different datasets to the intrinsic infrared  variability of these pre-main 
sequence sources which  has been found for stars in this cloud to have an average of
0.2 magnitudes at infrared wavelengths \citep{chs01}.
We note that the difference in the beamsize used for the FLWO and NTT photometry and by
the various other published data sets will also contribute a degree of added photometric
noise due to the presence of the strong nebular background, thus making the NTT photometry
preferable to the FLWO data for its higher angular resolution.

\subsection{Astrometry and the Electronic Catalog}
\label{app:catalog:astrometry}

Astrometry with reasonably high precision was performed by matching the XY pixel locations of 
a large number ($>\,50$) of  the observed sources to the equatorial positions of these sources listed
on the 2MASS world coordinate system and deriving full plate solutions using the IRAF CCMAP 
routine.  Mosaic positions of the 1997 and 1998 observations were shifted to fall onto a common 
XY  pixel  grid defined by the K band FLWO 1997 mosaic images. To create the common K 
band XY  grid, sources in the overlap regions between mosaic positions were matched and global 
offsets  calculated.  The two camera arrays of the FLWO STELIRcam instrument are not centered 
precisely on the sky and the J and H band coordinates were transformed using the IRAF 
GEOMAP routine into the K band XY coordinate grid.  The NTT positions were aligned to the 
NTT J band image.  For the FLWO plate solution, 161 2MASS sources  were matched to the 
FLWO XY  coordinates yielding  a plate scale of  0.596 ''/pixel and an astrometric solution with 
rms errors of $\sim\:0.10''$. An independent  solution of 82 NTT sources matched to the 2MASS 
database yielded a plate scale of 0.288 ''/pixel  and an astrometric solution having rms errors 
$\sim\,0.07''$. 

We construct the electronic version of the FLWO-NTT catalog based upon all the sources 
detected by our FLWO and NTT observations, and we compliment our electronic catalog by 
including sources identified in other catalogs and falling within our survey area, but 
that were saturated, undetected or unresolved by our observations.  Since our final catalog 
covers a substantially different area than comparable deep infrared surveys and includes 
numerous new sources, we chose to assign new source designations for our final catalog.  
These are based upon the IAU standard format that includes a catalog acronym, a source 
sequence, and source specifier.  For the catalog acronym, we chose the MLLA, based upon 
the initials of the last names of the authors. This acronym is currently unused in the 
Dictionary of Celestial Nomenclature.  We chose to sequence the catalog using a running 
number incremented from 00001 to 01010.  We use a specifier only where necessary to 
distinguish unresolved sources, typically employing the designations (A), (B), etc. 
NTT astrometry is preferentially used in the final catalog. For undetected or unresolved sources, 
we made every effort to include astrometry from the source's identifying catalog if the original 
catalog could be globally aligned to the FLWO-NTT catalog.  We list cross-references based on 
the most comprehensive or deep surveys; these include the \citet{lah97}, HC2000, \citet{luh2000} 
designations and \citet{ms94}.  For sources lacking cross-references in these catalogs, we list 
their 2MASS designations (circa the 2nd Incremental 2MASS Point Souce Catalog) where possible.  
The LR2000 designations are based on their derived equatorial coordinates and due to significant 
astrometric errors do not correspond to the positions we derive in the FLWO-NTT catalog. 
For example, we find off-sets of $-0.42\arcsec$ in RA and $0.44\arcsec$ in DEC between our 
positions and those of LR2000. After removing these offsets, we still find medain residuals 
of $0.44\,\arcsec$ between our coordinates and those of LR2000 with errors as large 
as $1\arcsec$; this is in contrast to the rms residuals of $0.1\,\arcsec$ between our catalog
and the 2MASS and HC2000 positions.  Hence we do not list the LR2000 position-dependent 
designations except where necessary to identify sources undetected by our catalog.

\placetable{tab:catalog}

\clearpage

\begin{deluxetable}{cccccc}
\tablecolumns{6}
\tablewidth{0pt}

\tabletypesize{\scriptsize}
\tablecaption{Three Power-Law IMF Parameters and Fits \tablenotemark{(a)}
\label{tab:imf_pars}
}

\tablehead{
\colhead{Parameter \tablenotemark{(b)}}&
\colhead{Range of Fits} &
\colhead{Best Fit \tablenotemark{(c)}} &
\colhead{($\pm$)}  &
\colhead{Best Fit \tablenotemark{(d)}} &
\colhead{($\pm$)} 
}

\startdata
\cutinhead{Model fit to K$\:=\:14.5$; $\;\;\chi^{2}\,\sim\,1$ \tablenotemark{(e)}}
 $\Gamma_{1}$  & $-1.0\:\longleftrightarrow\:-2.0$ & -1.16 & 0.16 & -1.24 & 0.20 \\[0.05cm]
 $\log\,m_{1}$ & $+0.1\:\longleftrightarrow\:-1.1$ & -0.17 & 0.10 & -0.19 & 0.13 \\[0.05cm]
 $\Gamma_{2}$  & $-0.4\:\longleftrightarrow\:+0.4$ & -0.24 & 0.07 & -0.16 & 0.15 \\[0.05cm]
 $\log\,m_{2}$ & $+0.1\:\longleftrightarrow\:-1.4$ & -1.05 & 0.05 & -1.00 & 0.13 \\[0.05cm]
 $\Gamma_{3}$  & $-0.4\:\longleftrightarrow\:+2.0$ &  1.10 & 0.25 &  1.08 & 0.38 \\[0.05cm]
\cutinhead{Model Fit to K$\:=\:15.0$; $\;\;\chi^{2}\,\sim\,1$}
 $\Gamma_{1}$  &  & -1.13 & 0.16 & -1.21 & 0.18 \\[0.05cm]
 $\log\,m_{1}$ &  & -0.19 & 0.11 & -0.22 & 0.11 \\[0.05cm]
 $\Gamma_{2}$  &  & -0.24 & 0.15 & -0.15 & 0.17 \\[0.05cm]
 $\log\,m_{2}$ &  & -1.00 & 0.10 & -0.92 & 0.13 \\[0.05cm]
 $\Gamma_{3}$  &  &  0.82 & 0.15 &  0.73 & 0.20 \\[0.05cm]
\cutinhead{Model Fit to K$\:=\:15.5$;  $\;\;\chi^{2}\,\sim\,0.3$}
 $\log\,m_{2}$ &  & -0.89 & ---- & -0.77 & ---- \\[0.05cm]
 $\Gamma_{3}$  &  &  0.30 & ---- &  0.30 & ---- \\[0.05cm]
\cutinhead{Model Fit to K$\:=\:16.5$; $\;\;\chi^{2}\,\sim\,0.3$}
 $\log\,m_{2}$ &  & ----- & ---- & -0.72 & ---- \\[0.05cm]
 $\Gamma_{3}$  &  & ----- & ---- &  0.23 & ---- \\
\enddata

\tablenotetext{(a)}{All tabulated fits derived using our standard set of PMS tracks (primarily
   from DM97).}
\tablenotetext{(b)}{The parameters $\Gamma_{i}$ are the power-law indices of the IMF 
   which here is defined as the number of stars per unit $\log ( \case{M}{M_{\sun}} )$.
   The parameters $m_{j}$ are the break masses in the power-law IMF and are in units of 
   $\log ( \case{M}{M_{\sun}} )$.}
\tablenotetext{(c)}{Model fits without Source Reddening.}
\tablenotetext{(d)}{Model fits accounting for Source Reddening.}
\tablenotetext{(e)}{$\chi^{2}$ given for fits with Source Reddening.}

\end{deluxetable}

\clearpage

\begin{deluxetable}{cclllclll}
\tablecolumns{9}
\tablewidth{0pt}

\tabletypesize{\small}
\tablecaption{Three Power-Law Trapezium Sub-Stellar IMF \tablenotemark{(a)}
\label{tab:ssimf_pars}
}

\tablehead{
\colhead{IMF Parameter} &
\colhead{ }   &
\multicolumn{3}{c}{Fit to K=16.0} &
\colhead{ }   &
\multicolumn{3}{c}{Fit to K=16.5} \\

\colhead{($\varphi$)} &
\colhead{ }   &
\colhead{$<\varphi>$} &
\colhead{$1\sigma$} &
\colhead{median($\varphi$)} &
\colhead{ }   &
\colhead{$<\varphi>$} &
\colhead{$1\sigma$} &
\colhead{median($\varphi$)}
}

\startdata
\cutinhead{Fits to $\mavlim$~KLF w/o Off-field Correction \tablenotemark{(b)}}
$ \Gamma_{1}$ &\phm{spc}& +1.39      & 0.22   & +1,30    &\phm{spc}& +1.36   & 0.18   & +1.30     \\[0.05cm]
$m_{1}$       &\phm{spc}& \phs0.0214 & 0.0022 & \phs0.020&\phm{spc}&  0.0220 & 0.0025 & \phs0.020 \\[0.05cm]
$ \Gamma_{2}$ &\phm{spc}& -4.87      & 1.52   & -5.00    &\phm{spc}& -5.50   & 2.01   & -5.00     \\[0.05cm]
$m_{2}$       &\phm{spc}& \phs0.0125 & 0.0012 & \phs0.012&\phm{spc}&  0.0137 & 0.0005 & \phs0.014 \\[0.05cm]
$ \Gamma_{3}$ &\phm{spc}& +3.70      & 2.60   & +4.00    &\phm{spc}& +5.70   & 1.59   & +6.00     \\[0.05cm]
\cutinhead{Fits to $\mavlim$, Off-field Corrected KLF \tablenotemark{(b)}}
$ \Gamma_{1}$ &\phm{spc}& +1.53      & 0.13   & +1,60    &\phm{spc}& +1.51   & 0.14   & +1.60     \\[0.05cm]
$m_{1}$       &\phm{spc}& \phs0.0217 & 0.0024 & \phs0.020&\phm{spc}&  0.0228 & 0.0025 & \phs0.025 \\[0.05cm]
$ \Gamma_{2}$ &\phm{spc}& -5.43      & 1.95   & -6.00    &\phm{spc}& -5.04   & 2.00   & -4.00     \\[0.05cm]
$m_{2}$       &\phm{spc}& \phs0.0130 & 0.0008 & \phs0.013&\phm{spc}&  0.0137 & 0.0008 & \phs0.014 \\[0.05cm]
$ \Gamma_{3}$ &\phm{spc}& +3.87      & 2.47   & +4.00    &\phm{spc}& +5.61   & 2.25   & +6.00     \\
\enddata
\tablenotetext{(a)}{Fits to Sub-Stellar Trapezium KLF using B97 tracks.}
\tablenotetext{(b)}{$\mavlim$~KLF has limits of $\av\,\le\,9$ and $\mbox{M}\,\ge\,0.01\,\solarmass$}
\end{deluxetable}

\clearpage

\begin{deluxetable}{lcccccllll}
\tablecolumns{10}
\tablewidth{0pt}
\rotate

\tabletypesize{\scriptsize}
\tablecaption{Summary of Theoretical Pre-Main Sequence Calculations
\label{tab:ml_comp}
}

\tablehead{
\colhead{Model \tablenotemark{(a)} } &
\colhead{$\tau_{min}$ \tablenotemark{(b)}} &
\colhead{$\mbox{M}_{max} $} &
\colhead{$\mbox{M}_{min} $} &
\colhead{Initial \tablenotemark{(c)} } &
\colhead{[D/H] \tablenotemark{(d)} } &
\colhead{Opacity \tablenotemark{(e)} } &
\colhead{EOS \tablenotemark{(f)} } &
\colhead{Convection \tablenotemark{(g)} } &
\colhead{Atmosphere \tablenotemark{(h)} } \\
\colhead{Name} &
\colhead{(Myrs)} &
\colhead{(\solarmass)} &
\colhead{(\solarmass)} &
\colhead{Conditions} &
\colhead{} &
\colhead{Table} &
\colhead{} &
\colhead{Model} &
\colhead{Model}
}

\startdata
 DM94   & 0.1 & 2.50 & 0.020 & Canonical & 2.0  & OPAL92+Alex89 & MHD      & FST-1   & Gray               \\[0.05cm]
 B97    & 0.1 & 0.15 & 0.001 & Canonical & 2.0  &  ?            & SC       &  ?      & Non-Gray           \\[0.05cm]
 DM97   & 0.1 & 3.00 & 0.017 & Canonical & 2.0  & OPAL92+Alex94 & MHD+OPAL & FST-2   & Gray               \\[0.05cm]
 BCAH98 & 1.0 & 1.40 & 0.020 & Canonical & 2.0  & OPAL96+Alex94 & SCVH     & MLT-1.0 & Non-Gray(NextGen)  \\[0.05cm]
 PS99   & 0.1 & 6.00 & 0.100 & Birthline & ---  & OPAL96+Alex94 & PTEH     & MLT-1.5 & Gray               \\[0.05cm]
 SDF00  & 0.1 & 7.00 & 0.100 & Canonical & 2.0  & OPAL96+Alex94 & PTEH(r)  & MLT-1.6 & Analytic Fit       

\enddata
\tablenotetext{a}{ All models used standard solar metallicity (Z=0.02).} 
\tablenotetext{b}{ Minimum Age (Myrs) listed by the models.}
\tablenotetext{c}{ Initial Physical Conditions from which the tracks are evolved. Canonical: the
    model stars are evolved from ``infinite'' spheroids; Birthline: the PS99 models begin with spherically
    accreting (single accretion rate, $10^{-5}\,\solarmass\,yr^{-1}$) protostars evolving along the
    birthline before beginning their PMS contraction phase.}
\tablenotetext{d}{ Deuterium abundance relative to hydrogen, in units of $10^{-5}$.
    The initial D/H for the PS99 models was $2.0\,\times\,10^{-5}$, however this is significantly modified
    by the burning of deuterium during the model's proto-stellar phase.}
\tablenotetext{e}{ Opacity Table (interior not atmosphere): Alex89 \citep{alex89}; OPAL92 \citep{opal92};
    Alex94 \citep{alex94}; OPAL96 \citep{opal96} }
\tablenotetext{f}{ Convection Model: FST (Full Spectrum Turbulence. FST-1: \citet[][]{can90,can92},
    FST-2: \citet[][]{can96}); MLT-1.X (Mixing Length Theory, 1.X = $\alpha$ = 1/H$_p$) }
\tablenotetext{g}{ Equation of State:  MHD  \citep{mhd88}; SC \citep{sc91, sc92};
    SCVH \citep{scvh95}; PTEH \citep[][r: Revised by SDF]{pols95}, OPAL \citep{rog96}  }
\tablenotetext{h}{ Treatment of Atmosphere: Analytic Fit (SDF00) is a 1D fit of $T(\tau)$ to atmosphere models;
    NextGen \citep[][]{hab99} }
\end{deluxetable}

\clearpage

\begin{deluxetable}{llcclllcccr}
\tablecolumns{11}
\tablewidth{0pt}
\rotate

\tabletypesize{\scriptsize}
\tablecaption{Comparison of  published Trapezium IMFs \tablenotemark{(a)}. 
\label{tab:imf_comp}}

\tablehead{
\colhead{Work}              &
\colhead{IMF}               &
\colhead{Distance}          &
\colhead{$\tau,\Delta\tau$ \tablenotemark{(b)}} &
\colhead{Extinction? \tablenotemark{(c)}}       &
\colhead{IR Excess?  \tablenotemark{(c)}}       &
\colhead{Field Star  \tablenotemark{(c)}}       &
\colhead{A$_{V}$ Limit? }   &
\colhead{IMF Peak }         &
\colhead{Area  \tablenotemark{(d)}} &
\colhead{Comments   \tablenotemark{(e)}} \\
\colhead{Name}              &
\colhead{Method}            &
\colhead{(pc), (m-M)}       &
\colhead{(Myr)}             &
\colhead{ }                 &
\colhead{ }                 &
\colhead{Correction?}       &
\colhead{ }                 &
\colhead{($\case{M}{M_{\sun}}$)} &
\colhead{($\sq\,\mbox{pc}$)}     &    
\colhead{ }
}

\startdata
 MLL2000 & Model KLF & 400, 8.00 & 0.8, 1.2 & Not      & Not      & Not      & None & 0.08 &0.34& Literature \\[-0.0cm]
         & Fit       &           &          & Included & Included & Included &      &      && K only     \\[0.5cm]

 HC2000  & (H-K)/K   & 480, 8.40 & 0.4, 1.0 & Derived & Empirical & Reddened & A$_{V}<2.5, 10,$& 0.15 &0.35&Keck \\[-0.0cm]
         &           &           &          &         &           & Galaxy Model & no limit    &      && H, K \\[0.5cm]

 LR2000 & M$_{J}\rightarrow\mbox{M}_{\sun}$ 
               & 440, 8.22 & 1.0, 0.0 & Derived & Assumed  & Assumed & (J-H)$<1.5$ & 0.40 &0.50& UKIRT   \\[-0.0cm]
        &      &           &          &         & None     & None    &             &      && I, J, H \\[0.5cm]

 Luhmen & IR Spectra & 450, 8.27 & 0.4, $\sim1$ & Derived & Assumed & Assumed & A$_{H}<1.4$ & 0.25 &0.073& NICMOS \\[-0.0cm]
 et al  &  + Colors  &           &              &         & None    & None    &             &      && F110, F160\\[0.5cm]

 This work & Model KLF & 400, 8.00 & 0.8, 1.2 & EPDF     & IXPDF    & Reddened  & A$_{V}<17$ & 0.10 &0.34& FLWO + NTT \\[-0.0cm]
           & Fit       &           &          & Derived & Derived & Obs. KLF  &            &      &(0.57)& J, H, K
\enddata

\tablenotetext{(a)}{All IMF derivations used the \citet{dm97} pre-main sequence models for masses less than
   1 \solarmass.}
\tablenotetext{(b)}{Cluster mean age and age spread used by authors. For \citet{luh2000}, 
   an empirical star forming history was used by the authors and those tabulated here are 
   approximate characterizations.}
\tablenotetext{(c)}{Listed if and how these quantities: extinction, excess and the contribution of
   background field stars, were included into that work's IMF derivation.}
\tablenotetext{(d)}{Size (area) of surveys in $\sq\,\mbox{parsec}$ assuming D = 400pc. The two
   values for this work are for the NTT/FLWO overlap region and the larger FLWO region. }
\tablenotetext{(e)}{Comments include location of observation(s) and broadband filters used.}

\end{deluxetable}

\clearpage

\begin{deluxetable}{llccl}
\tablecolumns{5}
\tablewidth{0pt}

\tabletypesize{\small}
\tablecaption{Summary of Near-Infrared Observations
\label{tab:obs}
}

\tablehead{
\colhead{Observatory\tablenotemark{(a)}}    &
\colhead{Date}  &
\colhead{Passband \tablenotemark{(b)} }     &
\colhead{Plate Scale \tablenotemark{(c)} }  &
\colhead{Comments} \\
\colhead{ }  &
\colhead{YYYY/MM/DD}  &
\colhead{ } &
\colhead{Beamsize} &
\colhead{ } 
}

\startdata
FLWO & 1997/12/14  &  H       & 0.596 / 3.58 & see also Lada et al. (2000) \\[0.05cm] 
FLWO & 1997/12/14  &  K       & 0.596 / 3.58 &  \\[0.05cm] 
FLWO & 1997/12/15  &  H       & 0.596 / 3.58 &  \\[0.05cm] 
FLWO & 1997/12/15  &  K       & 0.596 / 3.58 &  \\[0.05cm] 
FLWO & 1997/12/16  &  J       & 0.596 / 3.58 &  \\[0.05cm] 
FLWO & 1998/11/04  &  J       & 0.596 / 3.58 & see also Lada et al. (2000) \\[0.05cm] 
FLWO & 1998/11/04  &  H       & 0.596 / 3.58 &  \\[0.05cm] 
FLWO & 1998/11/04  &  L       & 0.596 / 3.58 &  \\[0.05cm] 
NTT  & 2000/03/14  & ~\Ks~    & 0.288 / 1.73 & see also Muench et al (2001) \\[0.05cm]
NTT  & 2000/03/14  &  H       & 0.288 / 1.73 &  \\[0.05cm]
NTT  & 2000/03/14  &  J       & 0.288 / 1.73 &  \\[0.05cm]
NTT  & 2000/03/14  & ~\Ks~    & 0.288 / 1.73 &  
\enddata

\tablenotetext{(a)}{FLWO: Fred Lawrence Whipple Observatory,
Smithsonian Astrophysical Observatory; NTT: New Technology Telescope, European Southern Observatory.}
\tablenotetext{(b)}{Filter central wavelengths $\lambda\,(\micron))$:
FLWO- J) $\;1.25\;$; H) $\;1.65\;$;   K)  $\;2.20\;$; L) $\;3.50\;$; .
NTT-  J) $\;1.25\;$; H) $\;1.65\;$; ~\Ks) $\;2.16\;$. }
\tablenotetext{(c)}{Plate scale: arcsec/pixel; Beamsize: the effective diameter of photometry beam (arcsec)}

\end{deluxetable}

\clearpage

\begin{deluxetable}{llllccccccccccccccrrrr}
\tablecolumns{22}
\tablewidth{0pt}
\rotate

\tabletypesize{\tiny}
\tablecaption{FLWO-NTT Near-Infrared Catalog
\label{tab:catalog}}

\tablehead{
\colhead{Seq.}  &
\colhead{Spec.} &
\colhead{R.A.}  &
\colhead{Dec.}  &
\multicolumn{4}{c}{FLWO (Mag)} &
\multicolumn{4}{c}{FLWO (Err)} &
\multicolumn{3}{c}{NTT  (Mag)} &
\multicolumn{3}{c}{NTT  (Err)} &
\colhead{Phot}   &
\colhead{H97}    &
\colhead{HC2000} &
\colhead{Other } \\

\colhead{}   &
\colhead{}     &
\colhead{(J2000)} &
\colhead{(J2000)} &
\colhead{J }    &
\colhead{H }    &
\colhead{K }    &
\colhead{L }    &
\colhead{J }    &
\colhead{H }    &
\colhead{K }    &
\colhead{L }    &
\colhead{J }    &
\colhead{H }    &
\colhead{\Ks~}  &
\colhead{J }    &
\colhead{H }    &
\colhead{\Ks~}  &
\colhead{Flag } &
\colhead{ID }   &
\colhead{ID }   &
\colhead{ID }
}
\startdata
 00001 &\phm{A}& 5 35 22.45 & -5 26 10.9 & 14.59 & 13.68 & 99.00 & 12.40 & 0.02 & 0.06 & -1.00 &  0.26 &         &       &       &       &       &       &   0  &           &   &   \\
 00002 &\phm{A}& 5 35 26.57 & -5 26 09.6 & 16.24 & 15.41 & 99.00 & 99.00 & 0.04 & 0.01 & -1.00 & -1.00 &         &       &       &       &       &       &   0  &           &   &   \\
 00003 &\phm{A}& 5 35 11.65 & -5 26 09.0 & 13.51 & 11.63 & 10.56 &  9.21 & 0.04 & 0.01 &  0.02 &  0.02 &         &       &       &       &       &       &   0  &           &   &   \\
 00004 &\phm{A}& 5 35 15.97 & -5 26 07.4 & 14.12 & 12.25 & 99.00 & 10.30 & 0.02 & 0.01 & -1.00 &  0.04 &         &       &       &       &       &       &   0  &           &   &   \\
 00005 &\phm{A}& 5 35 09.21 & -5 26 05.7 &       &       &       &       &      &      &       &       &   17.01 & 16.18 & 15.65 &  0.05 &  0.04 &  0.06 &   0  &           &   &   \\
 00006 &\phm{A}& 5 35 20.13 & -5 26 04.2 &       &       &       &       &      &      &       &       &   14.57 & 13.01 & 12.27 &  0.01 &  0.01 &  0.01 &   0  &           &   &  0535201-052604(4)  \\
 00007 &\phm{A}& 5 35 04.48 & -5 26 04.1 & 12.14 & 11.25 & 10.89 & 10.15 & 0.03 & 0.04 &  0.03 &  0.03 &         &       &       &       &       &       &   0  &           &   &  0535044-052604(4)  \\
 00008 &\phm{A}& 5 35 05.18 & -5 26 03.4 & 13.54 & 12.88 & 12.52 & 12.04 & 0.01 & 0.01 &  0.03 &  0.12 &         &       &       &       &       &       &   0  &      262  &   &  \\
 00009 &\phm{A}& 5 35 11.48 & -5 26 02.6 &  9.91 &  9.06 &  8.84 &  8.47 & 0.05 & 0.01 &  0.02 &  0.02 &         &       &       &       &       &       &   0  &      365  &   &  \\
 00010 &\phm{A}& 5 35 22.57 & -5 26 02.1 &       &       &       &       &      &      &       &       &   17.18 & 16.56 & 16.37 &  0.09 &  0.08 &  0.08 &   0  &           &   &  \\
 00011 &\phm{A}& 5 35 06.92 & -5 26 00.5 & 13.48 & 13.14 & 12.54 & 11.43 & 0.02 & 0.05 &  0.02 &  0.06 &   13.39 & 12.56 & 12.19 &  0.01 &  0.01 &  0.01 &   0  &      299  &   &  \\
 00012 &\phm{A}& 5 35 24.34 & -5 26 00.3 & 13.09 & 12.42 & 12.12 & 11.38 & 0.08 & 0.03 &  0.02 &  0.10 &   13.05 & 12.37 & 12.02 &  0.01 &  0.01 &  0.01 &   0  &     3101  &   &  \\
 00013 &\phm{A}& 5 35 10.48 & -5 26 00.3 & 13.01 & 12.25 & 11.82 & 11.11 & 0.03 & 0.03 &  0.02 &  0.05 &   12.92 & 12.07 & 11.66 &  0.01 &  0.01 &  0.01 &   0  &     3104  &   &  \\
 00014 &\phm{A}& 5 35 10.76 & -5 26 00.0 & 15.54 & 13.61 & 12.60 & 11.72 & 0.04 & 0.08 &  0.02 &  0.09 &   15.42 & 13.60 & 12.57 &  0.01 &  0.01 &  0.01 &   0  &           &   &  \\
 00015 &\phm{A}& 5 35 15.42 & -5 25 59.5 & 13.72 & 12.84 & 12.38 & 11.56 & 0.01 & 0.02 &  0.02 &  0.06 &   13.68 & 12.78 & 12.26 &  0.01 &  0.01 &  0.02 &   0  &     3103  &   &
\enddata

\tablecomments{The complete version of this table is in the electronic
edition of the Journal.  The printed edition contains only a sample.}
\tablecomments{This table is available only on-line as a machine-readable table.}

\end{deluxetable}

\clearpage
\newpage

\figcaption[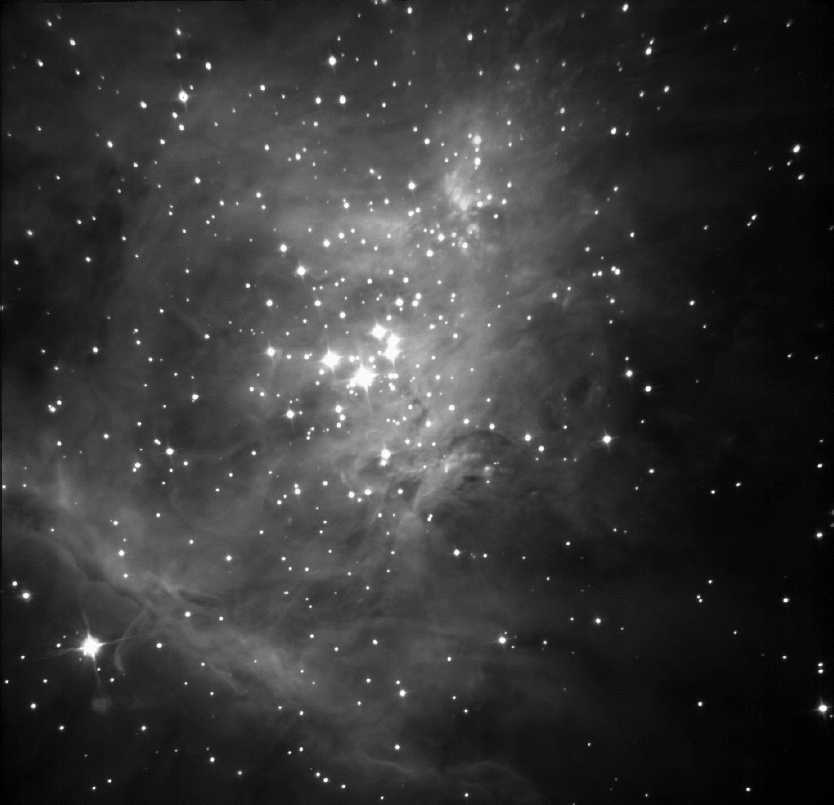]{
Color composite near-infrared  (JH\Ks) NTT image of the Trapezium Cluster.
Taken with SOFI at the ESO NTT telescope, La Silla, Chile, March 2000.
North is up and east is left and the field of view is $5\arcmin\,\times\,5\arcmin$.
Note the very red objects which trace the location of the parental molecular cloud
and lie in an arcing ridge from southwest of the namesake Trapezium stars through the
Orion South and the BN/KL regions and then directly north the Trapezium stars.
The NTT region is compared to that of other deep near-IR surveys in 
figure \ref{fig:trap_area}.
\label{fig:trap_color}
}

\figcaption[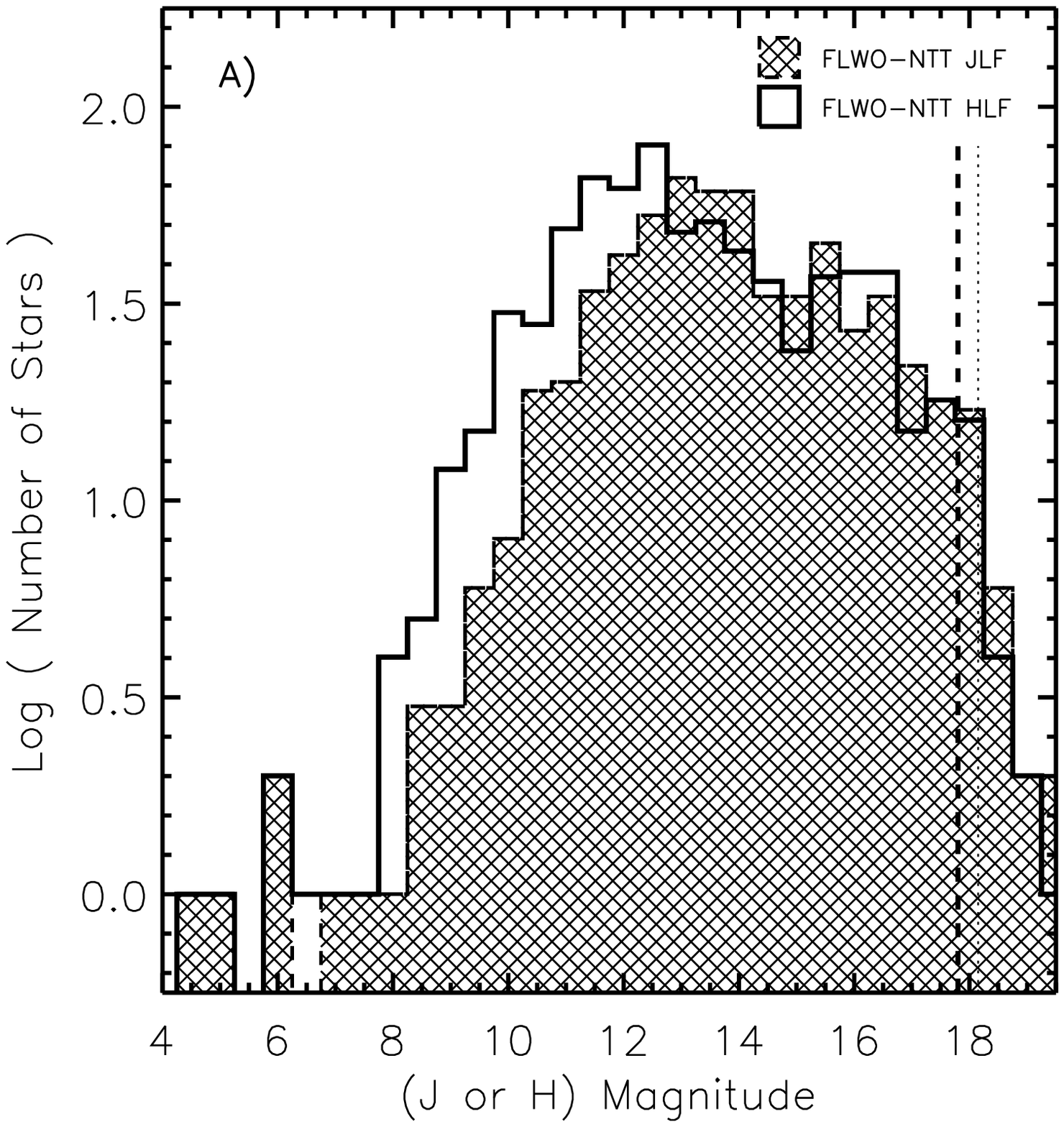,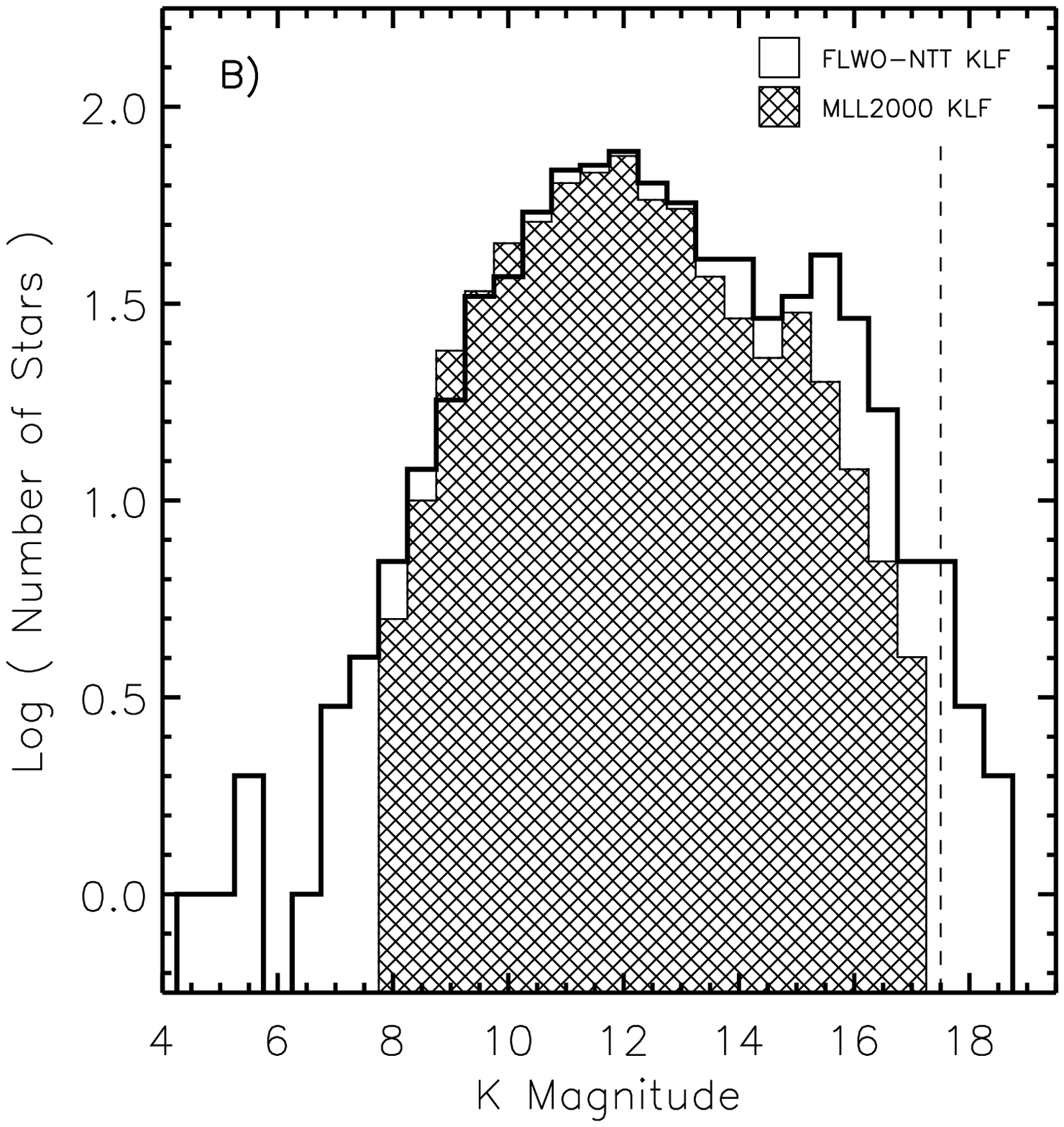]{
Raw near-infrared infrared luminosity functions.
a) Trapezium Cluster J and H band Luminosity Functions.  
The Trapezium HLF is the open histogram and the Trapezium JLF is the shaded histogram.
Completeness (90\%) limits are marked by a solid vertical line at 18.15 (J)
and a broken vertical line at 17.8 (H).
b) Trapezium Cluster K band Luminosity Function.
The Trapezium KLF constructed from the FLWO-NTT catalog is compared to the KLF
constructed in MLL2000.  The K=17.5 90\% completeness limit is demarked by a
vertical broken line.
\label{fig:jhklf}
}

\figcaption[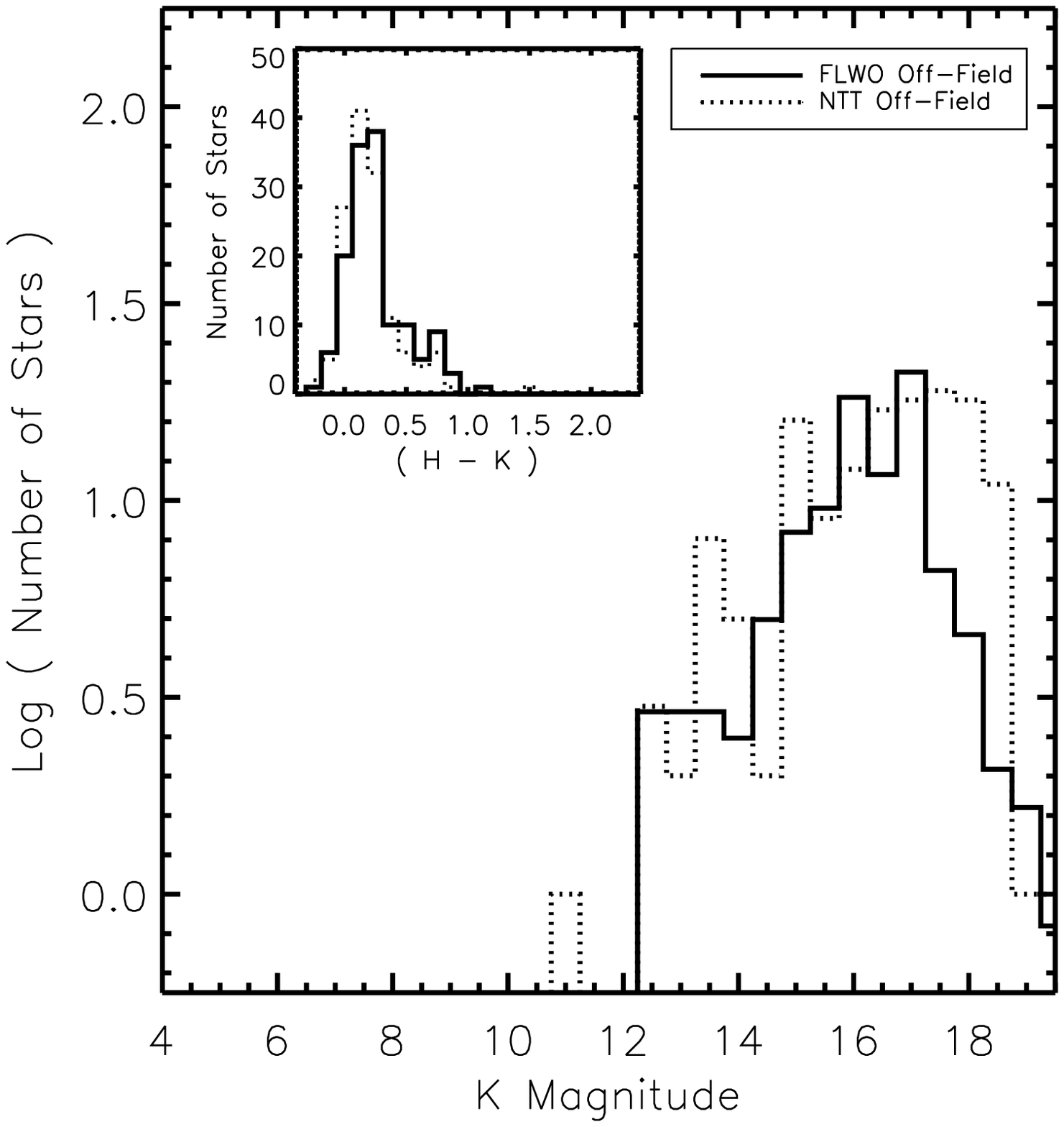,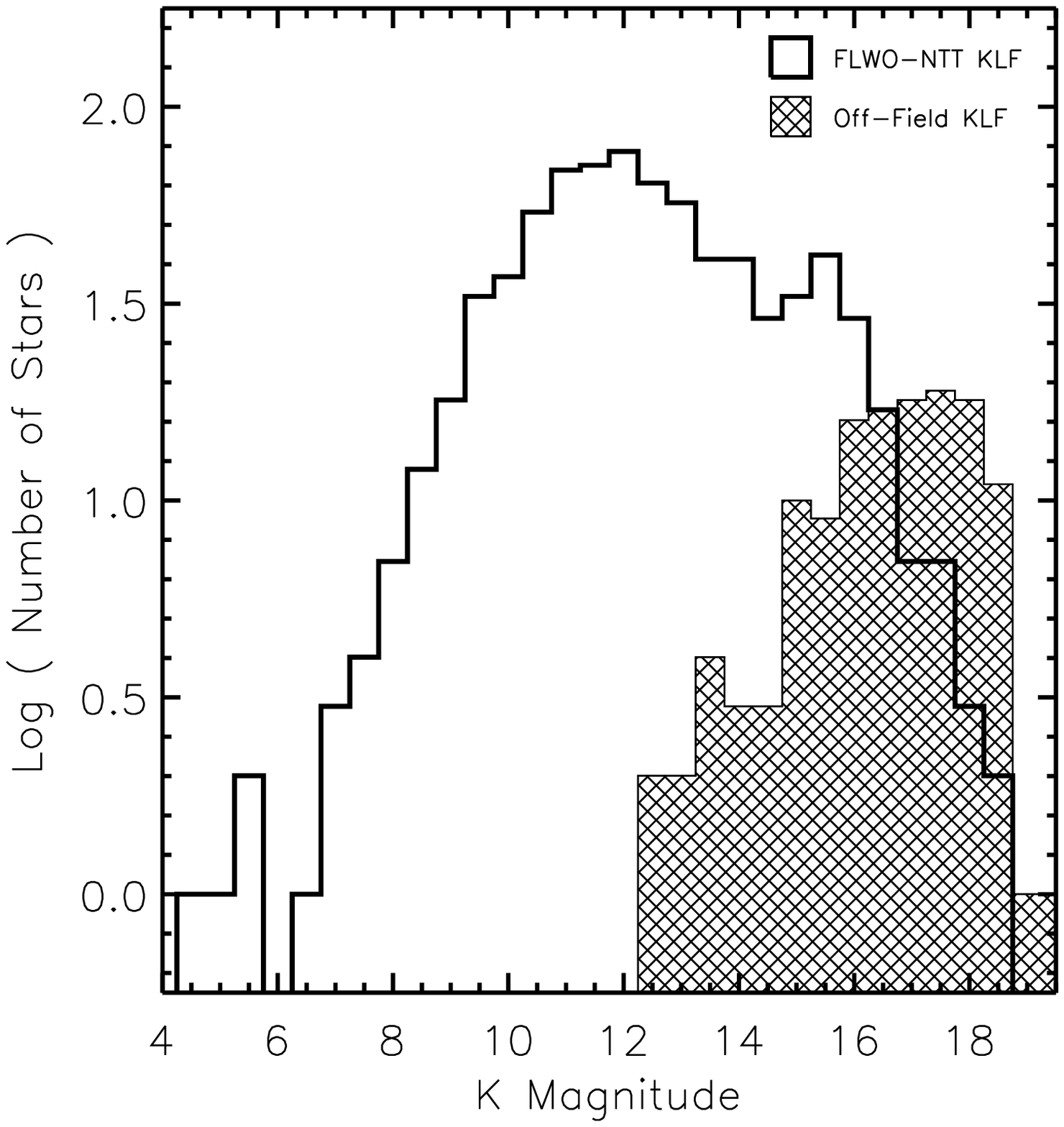]{
Observed control field KLF for the Trapezium Cluster.
a) The two histograms are the off-field KLFs obtained as part of the FLWO and NTT
observations. The NTT off-fields are aproximately 2 magnitudes deeper than
the FLWO off-fields, but the FLWO off-fields covered twice the area of the NTT off-fields.
Both are scaled to the size of the Trapezium NTT region.
The inset diagram shows the distribution of H-K colors for these two off-fields. 
Their similar narrow widths indicate they are free of interstellar extinction; 
b) The weighted average of the FLWO and NTT field stars KLFs is compared
to  the Trapezium Cluster KLF constructed in figure \ref{fig:jhklf}(b).
\label{fig:off_klf}
}

\figcaption[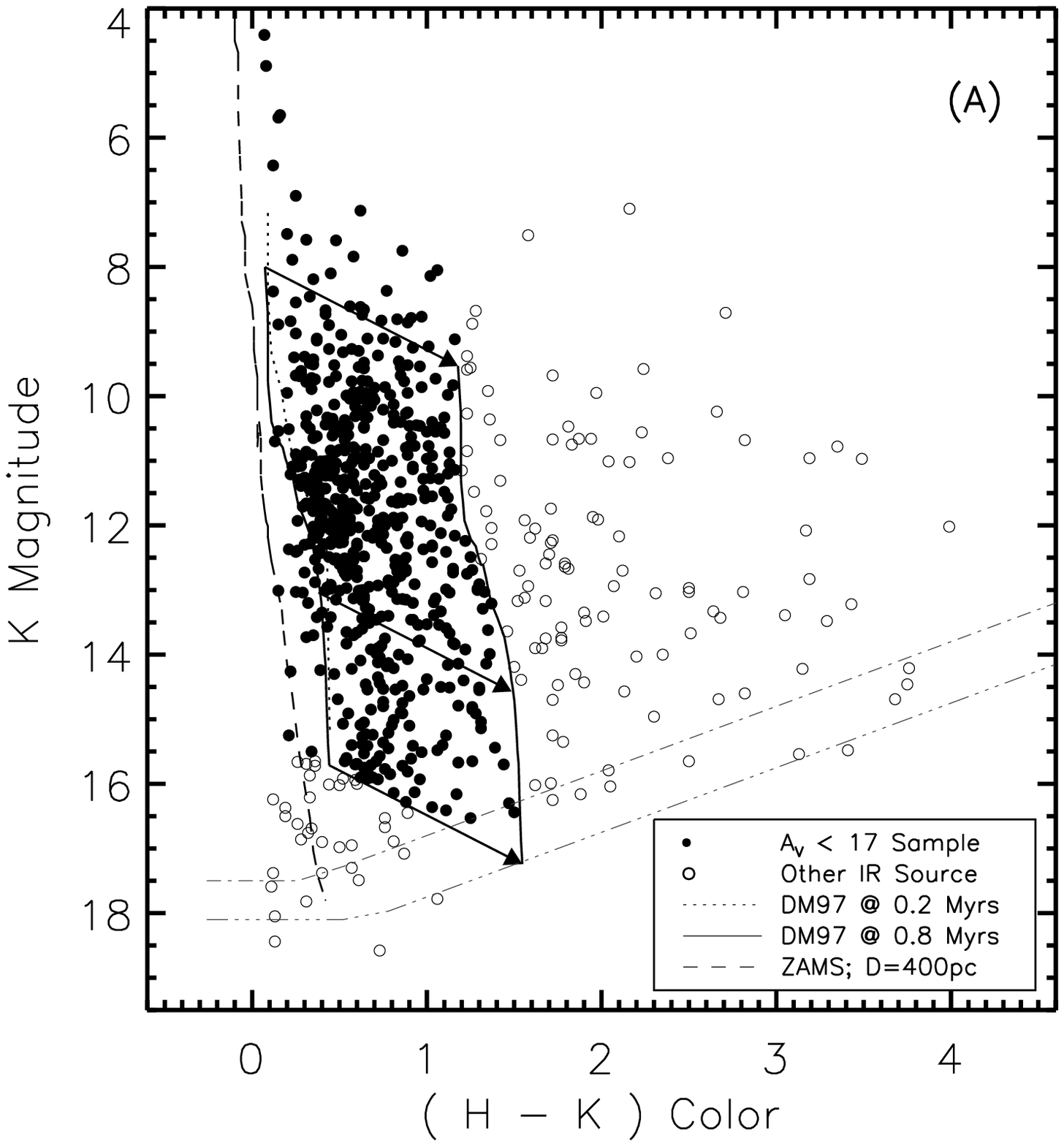, 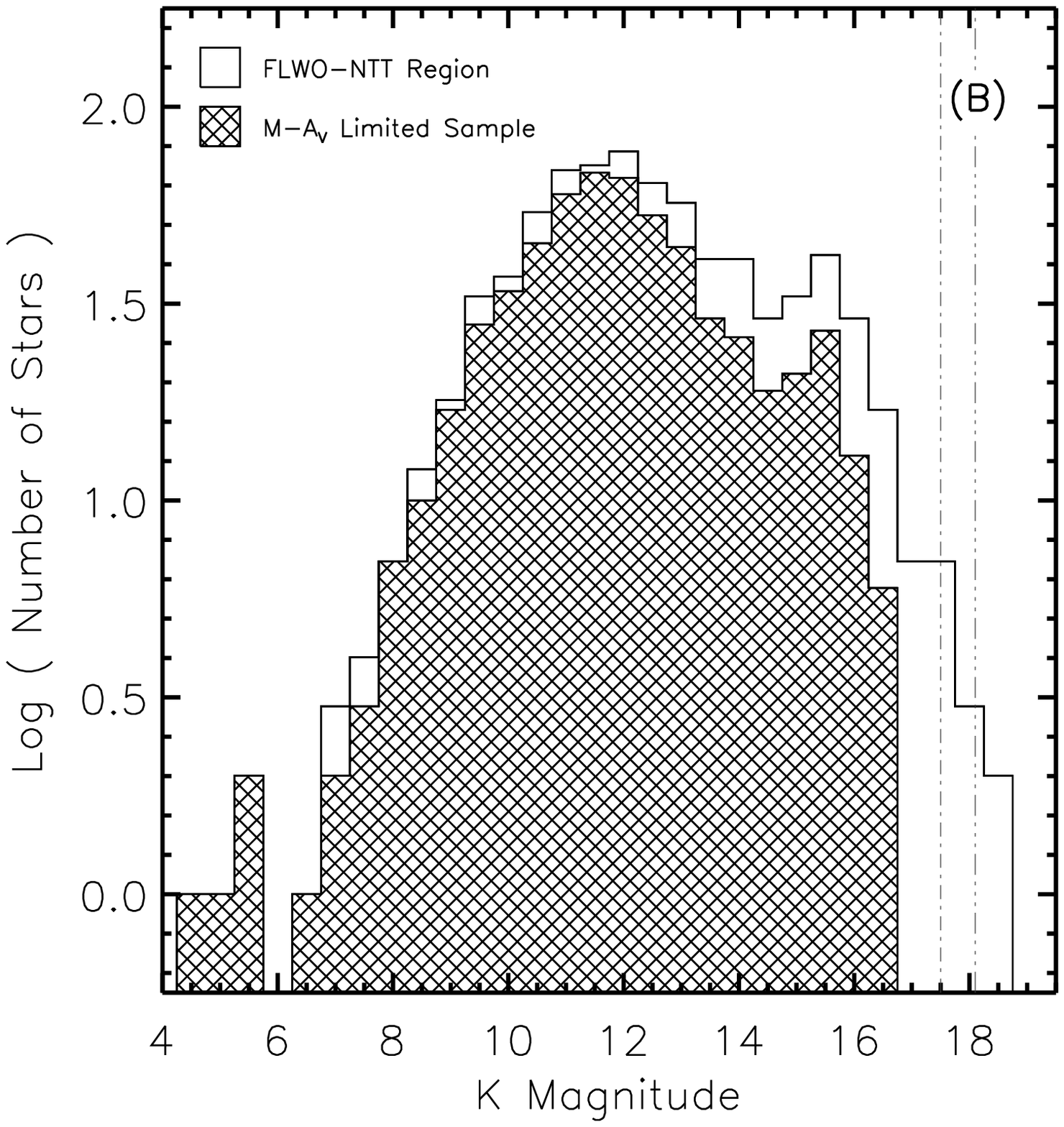]{
Deriving cluster KLF $\mavlim$~Completeness limits.
a) Trapezium Cluster (H - K) / K color-magnitude diagram.  Only stars falling within the
Trapezium NTT region are shown.  Stars selected to fall into our mass-limited, extinction-limited 
sample are indicated by filled circles.  The distribution of sources in this
color-magnitude space is compared to the location of the pre-main sequence 0.2 and 0.8 Myrs
isochrones from DM97.  Reddening vectors representing 17 magnitudes of visual extinction are
shown for 2.50, 0.08 and 0.02 \solarmass~stars at the mean age assumed for this cluster (0.8 Myrs).
The zero-age main sequence \citep{kh95,bes95} is shown for 03-M6.5 stars at a distance of 400pc.
b) Effects of mass/extinction limits on the cluster KLF. Comparison of the $\mavlim$~limited
KLF derived from (a) to the raw Trapezium KLF (see figure \ref{fig:jhklf}b). 
Sensitivity (K = 18.1) and completeness (K = 17.5) limits are shown as vertical broken lines. 
\label{fig:mavlim}
}

\figcaption[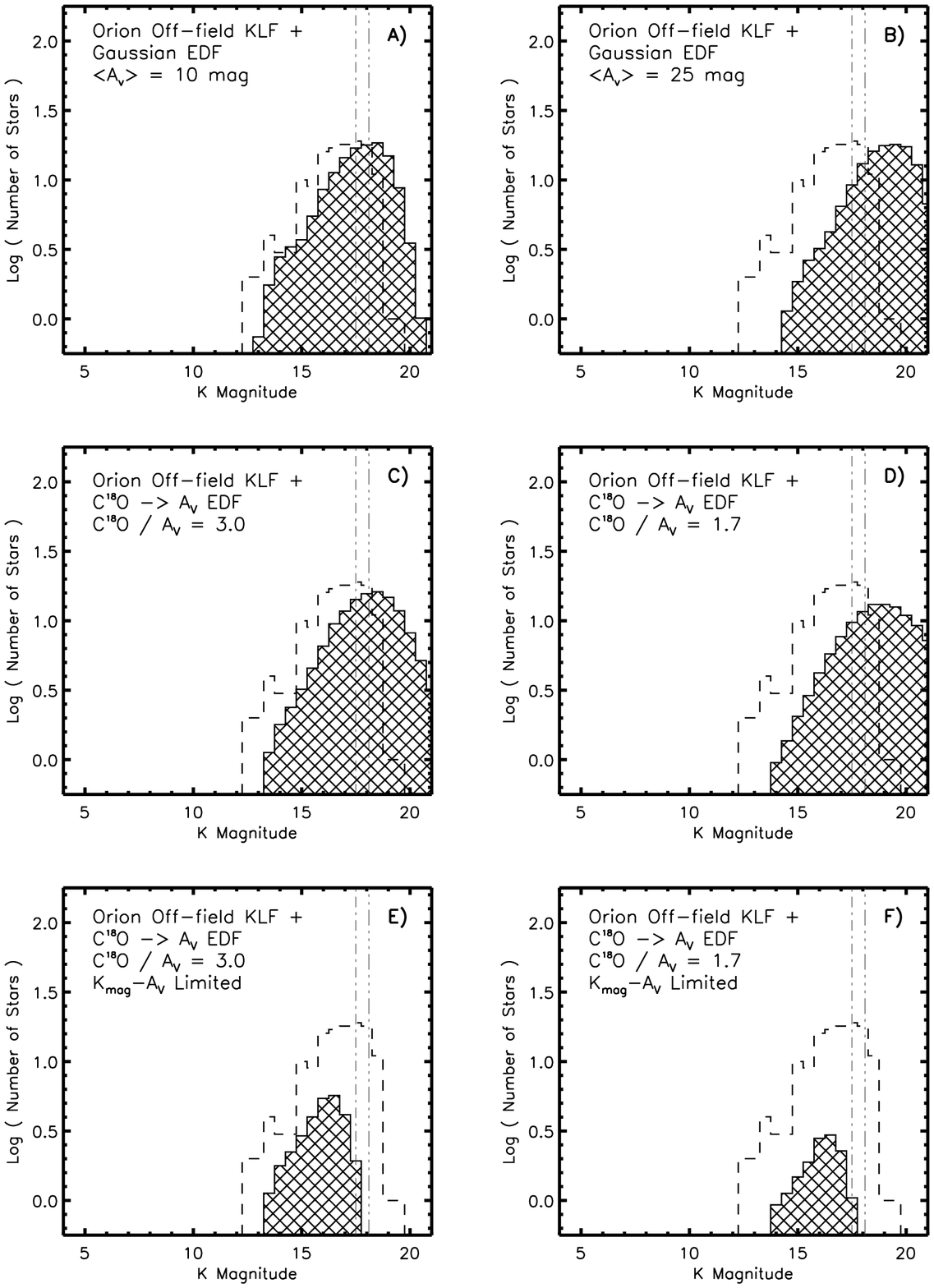]{
Testing the contribution of reddened field stars to the Trapezium Cluster KLF.  
Panels A \& B: The first pair of panels displays the observed off-field KLF (figure \ref{fig:off_klf}b )
reddened by ``background shields'' of extinction in the form of gaussian distributions
(of $\av$)~centered at 10 (panel A) and 25 (panel B) magnitudes; 
Panels C \& D: The second two panels are the observed off-field KLF reddened by an extinction map
converted from a $\cotracer$~map. The two panels represent the variation in the predicted
field star contamination as a function of the uncertainty in the $\cotracer$~to~$\av$~conversion;
Panels E \& F: The final panels are of the same experiment as performed in C \& D, but they have been
filtered to reflect the actual contribution due to the $\mavlim$~limits.
\label{fig:off-fraction}
}

\figcaption[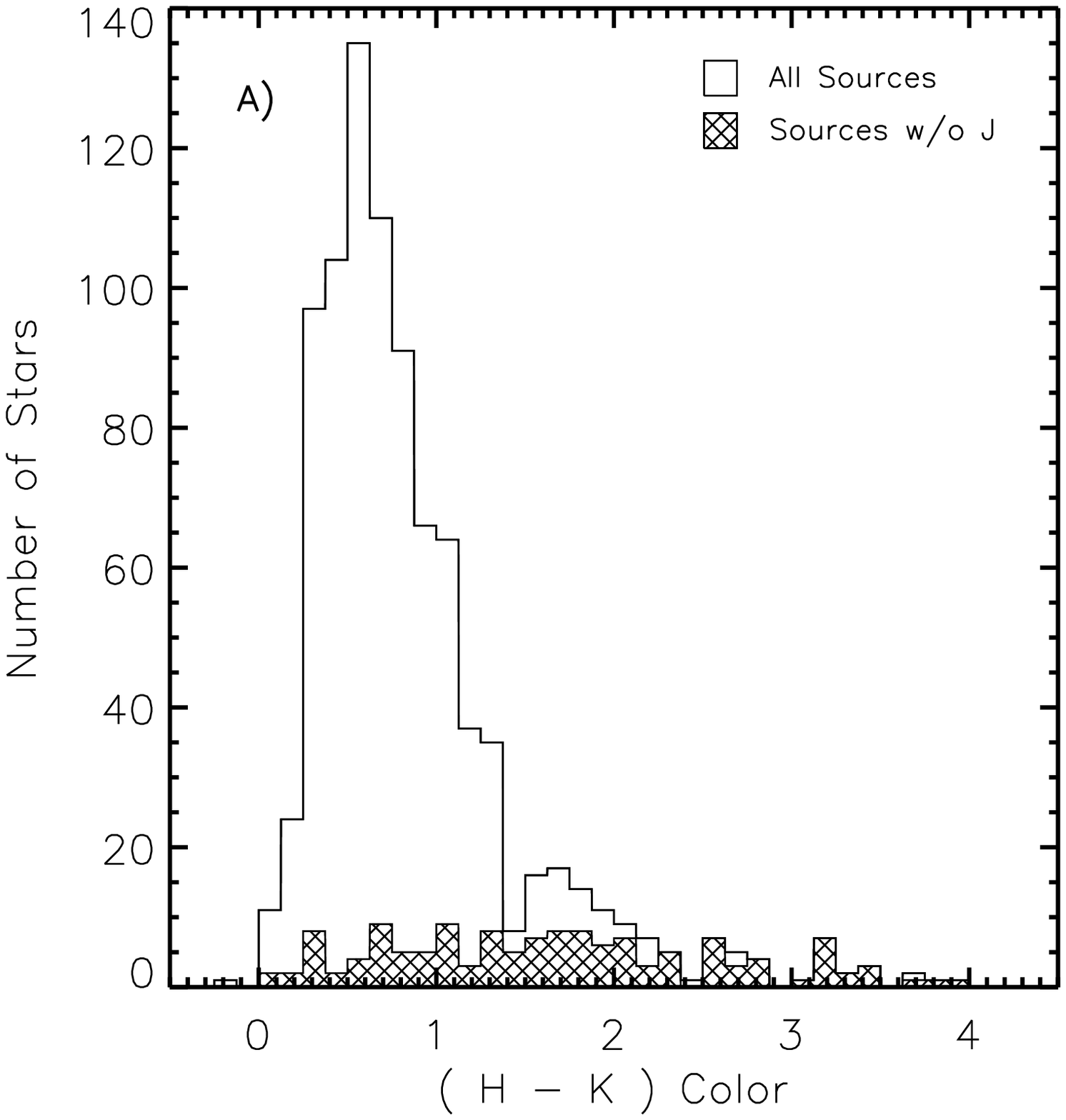,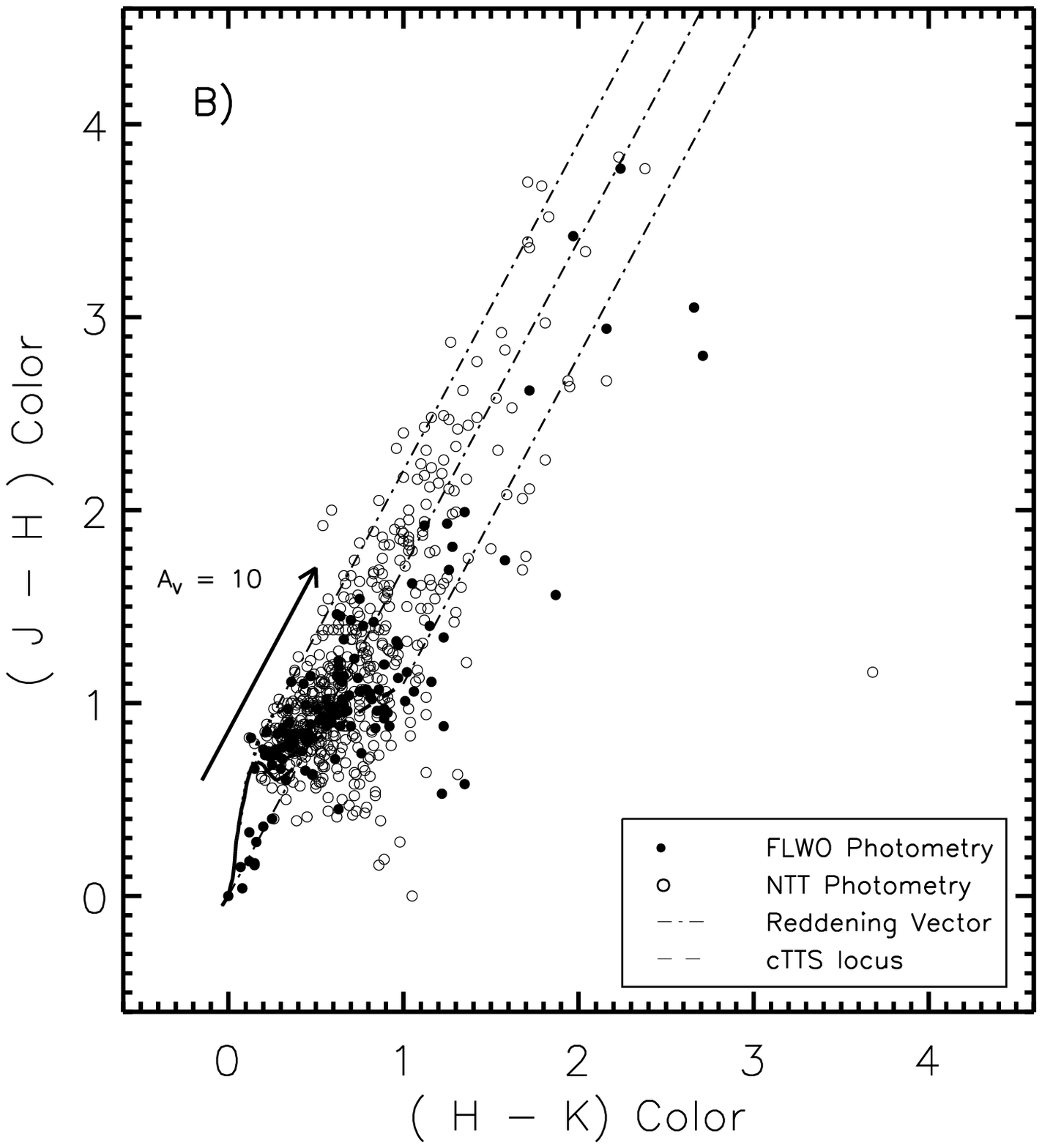]{
Infrared color distributions for the FLWO-NTT Trapezium observations.
a) Histogram of the observed (H - K) color for the FLWO-NTT Trapezium sources. 
The subset of these sources which lack J band measurements are indicated by the shaded histogram; 
b) Trapezium Cluster (H - K) vs (J - H) color-color diagram for the NTT region.
Symbols indicate if the source's colors were taken from the FLWO catalog (filled circles, JHK) or 
the NTT catalog (open circles, JH\Ks).
\label{fig:hk_jhk}
}

\figcaption[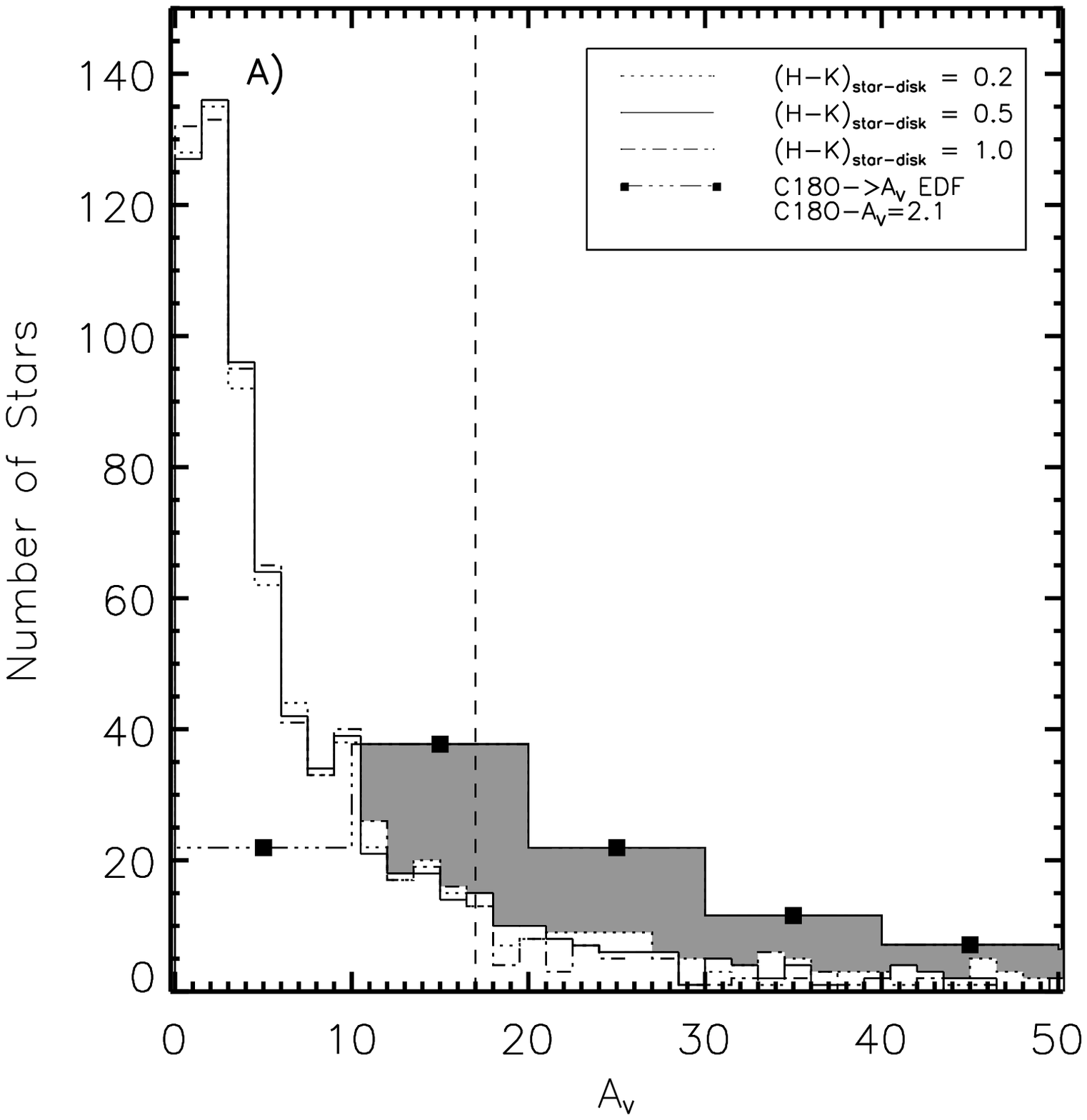,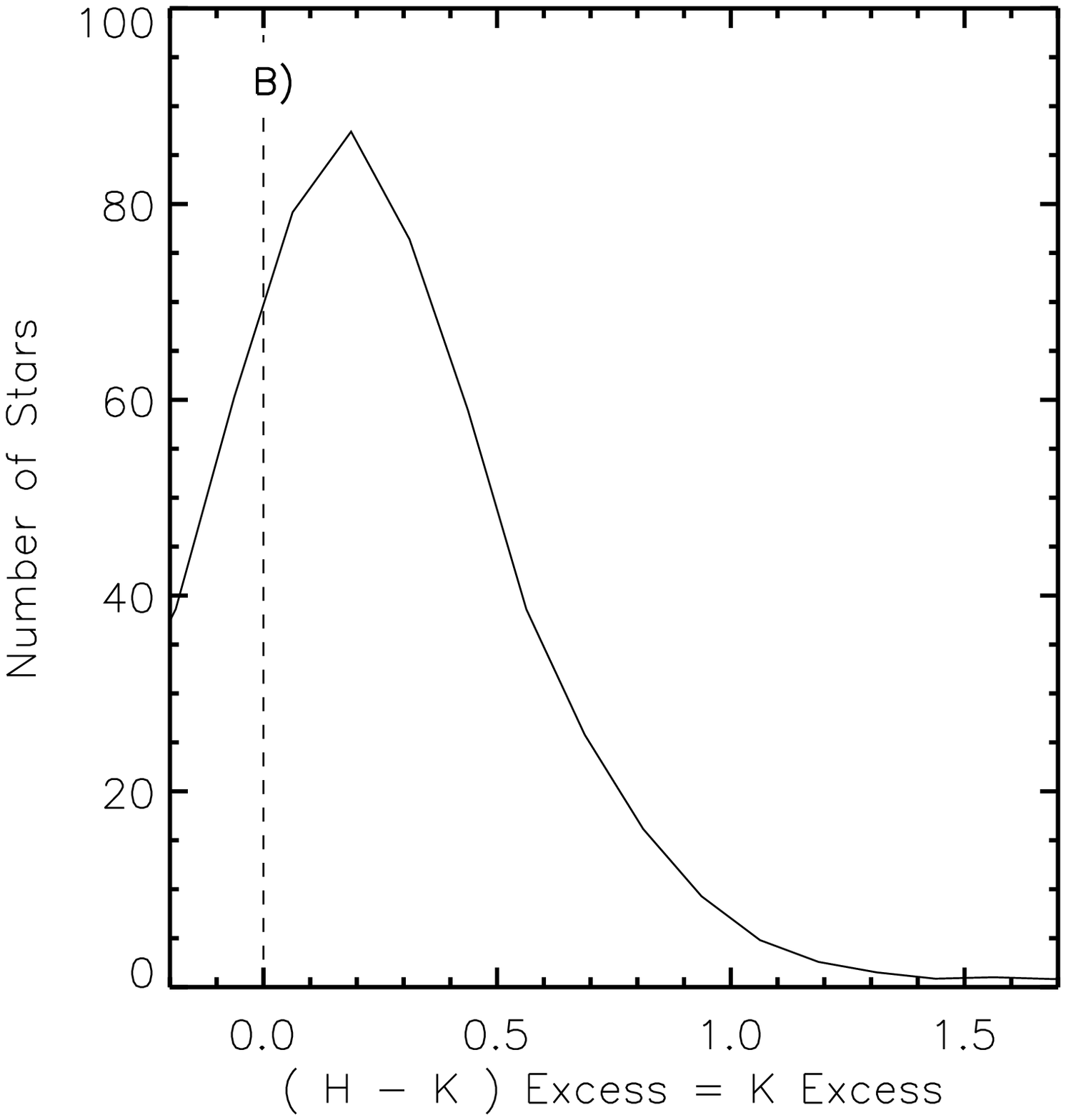]{
Probability distributions of source reddening for the Trapezium Clusters sources.
a) Extinction Probability Distribution Function (EPDF) for the Trapezium Cluster sources.
   Plotted are three variations in the EPDF under different assumptions
   of the typical (H-K)$_{\mbox{\small(star-disk)}}$ color for the 20\% of the stars
   lacking J band measurements. See section \ref{sec:imf:recipe_av} for derivation.
   It is compared to the extinction probability distribution function integrated
   from the $\cotracer\;\rightarrow\;\av$ map. Note that they are not well separated
   distributions.  A broken vertical line indicates the A$_{V}\:=\:17$, $\mavlim$~limit;
b) Infrared Excess Probability Distribution Function (IXPDF) for the Trapezium Cluster. 
   The derived H-\Ks~color excess distribution function is assumed to reflect a
   magnitude excess at K band alone.  See section \ref{sec:imf:recipe_irex} for derivation.
\label{fig:av_ixex}
}

\figcaption[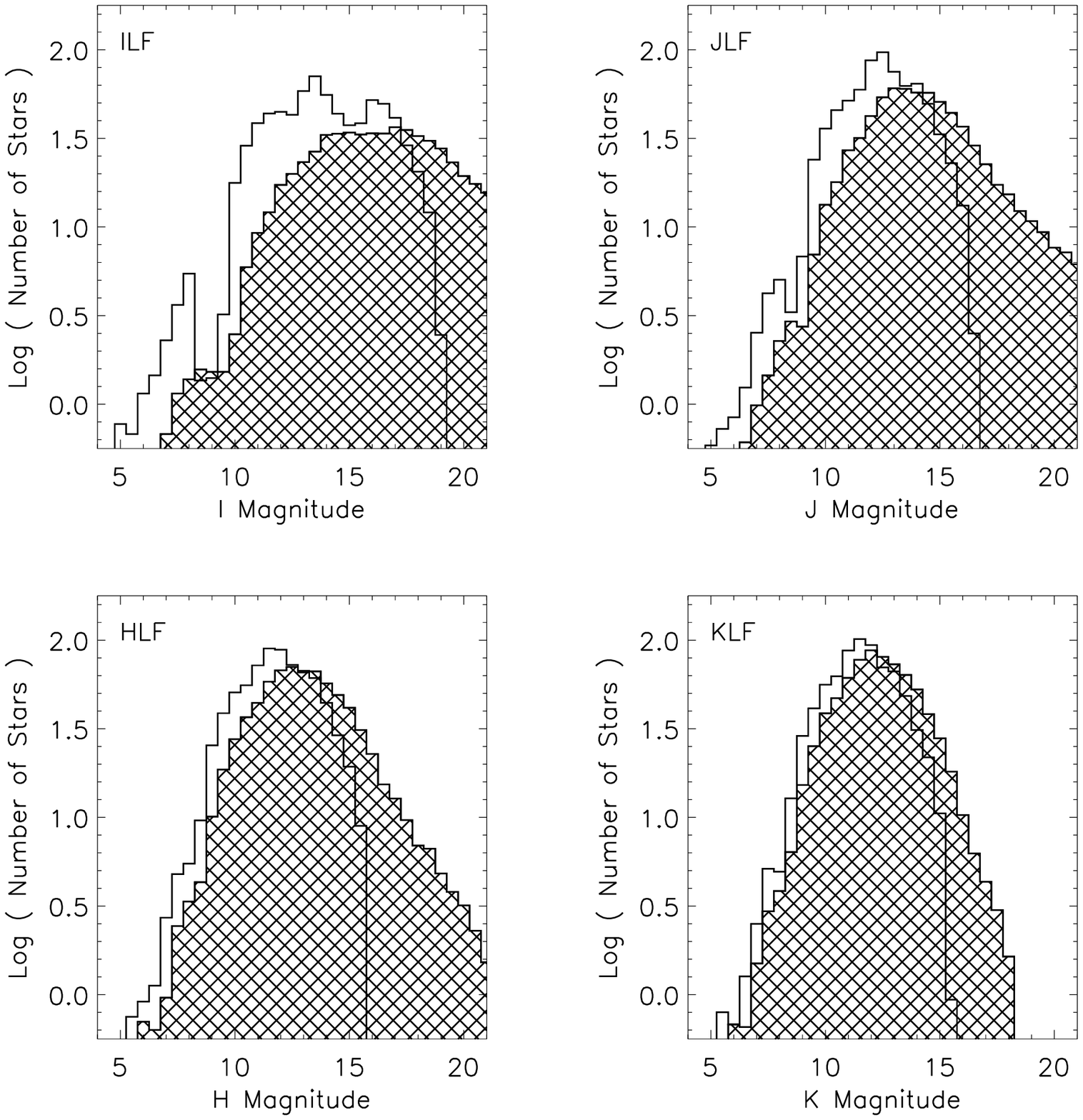]{
Effect of the Trapezium Cluster EPDF on cluster luminosity functions. 
A model luminosity function of the Trapezium (using the Trapezium IMF derived in MLL2000)
is convolved with the Trapezium Cluster EPDF at four different wavelengths.
Reddening effects are most significant at I and J bands and are minimized at K band.
\label{fig:lf_av}
}

\figcaption[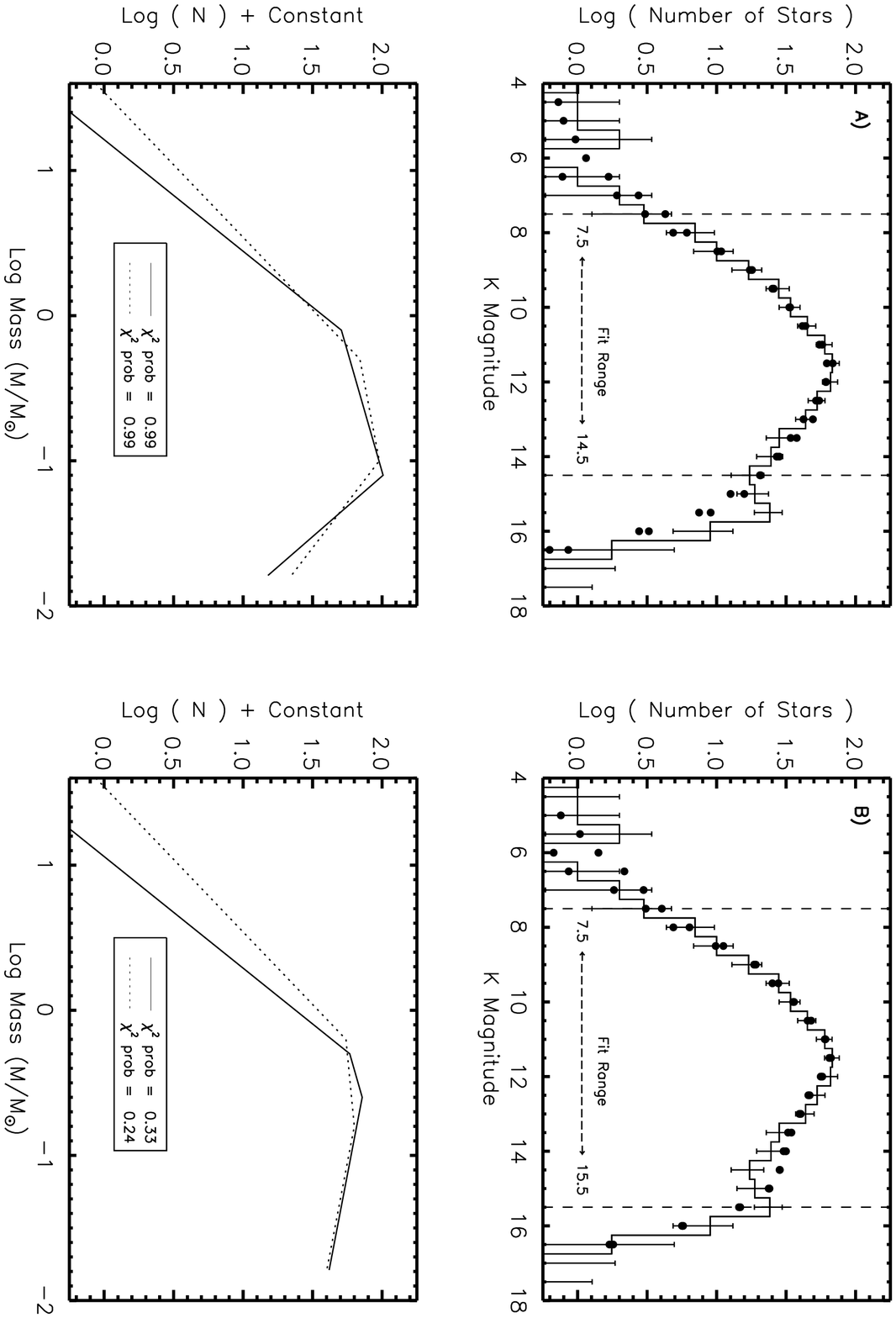]{
Best fitting model KLFs and corresponding cluster IMFs.
Top panels show the $\mavlim$~limited, background subtracted Trapezium KLF
(histogram) and best fit reddened model KLFs (unconnected filled circles).
Bottom panels display the resulting underlying IMFs and corresponding chi-sq probabilities.
Panel (a) shows models fit between K=7.5 and 14.5, the same range fit in MLL2000.
Panel (b) shows three power-law IMF fits to the secondary peak in the cluster KLF at K=15.5,
which correspond to low $\chi^{2}$ probability due to the presence of structure and the secondary 
KLF peak.
\label{fig:fit-best}
}

\figcaption[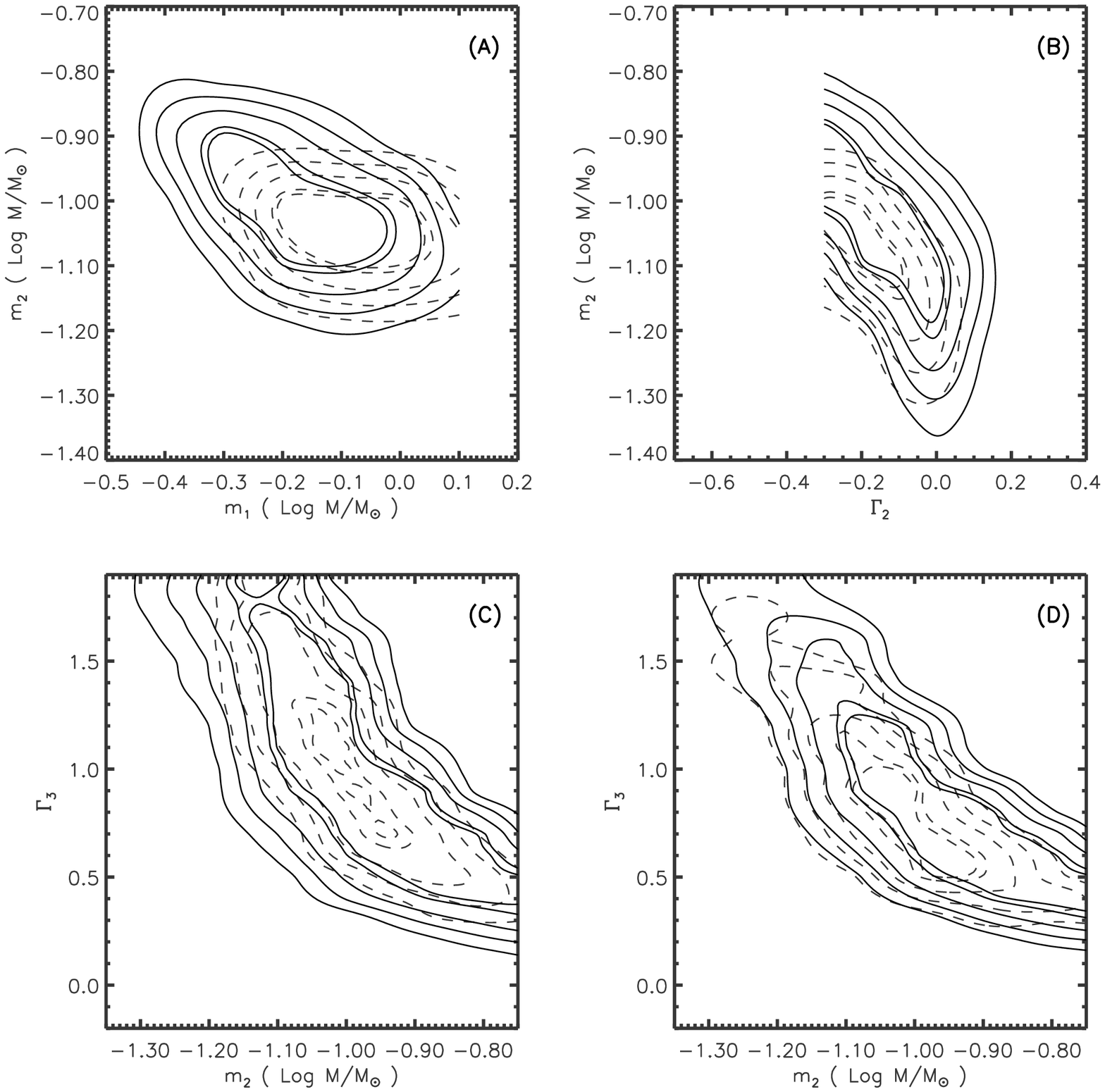]{
Contours of $\chi^2$ probability for the 5 parameters of the underlying three power-law IMF.
Two parameters are compared in each panel while fixing the other three to a best fit value.
Solid contours are best fit ranges from models that include source reddening.
Dashed contours are from best fit models without source reddening.
Contour levels are shown at intervals 95, 90, 70, 50 and 30\% confidence.
Panels (a) to (c) are shown for fits to K=14.5 and panel (d) is  shown for fits to K=15.
\label{fig:fit-range}
}

\figcaption[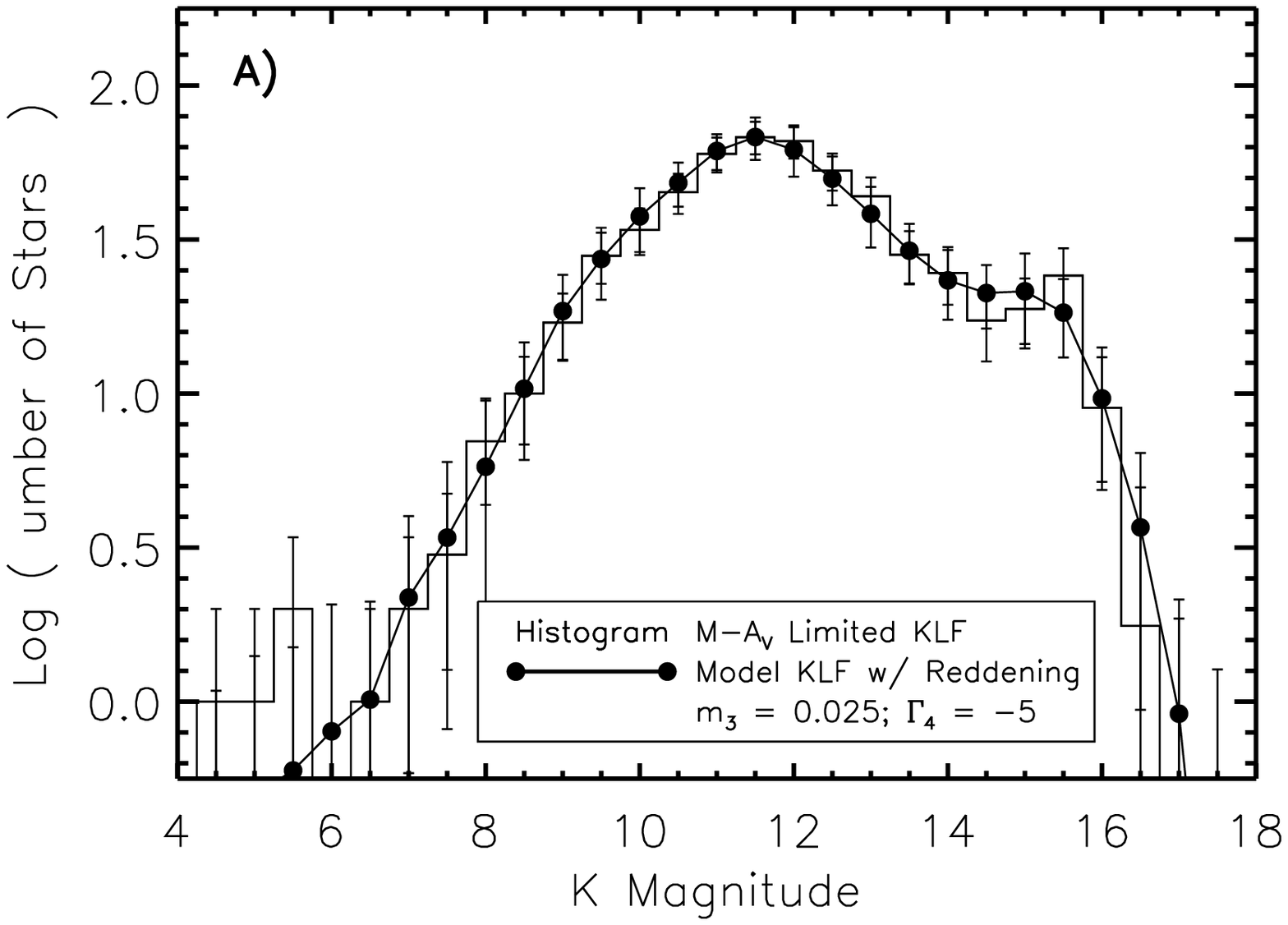, 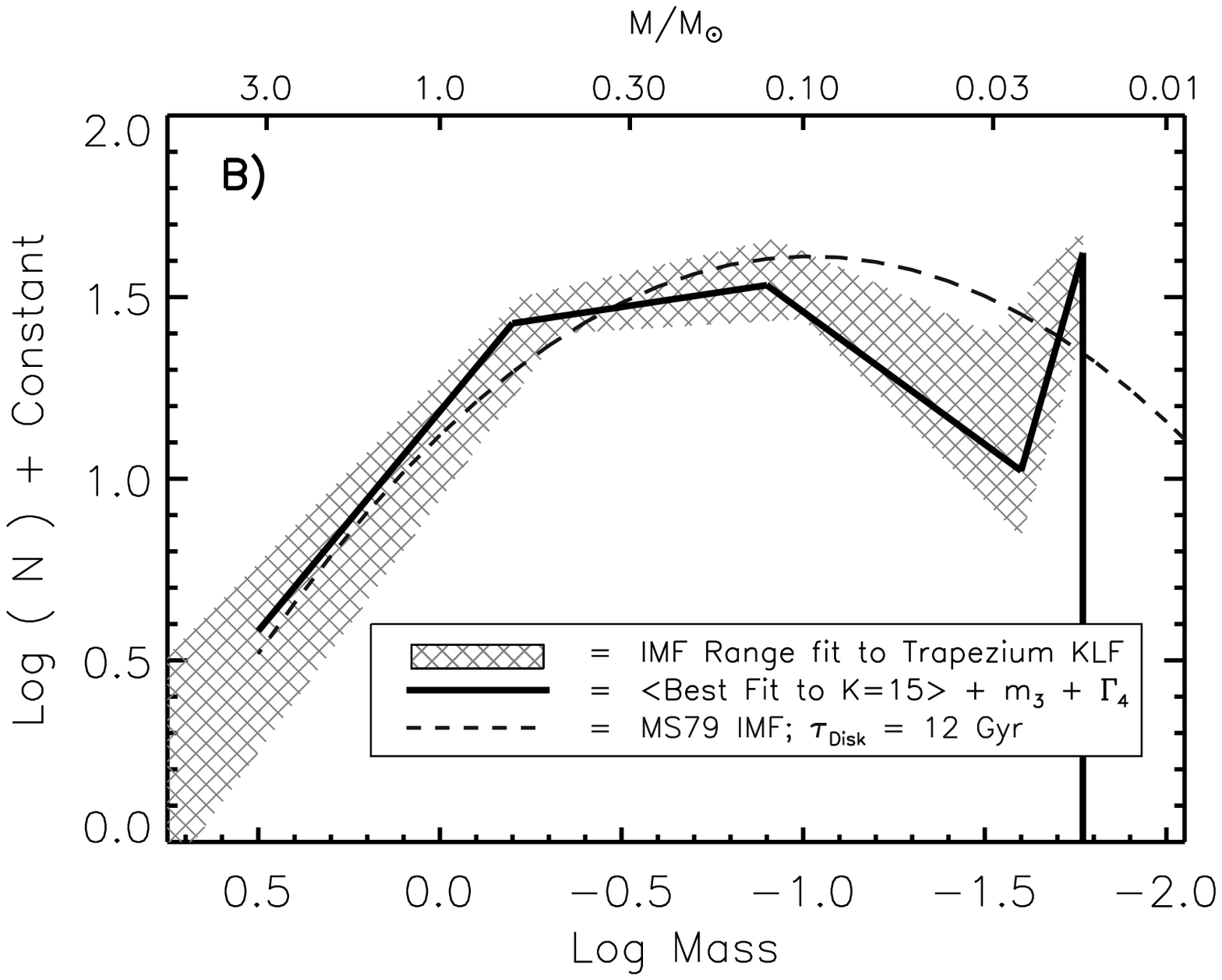]{Overall Trapezium IMF.
a) Trapezium $\mavlim$~limited KLF fit over the entire luminosity range by employing a four power-law
   IMF that truncates at the lower mass limit our standard merged PMS tracks ($\sim\,17\,\jupmass$).
   The brown dwarf IMF breaks from a steady decline ($\Gamma_{3}\,=\,0.73$)
   between 0.03 and 0.02 $\solarmass$~and rises to the mass limit of the PMS tracks.
   The rapid tailing off of the cluster KLF does imply that the sub-stellar Trapezium IMF is
   characterized by a truncation near 17~\jupmass~(see section \ref{sec:discuss:imf_klf:substar}). 
   Above the $m_{3}$ mass break the IMF is that described in table \ref{tab:imf_pars} for fits to K = 15.
b) Best fit Trapezium IMF and range of Trapezium IMFs derived from KLF modeling.
   Hatched areas are derived from the range of 90\% confidence contours for KLF fits.
   Solid line is the best fit Trapezium IMF listed in eq. \ref{eq:best_imf_dm97}.
   The Trapezium IMF is also compared to the log-normal \citet{ms79} field star IMF. 
\label{fig:imf_derived}
}

\figcaption[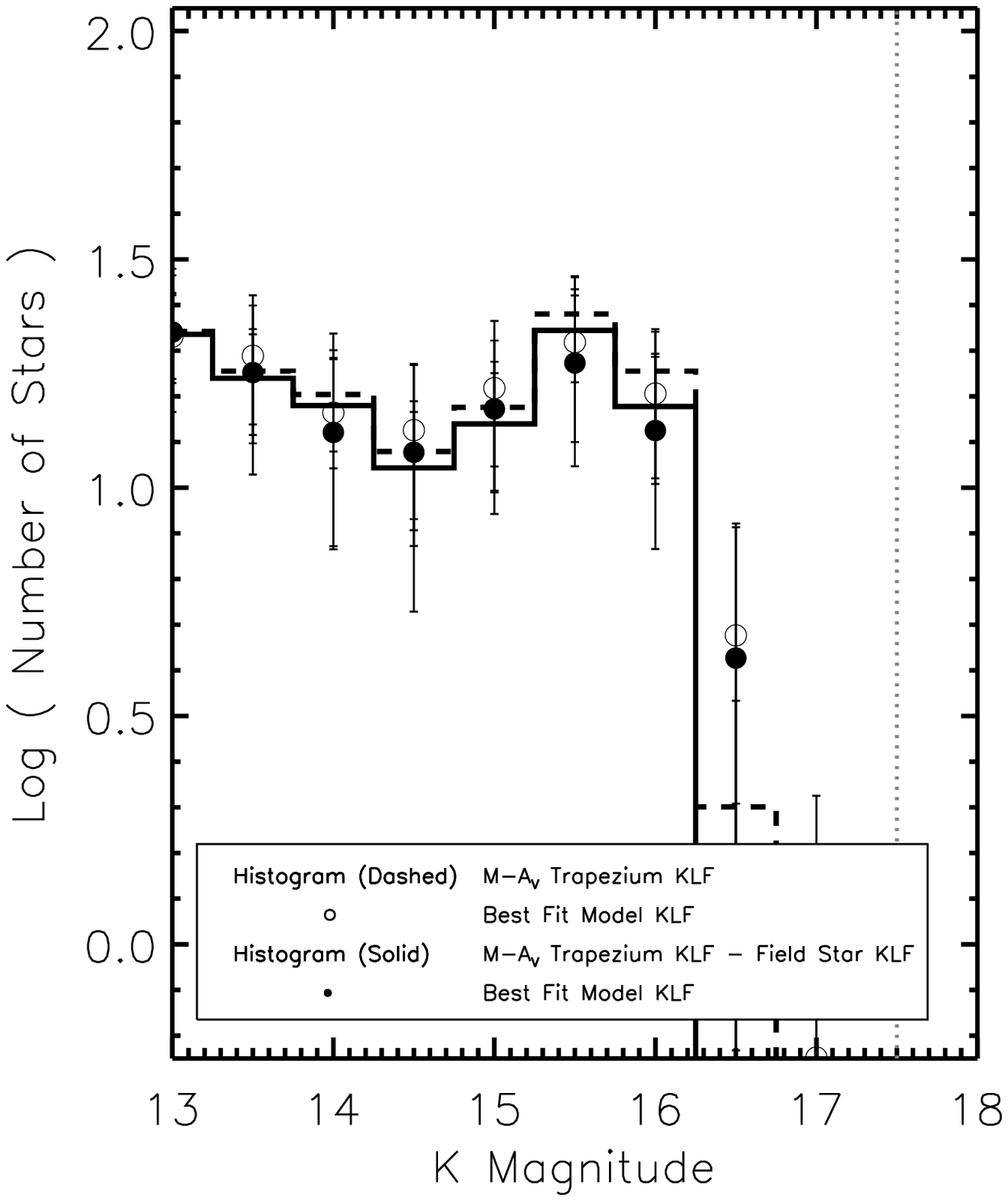,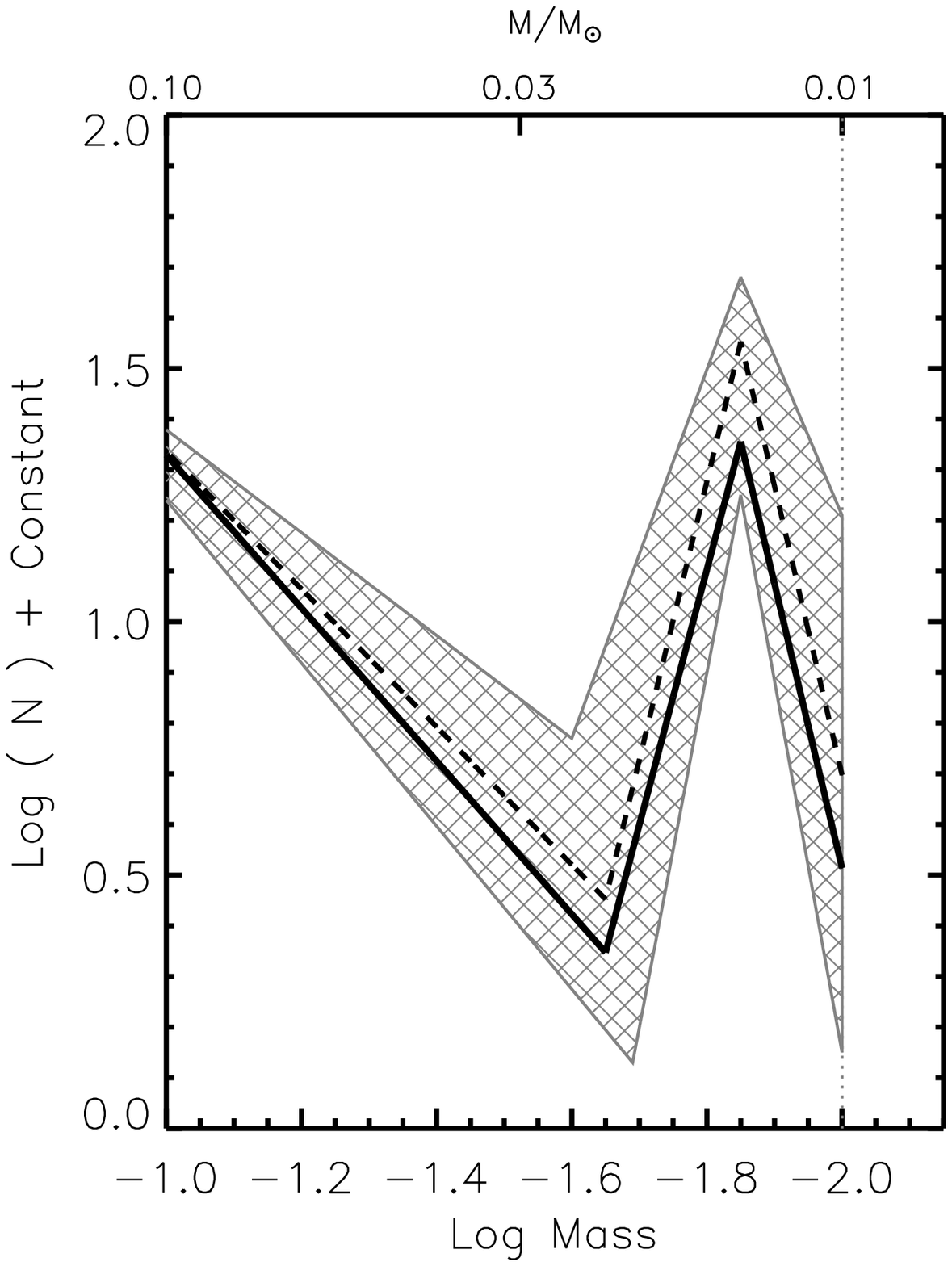]{The Trapezium Sub-Stellar IMF.
a) Best fit Model KLFs based on multi-power law IMFs are compared to the \mavlim~Trapezium KLF.
   The Trapezium Sub-Stellar \mavlim~KLF is shown with (solid histogram) and without
   (dashed histogram) correction for background field stars.
   Error bars are from counting statistics and the vertical dotted line demarks the 
   K=17.5 completeness limit.
   Model KLFs best fit to the $\mavlim$~KLF with (filled circles) and without (open circles)
   background correction are also shown, having error bars that corresponding to the
   $1\sigma$ variation of that KLF bin for 50 iterations of the model.
b) Derived Sub-Stellar Trapezium IMF using the B97 PMS tracks. Similar to the sub-stellar IMF
   derived using our standard PMS models (figure \ref{fig:imf_derived}b), the derived Trapezium
   brown dwarf IMF steadily declines, breaks and forms a significant secondary peak at
   0.014 - 0.013 \solarmass~before sharply declining below the deuterium burning limit.
   Our sample completeness limit (0.01\solarmass) is displayed as a vertical dotted line.
   The IMF fit range allowed by the KLF modeling is graphically displayed and are best
   fit 3 power-law IMFs from fits with and without background correction are shown
   (see table \ref{tab:ssimf_pars}).
\label{fig:bur97_fit}
}

\figcaption[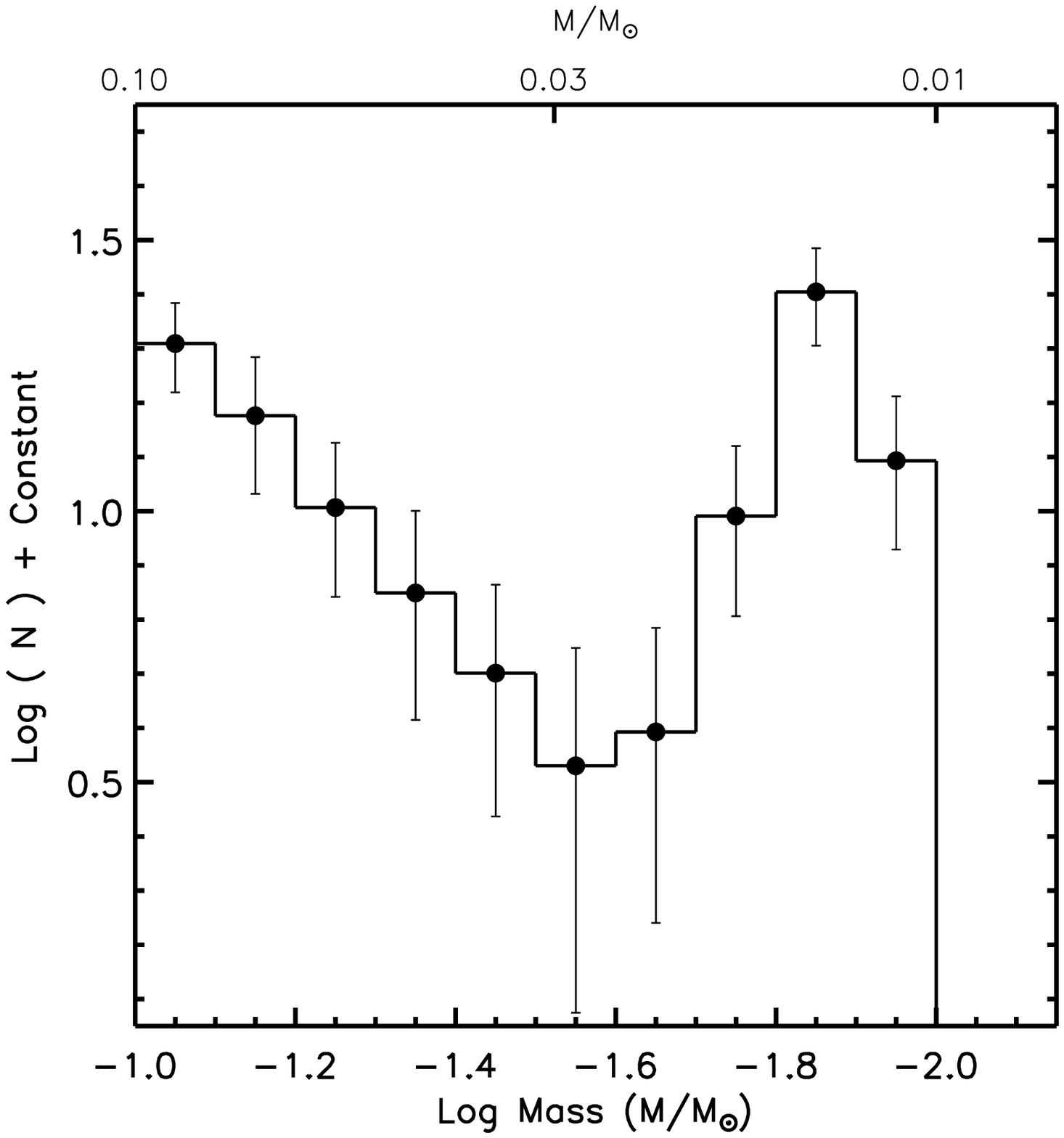]{A Monte Carlo simulation of the Trapezium Sub-Stellar IMF. Fifty (50) samples
of 150 brown dwarfs were drawn from the best fit Trapezium Sub-Stellar IMF derived from
B97 tracks, and each was binned into histogram form using equal sized log mass bins.
The plotted histogram represents the average of that bin for the iterations, and the
error bars represent the derived $1\sigma$ standard deviation of that IMF bin from
the Monte Carlo simulation.
\label{fig:ssimf}
}

\figcaption[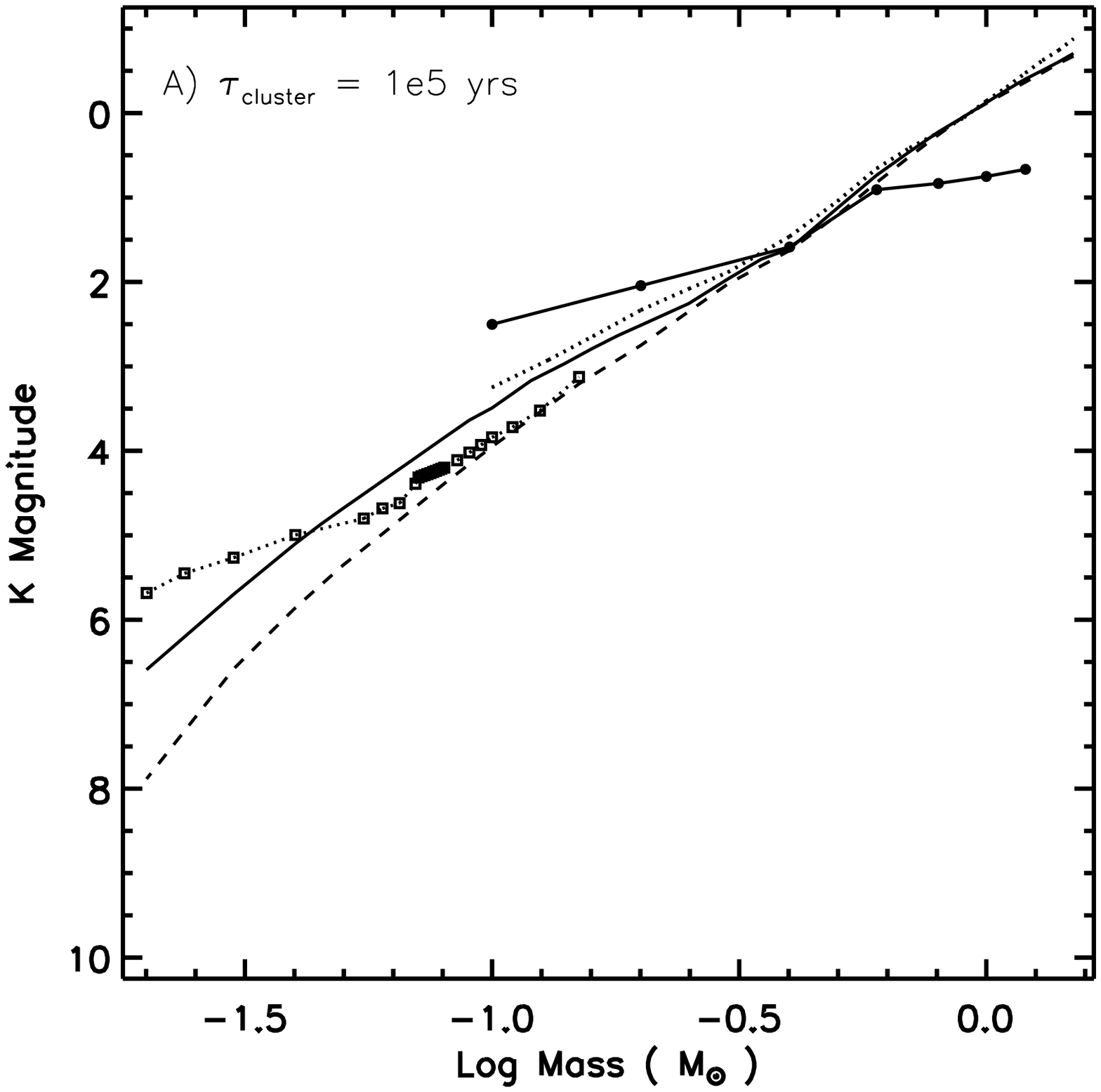,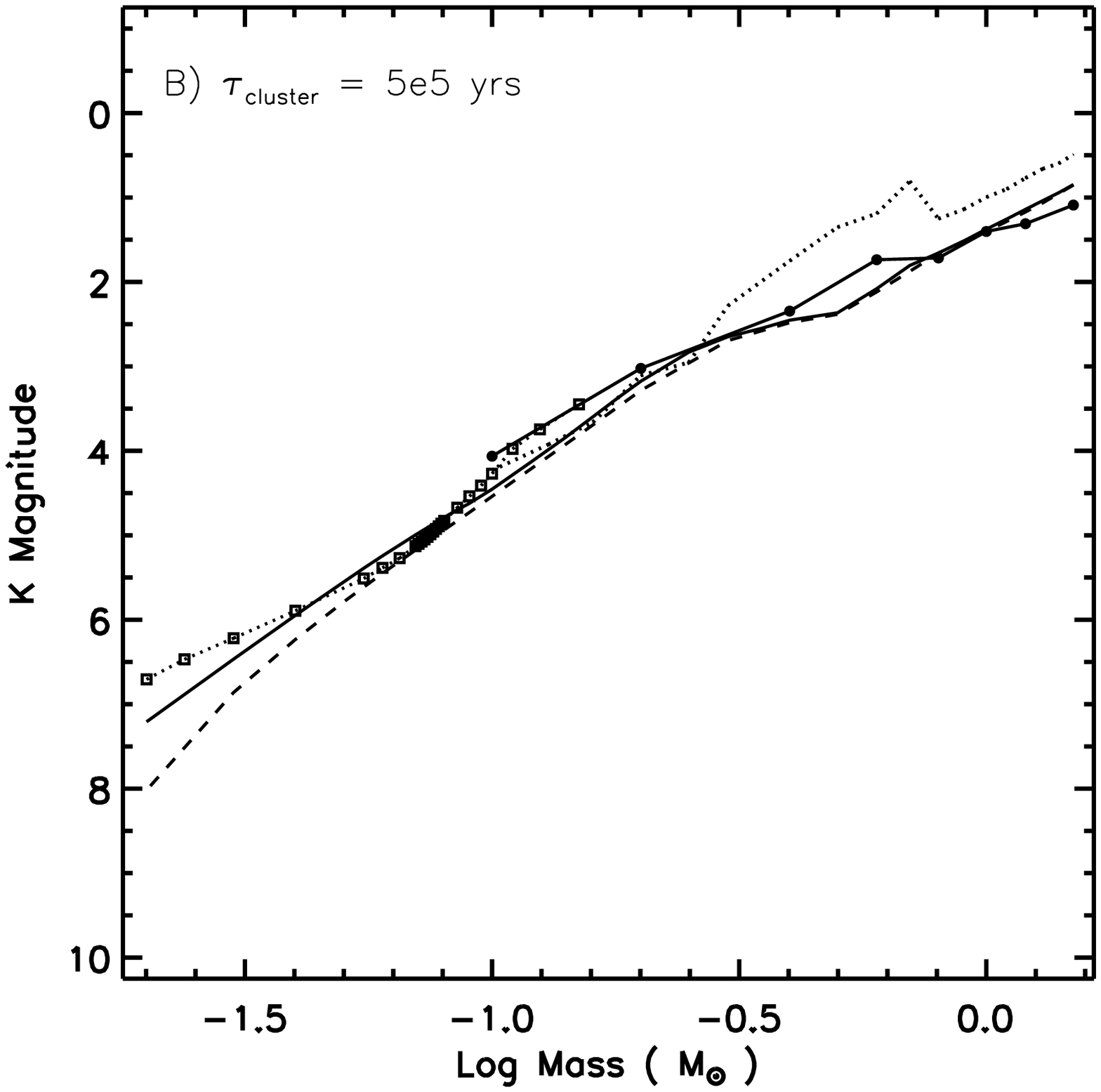,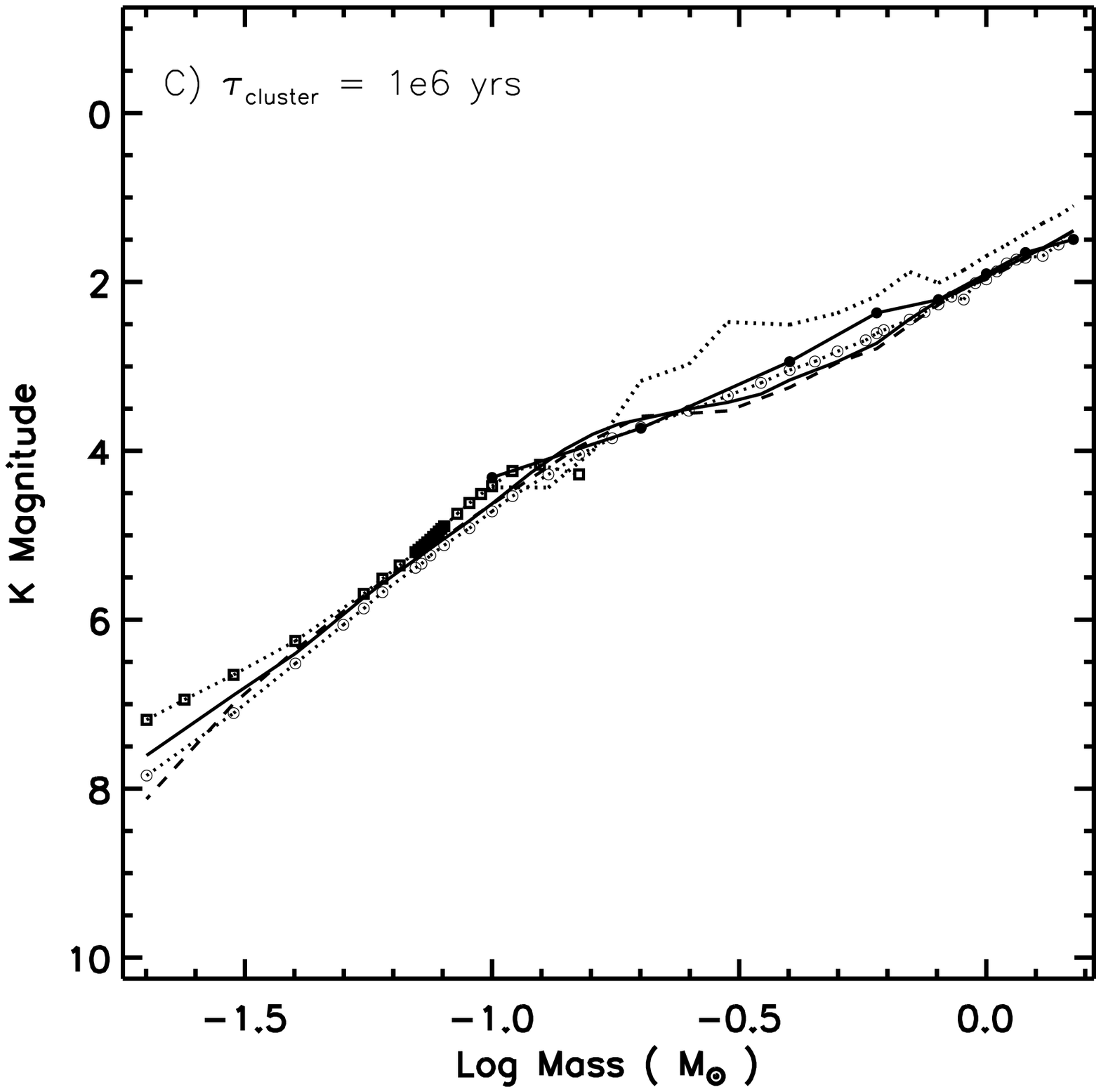,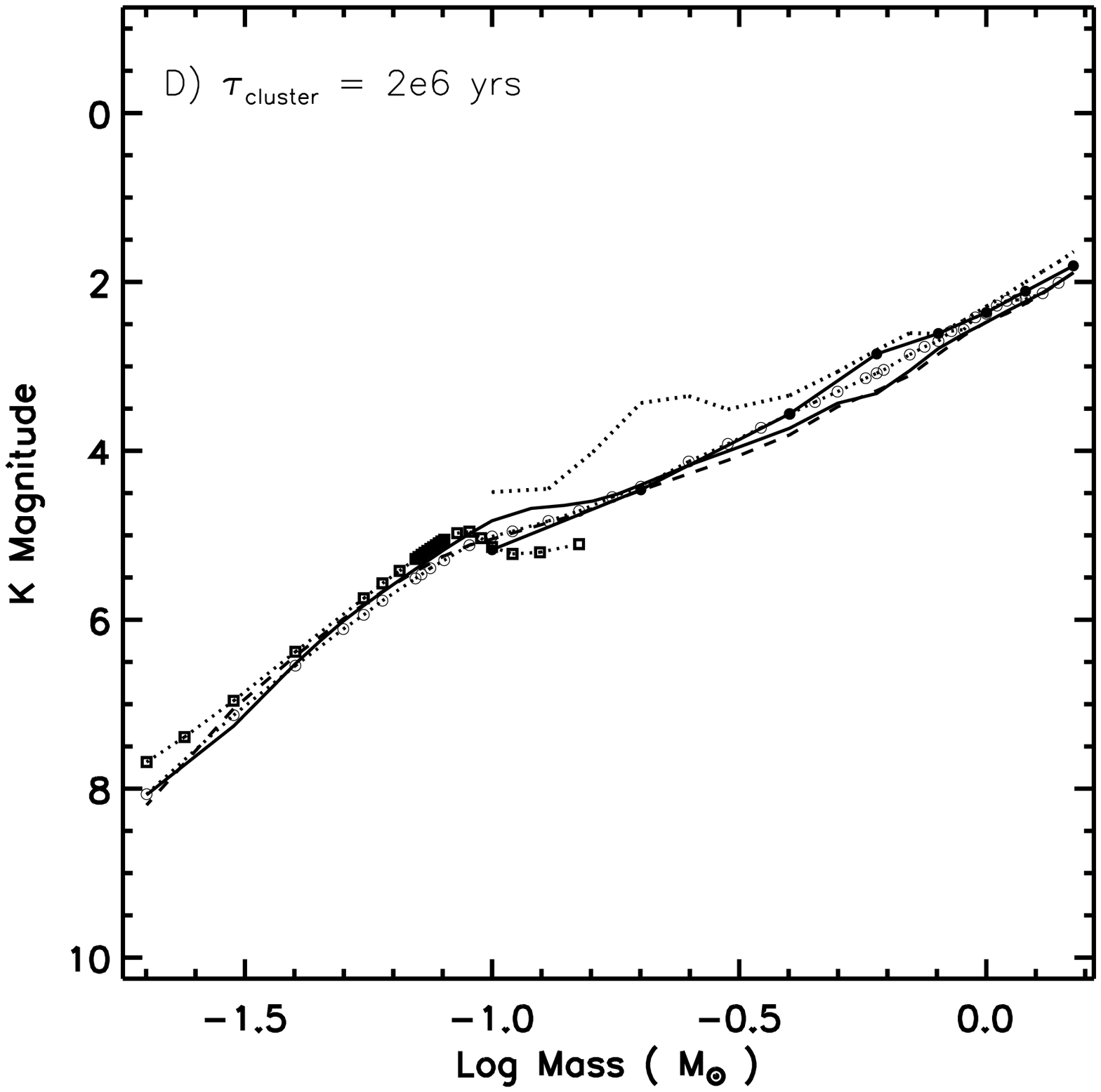,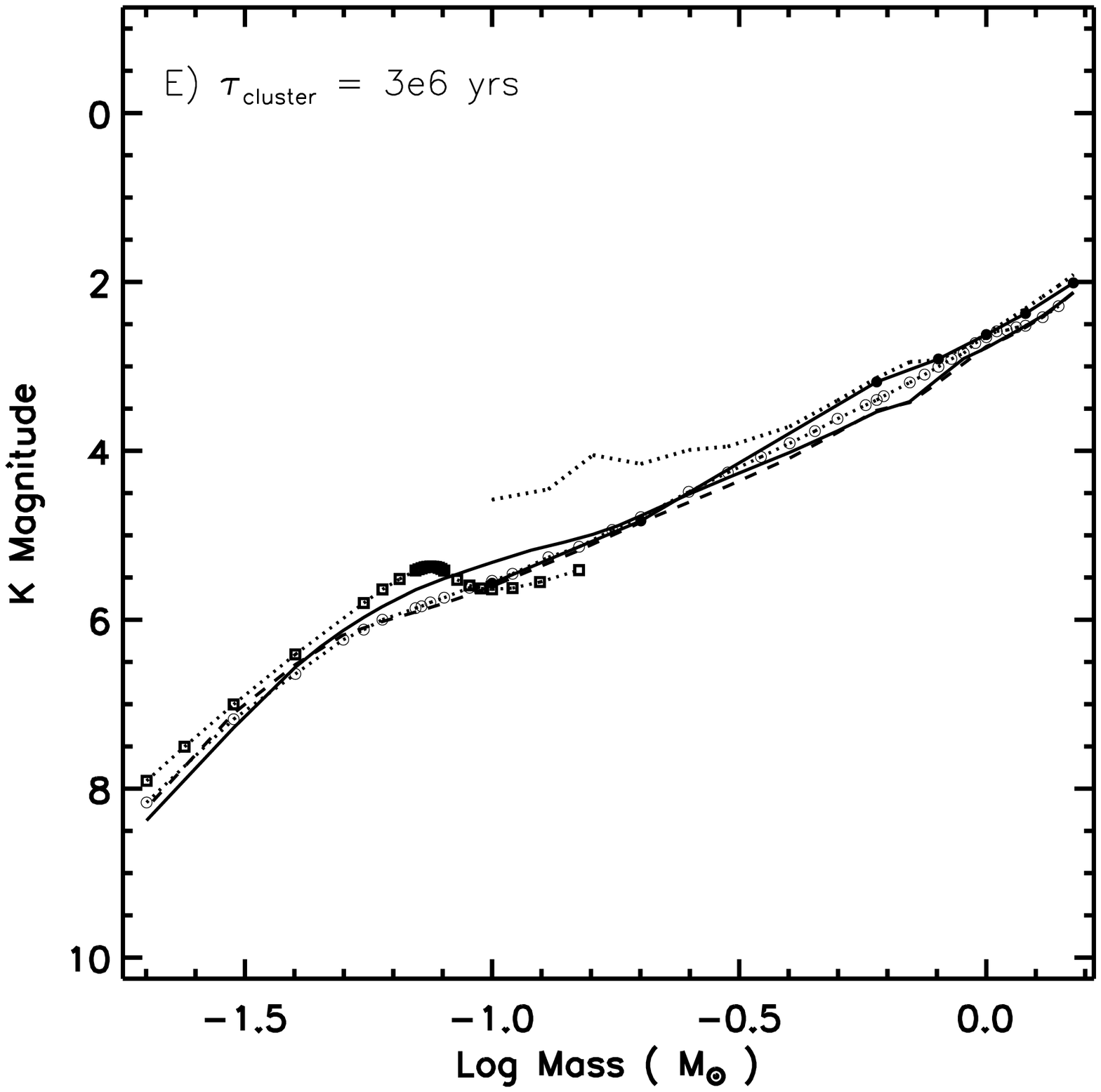,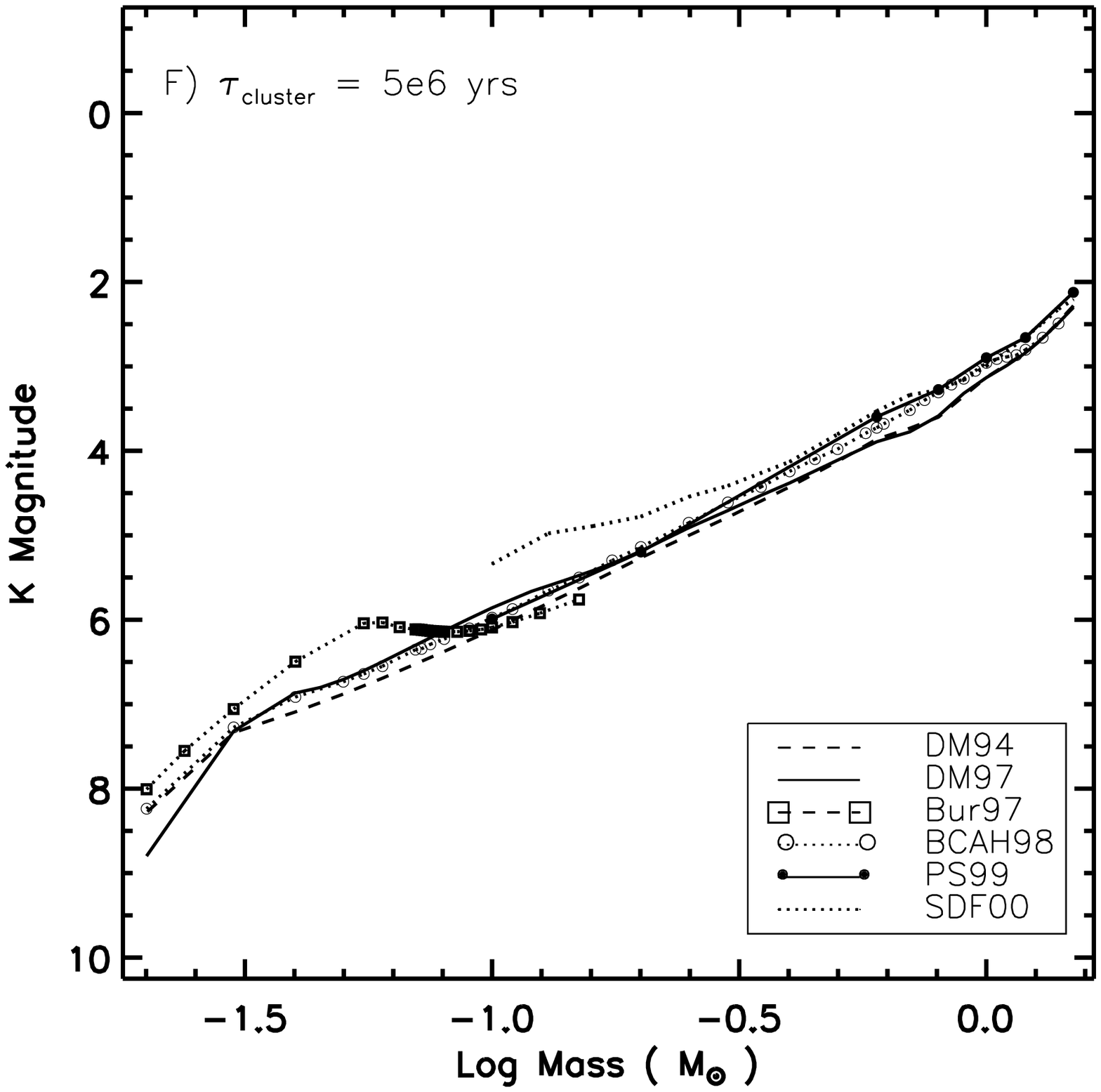]{
Comparison of theoretical mass-luminosity (infrared K magnitude)
relations.  Theoretical pre-main sequence mass-luminosity relations were extracted
from six different sets of evolutionary models (see table \ref{tab:ml_comp}).
In all cases the intrinsic model quantities (luminosity, effective temperature) were
converted to K magnitudes using a single $\mbox{T}_{eff}$-bolometric correction relationship.
Shown for 6 sets of cluster mean ages: 0.1 (a); 0.5 (b); 1 (c); 2 (d); 3 (e), and 5 (f) Myrs.
\citet{bcah98} models do not include models for ages less than 1 Myrs.
\label{fig:ml_comp}
}

\figcaption[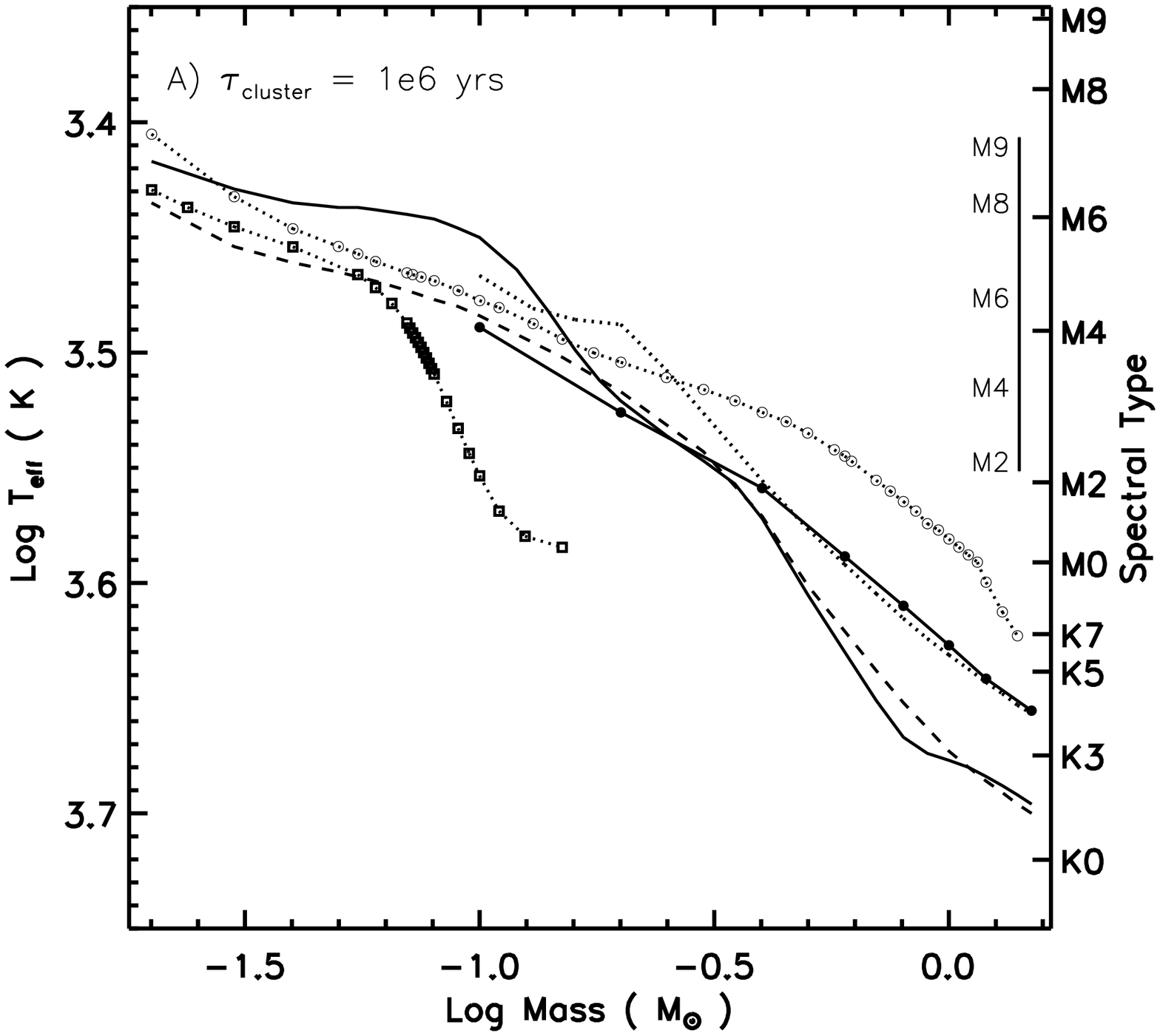,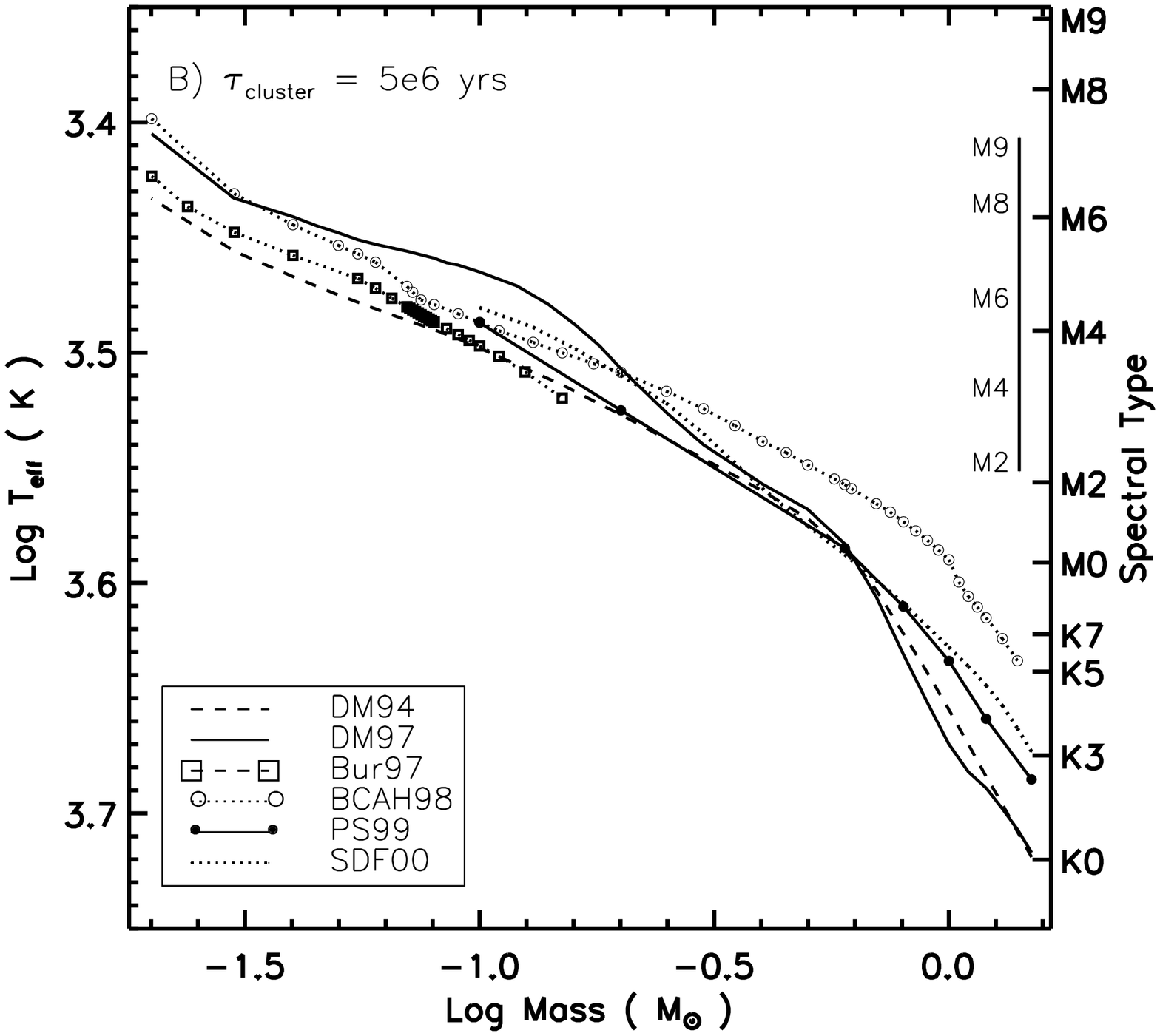]{Comparison of theoretical mass-$\mbox{T}_{eff}$/spectral type relations
\citep[figure adopted from][]{dant98}.
The spectral type - effective temperature - mass relationships were taken directly from the 6 sets
of PMS models at 1 (a) and 5 (b) Myrs.  Also shown is the gravity (dwarf vs sub-giant) dependence of
the spectral type to effective temperature calibration for late type PMS stars.
Because very young sub-solar mass stars and brown dwarfs are primarily on vertical Hayashi contraction
tracks in the HR diagram, there is theoretically a close correspondence between effective temperature and mass.
The effective temperature to spectral type scale (right hand y-axis) is a cool dwarf scale 
\citep[ summed from ][]{kh95,bes95,wgm99}. The inset spectral sequence is the hotter sub-giant
temperature-spectral scale tuned by \citet{luh99b}.
\label{fig:teff_comp}
}

\figcaption[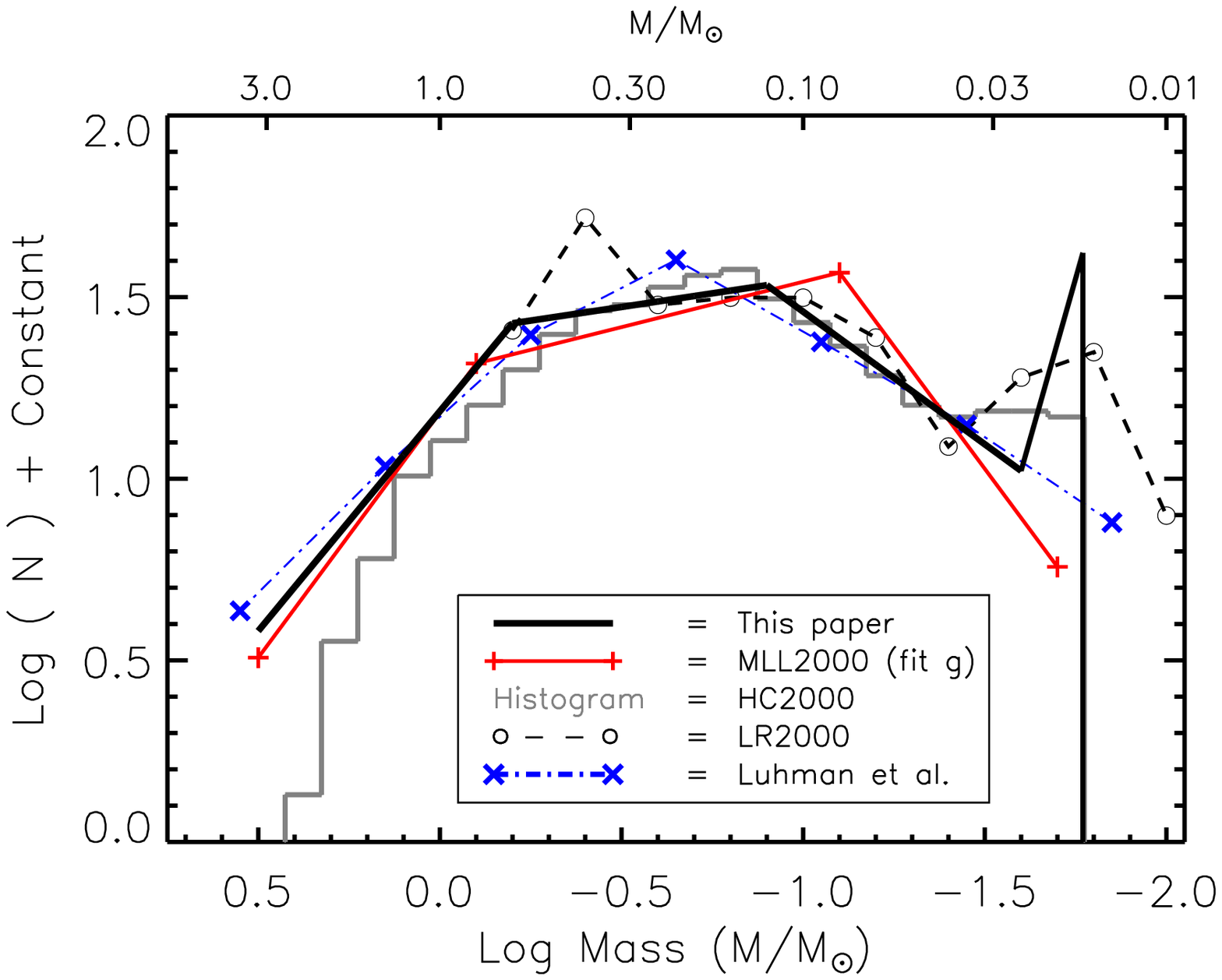]{Comparison of five published Trapezium IMFs derived from IR photometry.
   All derivations used the DM97 PMS tracks for masses less than 1 \solarmass. 
   Table \ref{tab:imf_comp} summarizes differences among the derivation methods.  
   The HC2000 Trapezium IMF corresponds to their A$_V\:<\:10$ limited sample.  
   The MLL2000 IMF is fit (g) from their table 2.
\label{fig:imf_comp}
}

\figcaption[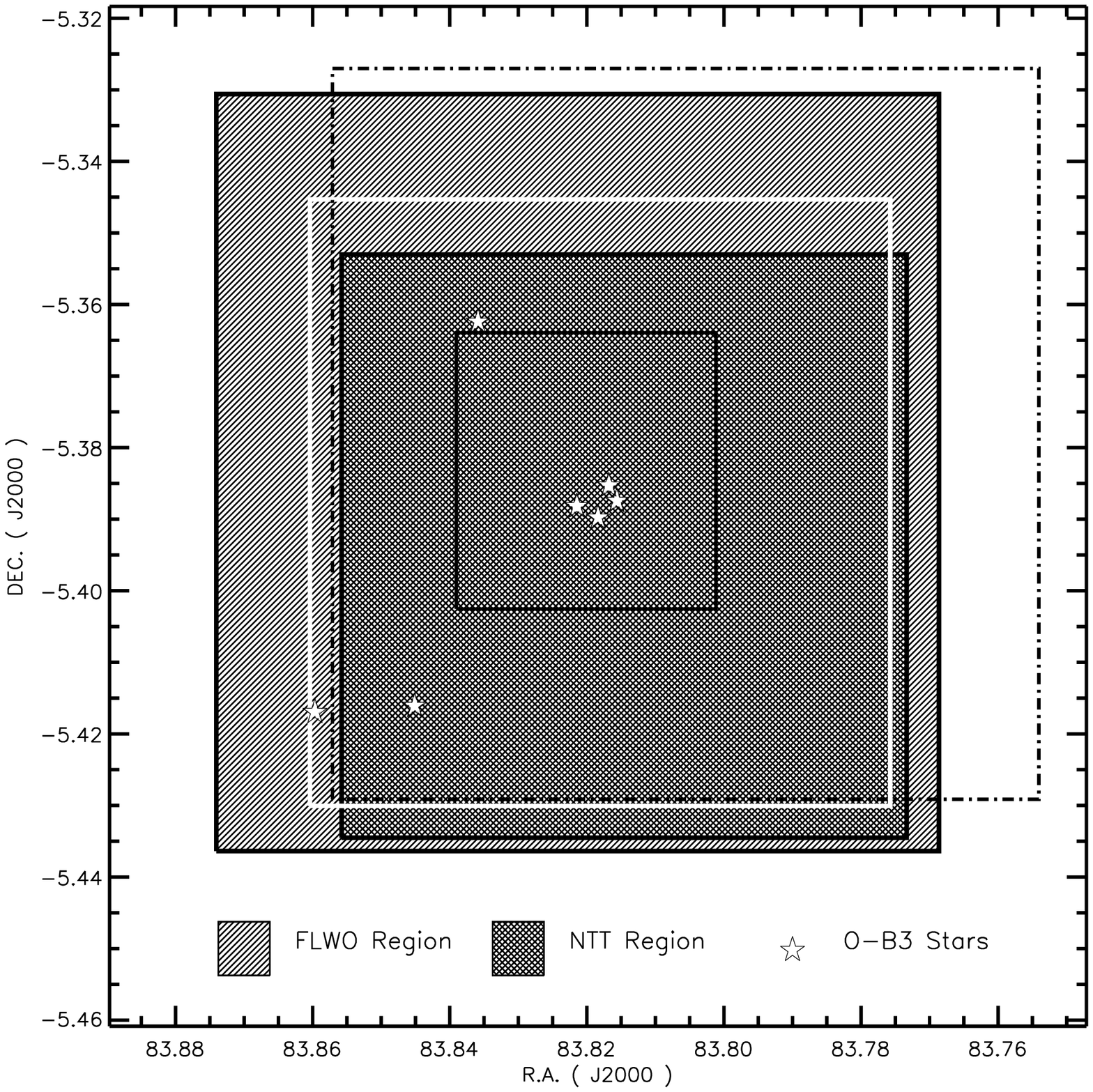]{
Comparison of the survey areas of recent deep infrared studies of the Trapezium Cluster.
The two shaded regions represent the $6\farcm5\,\times\,6\farcm5$ 
FLWO survey and the $5\arcmin\,\times\,5\arcmin$ NTT survey presented in this work.
Also shown are the NICMOS survey \citep[ solid black border]{luh2000}, the Keck survey 
\citep[solid white border]{hc2000},and the UKIRT survey \citep[ broken black border]{lr2000}.  
The locations of luminous cluster members, spectral types B3 and 
earlier, are shown as white stars.
\label{fig:trap_area}
}

\clearpage
\newpage

\pagestyle{myheadings}
\pagenumbering{arabic}
\setcounter{page}{1}

\clearpage
\newpage
\markright{Figure 1} 
\begin{figure}[h]
\epsscale{0.85}
\end{figure}

\clearpage
\newpage
\setcounter{page}{2}
\markright{Figure 2a} 
\begin{figure}[h]
\epsscale{1.00}
\plotone{figure02a.eps}
\end{figure}

\clearpage
\newpage
\setcounter{page}{2}
\markright{Figure 2b} 
\begin{figure}[h]
\plotone{figure02b.eps}
\end{figure}

\clearpage
\newpage
\setcounter{page}{3}
\markright{Figure 3a} 
\begin{figure}[h]
\plotone{figure03a.eps}
\end{figure}

\clearpage
\newpage
\setcounter{page}{3}
\markright{Figure 3b} 
\begin{figure}[h]
\plotone{figure03b.eps}
\end{figure}

\clearpage
\newpage
\setcounter{page}{4}
\markright{Figure 4a} 
\begin{figure}[h]
\plotone{figure04a.eps}
\end{figure}

\clearpage
\newpage
\setcounter{page}{4}
\markright{Figure 4b} 
\begin{figure}[h]
\plotone{figure04b.eps}
\end{figure}

\clearpage
\newpage
\setcounter{page}{5}
\markright{Figure 5} 
\begin{figure}[!ht]
\epsscale{0.85}
\plotone{figure05.eps}
\end{figure}

\clearpage
\newpage
\setcounter{page}{6}
\markright{Figure 6a} 
\begin{figure}[h]
\epsscale{1.00}
\plotone{figure06a.eps}
\end{figure}

\clearpage
\newpage
\setcounter{page}{6}
\markright{Figure 6b} 
\begin{figure}[h]
\plotone{figure06b.eps}
\end{figure}

\clearpage
\newpage
\setcounter{page}{7}
\markright{Figure 7a} 
\setcounter{page}{7}
\begin{figure}[h]
\plotone{figure07a.eps}
\end{figure}

\clearpage
\newpage
\setcounter{page}{7}
\markright{Figure 7b} 
\begin{figure}[h]
\plotone{figure07b.eps}
\end{figure}

\clearpage
\newpage
\setcounter{page}{8}
\markright{Figure 8} 
\begin{figure}[h]
\plotone{figure08.eps}
\end{figure}

\clearpage
\newpage
\setcounter{page}{9}
\markright{Figure 9} 
\begin{figure}[!ht]
\epsscale{0.85}
\plotone{figure09.eps}
\end{figure}

\clearpage
\newpage
\markright{Figure 10} 
\begin{figure}[h]
\epsscale{1.}
\plotone{figure10.eps}
\end{figure}

\clearpage
\newpage
\setcounter{page}{11}
\markright{Figure 11a}
\begin{figure}[h]
\epsscale{0.75}
\plotone{figure11a.eps}
\end{figure}

\clearpage
\newpage
\setcounter{page}{11}
\markright{Figure 11b}
\begin{figure}[h]
\epsscale{0.75}
\plotone{figure11b.eps}
\end{figure}

\clearpage
\newpage
\setcounter{page}{12}
\markright{Figure 12a} 
\begin{figure}[h]
\epsscale{0.70}
\plotone{figure12a.eps}
\end{figure}

\clearpage
\newpage
\thispagestyle{myheadings}
\setcounter{page}{12}
\markright{Figure 12b} 
\begin{figure}[h]
\epsscale{0.70}
\plotone{figure12b.eps}
\end{figure}
\clearpage
\pagebreak

\clearpage
\newpage
\setcounter{page}{13}
\thispagestyle{myheadings}
\markright{Figure 13} 
\begin{figure}[h]
\epsscale{0.70}
\plotone{figure13.eps}
\end{figure}
\clearpage
\pagebreak

\clearpage
\newpage
\setcounter{page}{14}
\thispagestyle{myheadings}
\markright{Figure 14a} 
\begin{figure}[h]
\epsscale{1.00}
\plotone{figure14a.eps}
\end{figure}

\clearpage
\newpage
\setcounter{page}{14}
\markright{Figure 14b} 
\begin{figure}[h]
\epsscale{1.00}
\plotone{figure14b.eps}
\end{figure}

\clearpage
\newpage
\setcounter{page}{14}
\markright{Figure 14c} 
\begin{figure}[h]
\plotone{figure14c.eps}
\end{figure}

\clearpage
\newpage
\setcounter{page}{14}
\markright{Figure 14d} 
\begin{figure}[h]
\plotone{figure14d.eps}
\end{figure}

\clearpage
\newpage
\setcounter{page}{14}
\markright{Figure 14e} 
\begin{figure}[h]
\plotone{figure14e.eps}
\end{figure}

\clearpage
\newpage
\setcounter{page}{14}
\markright{Figure 14f} 
\begin{figure}[h]
\plotone{figure14f.eps}
\end{figure}

\clearpage
\newpage
\setcounter{page}{15}
\markright{Figure 15a} 
\begin{figure}[h]
\plotone{figure15a.eps}
\end{figure}

\clearpage
\newpage
\setcounter{page}{15}
\markright{Figure 15b} 
\begin{figure}[h]
\plotone{figure15b.eps}
\end{figure}

\clearpage
\newpage
\markright{Figure 16} 
\begin{figure}[h]
\plotone{figure16.eps}
\end{figure}

\clearpage
\newpage
\markright{Figure A-1} 
\begin{figure}[h]
\plotone{figure17.eps}
\end{figure}

\end{document}